\documentclass[aps,pra,reprint,superscriptaddress]{revtex4-2}

\usepackage[english]{babel}
\usepackage{graphics}
\usepackage{array}
\usepackage{graphicx}
\usepackage{amsfonts,amssymb,amsmath,braket,xcolor}

\usepackage{color}
\usepackage{todonotes}
\usepackage{threeparttable}
\usepackage{longtable}
\usepackage{nicematrix}
\usepackage{multirow}
\usepackage[separate-uncertainty,per-mode=fraction]{siunitx}
\usepackage{booktabs}

\definecolor{darkblue}{rgb}{0.0,0.0,0.4}
\definecolor{darkgreen}{rgb}{0.0,0.4,0.0}
\definecolor{darkred}{rgb}{0.6,0.0,0.0}
\usepackage[bookmarksopen]{hyperref}
\hypersetup{colorlinks,linkcolor=darkblue,citecolor=darkblue,urlcolor=darkblue}

\newcommand{\BaFfour}{{${}^{134}$BaF}} 
\newcommand{\BaFfive}{{${}^{135}$BaF}} 
\newcommand{\BaFsix}{{${}^{136}$BaF}} 
\newcommand{\BaFseven}{{${}^{137}$BaF}} 
\newcommand{\BaFeight}{{${}^{138}$BaF}}

\newcommand{\gs}{$\mathrm{X}^2\Sigma^+$}
\newcommand{\exs}{$\mathrm{A}^2\Pi_{1/2}$}
\newcommand{\exss}{$\mathrm{A}^2\Pi_{3/2}$}

\newcolumntype{M}[1]{>{\centering\arraybackslash}m{#1}}

\usepackage{titlesec}

\titleformat{\paragraph}[block]{\normalfont\normalsize\bfseries}{\theparagraph}{1em}{}
\titlespacing*{\paragraph}{0pt}{1ex plus 0.5ex minus .2ex}{0.8ex plus .2ex}

\newcommand{\sixjsymbol}[6]{
\begin{Bmatrix}
#1 & #2 & #3 \\
#4 & #5 & #6
\end{Bmatrix}
}

\newcommand{\threejsymbol}[6]{
\begin{pmatrix}
#1 & #2 & #3 \\
#4 & #5 & #6
\end{pmatrix}
}

\begin{document}
\title{High-resolution spectroscopy of barium monofluoride:\\Odd isotopologues, hyperfine structure and isotope shifts}
\author{Felix Kogel}
\affiliation{5. Physikalisches  Institut  and  Center  for  Integrated  Quantum  Science  and  Technology,Universit\"at  Stuttgart,  Pfaffenwaldring  57,  70569  Stuttgart,  Germany}

\author{Yuly Chamorro}

\affiliation{University of Groningen, 9747AG Groningen, The Netherlands}

\author{Mangesh Bhattarai}
\affiliation{Department of Physics, University of Chicago, Chicago, Illinois 60637, USA}
\affiliation{Physics Division, Argonne National Laboratory, Lemont, IL 60439, USA}

\author{Marian Rockenh\"auser} 
\affiliation{5. Physikalisches  Institut  and  Center  for  Integrated  Quantum  Science  and  Technology,Universit\"at  Stuttgart,  Pfaffenwaldring  57,  70569  Stuttgart,  Germany}

\affiliation{Vienna Center for Quantum Science and Technology,
Atominstitut, TU Wien, Stadionallee 2, 1020 Vienna, Austria}

\author{Tatsam Garg} 
\affiliation{5. Physikalisches  Institut  and  Center  for  Integrated  Quantum  Science  and  Technology,Universit\"at  Stuttgart,  Pfaffenwaldring  57,  70569  Stuttgart,  Germany}

\affiliation{Vienna Center for Quantum Science and Technology,
Atominstitut, TU Wien, Stadionallee 2, 1020 Vienna, Austria}

\author{David DeMille}
\affiliation{Department of Physics and Astronomy, Johns Hopkins University, Baltimore, MD 21218, USA }
\affiliation{Department of Physics, University of Chicago, Chicago, Illinois 60637, USA}
\affiliation{Physics Division, Argonne National Laboratory, Lemont, IL 60439, USA}

\author{Anastasia Borschevsky}

\affiliation{University of Groningen, 9747AG Groningen, The Netherlands}

\author{Tim Langen}
\email{tim.langen@tuwien.ac.at}

\affiliation{5. Physikalisches  Institut  and  Center  for  Integrated  Quantum  Science  and  Technology,Universit\"at  Stuttgart,  Pfaffenwaldring  57,  70569  Stuttgart,  Germany}

\affiliation{Vienna Center for Quantum Science and Technology,
Atominstitut, TU Wien, Stadionallee 2, 1020 Vienna, Austria}

\begin{abstract} 
Barium monofluoride (BaF) is a promising molecular species for precision tests of fundamental symmetries and interactions. We present a combined theoretical and experimental study of BaF spectra and isotope shifts, focusing in particular on the poorly understood odd isotopologues \BaFseven\ and \BaFfive. By comparing state-of-the-art ab initio calculations with high-resolution fluorescence and absorption spectroscopy data, we provide a benchmark for electronic structure theory and disentangle the hyperfine and rovibrational spectra of the five most abundant isotopologues, from \BaFeight\ to \BaFfour. The comprehensive knowledge gained enables a King plot analysis of the isotope shifts that reveals the odd-even staggering of the barium nuclear charge radii. It also paths the way for improved laser cooling of rare BaF isotopologues and crucially supports future measurements of nuclear anapole and Schiff moments.
\end{abstract}

\maketitle

\section{Introduction}
Molecules have recently been established as powerful precision sensors for the behavior of fundamental particles and interactions~\cite{Safronova2018,DeMille2023}. Their internal structure makes them highly sensitive to the effects of a hypothetical electron electric dipole moment~\cite{Hudson2002,ACME2018,Roussy2023}, nuclear Schiff and magnetic quadrupole moments~\cite{Grasdijk2021,Flambaum2014} and the variation of fundamental constants~\cite{Shelkovnikov2008, Bethlem2009,Truppe2013, Schiller2014,Kobayashi2019,Leung2023}. Moreover, their spectra can be sensitive to the shape and size of the nuclear charge and magnetization distributions~\cite{Athanasakis2023, Wilkins2023,Udrescu2024}, and molecules containing nuclei with unpaired nuclear spins can be used to probe nuclear-spin-dependent parity violation, anapole or Schiff moments~\cite{Demille2008,Cahn2014,Norrgard2019}. 

It is also highly desirable to further increase the sensitivity of such measurements using laser cooling~\cite{Lim2018,Fitch2020methods,Kozyryev2017,Aggarwal2018} and, recently, progress in this direction has been made for many molecular species~\cite{Fitch2021,Langen2023}. Notably, the detailed spectroscopic information required to enable laser cooling is closely linked to the knowledge needed to carry out precision measurements for the fundamental physics goals. In turn, successful laser cooling can dramatically enhance spectroscopic sensitivity by increasing interrogation times and reducing thermal broadening.

The list of species that are both sensitive to fundamental physical effects of interest and suitable for laser cooling includes barium monofluoride (BaF), where high-resolution spectroscopy~\cite{Rockenhaeuser2023} has enabled the realization of optical cycling~\cite{Albrecht2020,Chen2017}, low noise detection~\cite{Rockenhaeuser2023}, Sisyphus laser cooling~\cite{Rockenhaeuser2024,Kogel2024serrodynes} and magneto-optical trapping of the most abundant isotopologue \BaFeight~\cite{Zeng2024}, as well as laser cooling of the rarer isotopologues \BaFsix~\cite{Kogel2024isotope} and \BaFseven~\cite{Kogel2025lasercooling137}. BaF distinguishes itself by
the existence of a total of five isotopologues with abundance greater than $1\%$, which show significant enhancements for searches for the electron electric dipole moment~\cite{Aggarwal2018,Boeschoten2024} and parity violation~\cite{Demille2008,Altuntas2018}. The barium nuclei in certain unstable isotopologues are also expected to exhibit significant octupole deformations~\cite{Phillips1986}, which may enhance sensitivity to Schiff moments arising from charge-parity violating hadronic physics~\cite{ArrowsmithKron2024}. These observations collectively establish BaF as a compelling platform for exploring a wide range of fundamental phenomena via spectroscopy.

However, while the most abundant, even isotopologues of BaF have previously been studied extensively~\cite{Rockenhaeuser2023,Bu2022,Ryzlewicz1980,Effantin1990,Ernst1986}, precise knowledge about the odd---or fermionic---isotopologues of BaF, and in particular their hyperfine structure, has remained scarce~\cite{Steimle2011,Preston2025}, rendering the interpretation of spectra and assignments of lines of all isotopologues, including even ones with low abundance, challenging. This missing knowledge has, for example, limited achievable laser cooling forces for \BaFseven~\cite{Kogel2025lasercooling137, Kogel2024serrodynes} and leaves the transition frequencies of short-lived isotopologues poorly constrained and uncertain.

Here, we systematically study the chain of stable BaF isotopologues from \BaFfour\, to \BaFeight\, using both fluorescence and absorption spectroscopy. We support our experimental results obtained in two independent and complementary experimental setups with state-of-the-art ab initio calculations of the excited-state hyperfine structure of the odd isotopologues, allowing us to benchmark such calculations and to reliably assign a large number of transitions. This enables us to systematically explore the isotope shifts in the BaF molecule with high precision using a King plot analysis. Our observations reveal the odd-even staggering of the nuclear charge radius of the barium atom, as we move from isotopes with odd numbers of neutrons to isotopes with even numbers of neutrons. Moreover, the extracted information provides important spectroscopic input for laser cooling and efficient detection of the odd isotopologues of barium monofluoride~\cite{Kogel2021}, which will be important ingredients for precision studies of nuclear spin-dependent parity violation and Schiff moments~\cite{Altuntas2018,AltuntasPRA}.

\begin{figure}[tb]
    \centering
    \includegraphics[width=0.91\columnwidth]{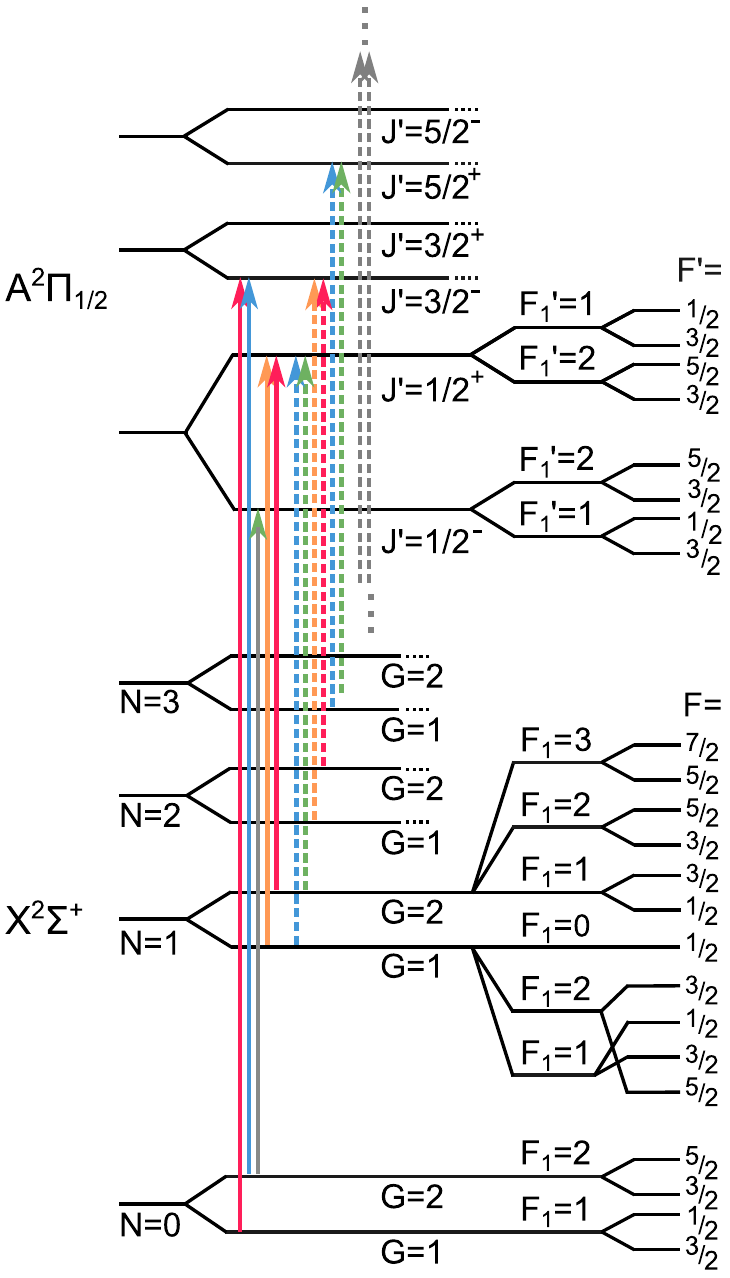}
    \caption{Level structure of the 
    \gs$(\nu)\rightarrow$\exs$(\nu')$ transition in the odd isotopologues of BaF. Here, $N$ denotes the rotational, $J$ the total angular momentum without nuclear spins, and $G$, $F_1$ and $F$ additional intermediate and hyperfine angular momentum quantum numbers described in the text. Solid arrows indicate
    transitions probed by fluorescence spectroscopy, which resolves the full hyperfine structure, as shown in Fig.~\ref{fig:fluorescencespectroscopy}. Dashed arrows indicate transitions investigated via absorption spectroscopy, as shown in Fig.~\ref{fig:absorptionspectroscopy}, with the dash-dotted arrows summarizing any higher transitions probed. Absorption spectroscopy is performed for transitions between different vibrational quantum numbers $\nu$ and $\nu'$, which are of important interest as vibrational repumpers in laser cooling. Dots indicate substructures not shown, energy spacings are not to scale, and colors are used for visual clarity. A version of this figure containing all energy spacings extracted in this work can be found in the appendix.}
\label{fig:levelstructure}
\end{figure}

\section{Level structure of the BaF molecule}
The level structure of even isotopologues of BaF, which feature only the nuclear spin of the fluorine atom $I^{(F)}=1/2$, is similar to that of other alkaline-earth monofluorides used in laser cooling experiments~\cite{Fitch2021} and has been described in detail in previous work~\cite{Rockenhaeuser2023,Steimle2011}. 

In contrast to this, the odd isotopologues, including both \BaFfive\, and \BaFseven\, that will be studied in detail in the following, feature two nuclear spins and thus exhibit a complex hyperfine structure. A summary of the level structure of the lowest rotational levels in the $X ^2\Sigma^{+}$ state and the $A ^2\Pi_{1/2}$ state, is given in Fig.~\ref{fig:levelstructure}. The full Hamiltonian and matrix elements describing these states are summarized in the appendix.  

In short, in both \BaFseven\, and 
\BaFfive, the angular momentum structure of the odd isotopologues is shaped by the nuclear spin $I^{(\mathrm{Ba})} = 3/2$ of the respective barium nuclei. The ground state \gs\, is described by Hund's case $\left(b_{\beta S}\right)$~\cite{Steimle2011}, where the electron spin $\mathbf{S}$ strongly couples with the barium nuclear spin $\mathbf{I}^{(\mathrm{Ba})}$ instead of the rotational angular momentum $\mathbf{N}$, reflected in the molecular constants $b_F \gg N\times\gamma$. Here, $b_F$ and $\gamma$ describe the Fermi contact interaction and the coupling strength between electron spin and rotation, respectively, and $N$ is the rotational quantum number associated with $\mathbf{N}$. This coupling leads to an intermediate angular momentum $\mathbf{G} = \mathbf{S} + \mathbf{I}^{(\mathrm{Ba})}$ creating two well-separated manifolds for each rotational level, characterized by the quantum numbers $G = 1$ and $G = 2$.
These $G$ manifolds are split by energies comparable to rotations, and interact with these rotations $\mathbf{N}$ to produce another intermediate angular momentum $\mathbf{F}_1 = \mathbf{N} + \mathbf{G}$. The fluorine nuclear spin contributes a second, weaker hyperfine interaction. This weak coupling between $\mathbf{F}_1$ and $\mathbf{I}^{(\mathrm{F})}$ is described by the total angular momentum $\mathbf{F} = \mathbf{F}_1 + \mathbf{I}^{(\mathrm{F})}$. These interactions collectively produce a dense hyperfine level structure, as shown in Fig.~\ref{fig:levelstructure}. 

The excited state \exs\ is described by Hund's case $\left(a_{\beta J}\right)$~\cite{Steimle2011}. Here, the levels of the rotational ladder, originating from the total angular momentum $\mathbf{J}$ without nuclear spins, split into $\Lambda$-doublets $J^P$ with opposite well-defined parity $P=\pm$. These states couple to $\mathbf{I}^{(\mathrm{Ba})}$, forming an intermediate angular momentum $\mathbf{F}_1$. Subsequently, $\mathbf{F}_1$ interacts with $\mathbf{I}^{(\mathrm{F})}$, resulting in the total angular momentum $\mathbf{F} = \mathbf{F}_1 + \mathbf{I}^{(\mathrm{F})}$. The magnetic hyperfine interaction for the \exs\ state is expressed with the spectroscopic constants $d$ and $h_{1/2}=a- b_F/2 -c/3$, where $d$ corresponds to an off-diagonal correction dominated by the $\Lambda$-doubling, and $a$ and $c$ describe the magnetic dipole coupling of the nuclear magnetic moment to spin and orbital momentum of the electron. Additionally, the electrostatic hyperfine interaction between the nuclear quadrupole moment and the electric field gradient is reflected in the electric quadrupole parameter $eq_0Q$. Although hyperfine splitting due to $\mathbf{I}^{(\mathrm{F})}$ is relatively weak in the excited state, it has recently been shown to be resolved in the even isotopologues~\cite{Rockenhaeuser2023, Denis2022}, while the hyperfine structure due to $\mathbf{I}^{(\mathrm{Ba})}$ remains largely unexplored, in particular for excited states. 

\begin{figure*}[tb]
    \centering
    \includegraphics[width=\textwidth]{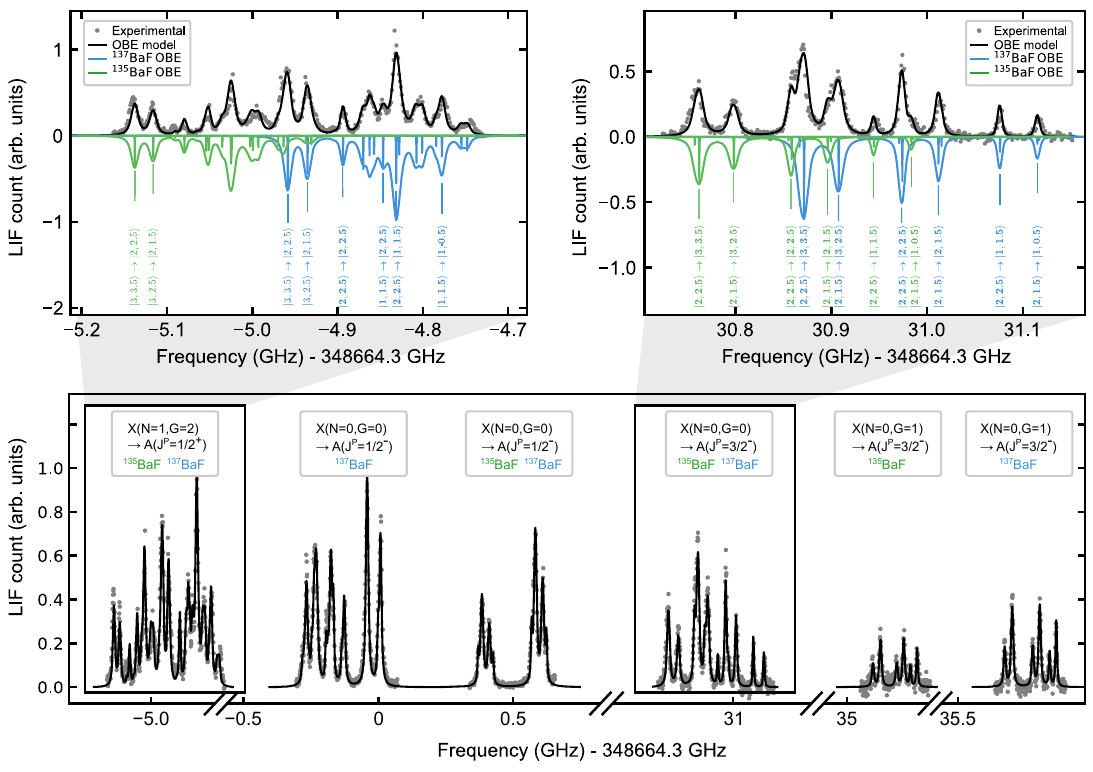}
    \caption{Fluorescence spectroscopy. Experimental (gray points) and modeled (black curve) spectra spanning all measured transitions arising from different $N$ and $G$ levels in the \BaFseven\, and \BaFfive\, \gs\, ground states, and $J^P$ levels in the \exs\, states. Each experimental data point in a spectrum scan is typically averaged over $10$ repetitions, each taken at a different spots on the ablation target to minimize drifts in molecular signal. With $\approx\! 100~\mu$W of incident laser light, signal strengths of $100-1400\,$counts/pulse are observed, on a background of about $280$/pulse from scattered laser light. Backgrounds from scattered laser light are determined from signals outside the time window of the molecular pulse and subtracted. The insets show zoomed in example scans of the $X(N=1,G=2)\rightarrow A(J^p=1/2^+)$ and $X(N=0,G=2)\rightarrow A(J^p=3/2^-)$ transitions, the former being one of the transitions relevant for laser cooling~\cite{Kogel2025lasercooling137}. These plots also show two types of simulated spectra.  The first is a "stick plot", showing narrow lines with heights set by the square of the associated transition dipole matrix element.   The second is a complete model of the expected spectrum that uses the optical Bloch equations to determine the height and power-broadened width of each line (see appendix for details). In the simulated spectra, relative line heights are scaled by the natural abundance of the respective isotopologue, and line positions and dipole matrix elements are set by the final parameters determined in this work. Hyperfine Hamiltonian parameters were determined, for each isotope separately, from splittings between pairs of lines within each continuous spectrum. Separations between the hyperfine centers-of-gravity for the two isotopes within each continuous spectrum, as well as separations between the disconnected black traces, were varied to minimize the overall goodness of fit. The resulting separations were consistent, within uncertainties, with expectations based on non-hyperfine parameters extracted from the absorption spectroscopy reported in Fig.~\ref{fig:absorptionspectroscopy}.}
    \label{fig:fluorescencespectroscopy}
\end{figure*}

\section{Fluorescence spectroscopy}

To explore this hyperfine structure, we perform laser-induced fluorescence spectroscopy on a cryogenic buffer gas-cooled molecular beam \cite{Hutzler2011,Hutzler2012,Bu2017,Albrecht2020} of BaF. We use a neon buffer gas at a temperature of $15\,$K.  Following laser ablation of a Ba metal target in the buffer gas cell, BaF molecules are formed via chemical reaction with a small admixture of SF$_6$ to the buffer gas.  The source is operated at a repetition rate of $10\,$Hz  and produces beams with a mean forward velocity of $180$\,m/s. Expansion of the buffer gas at the source exit cools the molecules to a rotational temperature of approximately $4$K. 

Spectroscopy is carried out approximately $60\,$cm downstream from the source, where a frequency-tunable laser beam with a wavelength of around $860\,$nm---stabilized via a transfer cavity lock referenced to a stabilized HeNe laser~\cite{Zhao1998, Gomez2004}---intersects the molecular beam orthogonally. We calibrate the non-linearity of the piezo scan of the transfer cavity using a second laser beam, offset by a known radio-frequency interval relative to the primary beam using an acousto-optic modulator.  This yields an uncertainty in the relative laser frequency, within a single frequency scan, of $0.3$ MHz. The frequency separation between different scans is known to $\pm 100$ MHz, from a commercial wavemeter. Molecular fluorescence at the same wavelength as the excitation light is collected in a photomultiplier using mirrors and a light pipe \cite{DaveRahmlow,Shimizu1983} and the resulting signals are recorded using standard photon-counting methods. 

From prior studies \cite{Ryzlewicz1982,Preston2025,Steimle2011} and the absorption spectroscopy and theory calculations reported below, the electronic and rotational energies, isotope shifts, and ground state hyperfine structure were sufficiently well known to determine the ground state quantum numbers $N$ and $G$, and excited state quantum numbers $J'^{P}$, for each of the transitions observed. Specifically, our analysis covered several groups of transitions within the $X(N=0)\to A(J'^{P} = 1/2^{-})$, $X(N=0)\to A(J'^{P} = 3/2^{-})$, and $X(N=1)\to A(J'^{P} = 1/2^{+})$ manifolds. Example results are summarized in Fig.~\ref{fig:fluorescencespectroscopy}. Each spectrum spanned a laser frequency range of $350-400\,$MHz, which was scanned in steps of $1\,$MHz. During this procedure the laser frequency was alternately ramped up or down to average over effects of drift or hysteresis. Each of the transitions observed was carefully assigned and fitted, as detailed in the appendix. We note that in an earlier paper that analyzed the rotational and hyperfine structure of the \exs\, state in \BaFseven, and \BaFfive~\cite{Steimle2011}, only higher-lying rotational levels were analyzed, and the Hamiltonian employed there did not include the quadrupole contribution, the contributions with coefficient $h_{1/2}$ for either Ba or F, or the contributions with coefficient $d$ for F.  In our analysis of the hyperfine structure, the values of $d^{(F)}$ and $h_{1/2}^{(F)}$ were fixed to those determined in a recent study of the structure of the \exs\, levels in $^{138}$BaF \cite{Denis2022}, and the corresponding values $d^{(Ba)}$ and $h_{1/2}^{(Ba)}$ were determined for the two different odd isotopologues \BaFfive\, and \BaFseven. The values and associated uncertainties determined for the parameters describing the \exs\ hyperfine structure are summarized in Table~\ref{table:HFSparameters}.

Since the fits were performed independently for the two isotopologues, it is useful to check the consistency of the extracted molecular parameters against the expected scaling of the intrinsic nuclear parameters, specifically the nuclear magnetic moments (for $d$ and $h_{1/2}$) and nuclear electric quadrupole moments (for $eq_0Q$). Both types of moments are known for both isotopes from prior atomic spectroscopic measurements~\cite{Hay1941,Pekka2008}.  Tab.~\ref{table:scaling} presents this comparison, showing excellent agreement.

\begin{table}[tb]
\begin{tabular}{ccccc} 
\toprule
  & $^{137}$BaF& $^{137}$BaF& $^{135}$BaF \\ 
  Parameter& Theory&Experiment&Experiment\\
 \midrule
  $d^{\rm (Ba)}$ &$239.1 \pm 14.4$ & $254.3 \pm 0.5$ & $227.7 \pm 1.2$\\ 
  $h_{1/2}^{\rm (Ba)}$ & $199.5 \pm 11.4$  & $206.6 \pm 0.3$ &$185.9 \pm 0.8$\\
  $e q_0 Q^{\rm (Ba)}$ & - & $-89.0 \pm 0.7$& $-56.0 \pm 3.1$\\
 \bottomrule
\end{tabular}
\caption{Values of the magnetic hyperfine and electric quadrupole parameters extracted in the current study, all expressed in megahertz (MHz).}
\label{table:HFSparameters}
\end{table}

\begin{table}[tb]
\begin{tabular}{ccc|cc} 
\toprule
Ref.~\cite{Hay1941} & \multicolumn{2}{c|}{This work} & Ref.~\cite{Pekka2008} & This work\\
\midrule
 $\dfrac{\mu^{(137)}}{\mu^{(135)}}$ & $\dfrac{d^{(137)}}{d^{(135)}}$ & $\dfrac{h^{(137)}}{h^{(135)}}$ & $\dfrac{Q^{(137)}}{Q^{(135)}}$ & $\dfrac{eq_0Q^{(137)}}{eq_0Q^{(135)}}$\\ 
 \midrule
  $1.118(5)$ & $1.12(1)$ & $1.11(1)$ & $1.53(4)$ & $1.59(11)$ \\ 
\bottomrule
\end{tabular}
\caption{Comparison of ratios of extracted magnetic dipole (electric quadrupole) hyperfine parameters with the ratios of nuclear magnetic dipole (electric quadrupole) moments of the two isotopes $^{135}$Ba and $^{137}$Ba~\cite{Hay1941,Pekka2008}.}
\label{table:scaling}
\end{table}

\begin{figure*}[tb]
    \centering
    \includegraphics{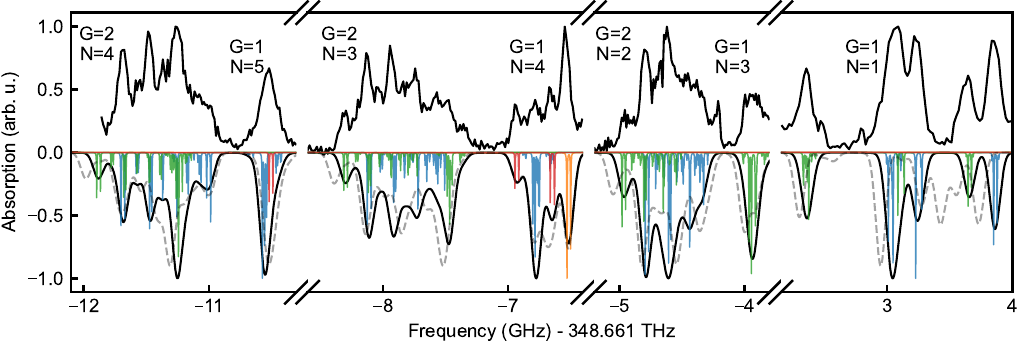}
    \caption{Example absorption spectroscopy of BaF. The lines shown belong to the $X(\nu=0)\rightarrow A(\nu'=0)$ transition, which is the main cooling transition for laser cooling. Experimental data is shown in the top half of the plot, a prediction based the constants determined in this work in the bottom half. Labels denote the respective ground state angular momenta $G$ and $N$ involved in the transitions. Example transitions for $N=1,2,3,\ldots$ and $G=1,2$ are indicated in Fig.~\ref{fig:levelstructure}.
    Individual predicted lines for \BaFseven\, through \BaFfour\, are shown in blue, orange, green and red, respectively. The envelope shown as a solid black line takes into account the Doppler broadening at $3.5\,$K, using the molecular constants determined in this work. The dashed line is based on the previous best set of constants~\cite{Steimle2011}, which predict significantly shifted transitions and even transitions that were not observed. This discrepancy makes accurate line assignments challenging, underscoring the importance of our fluorescence spectroscopy and theoretical results for reliable assignments. For clarity, the amplitude of the data in every x-axis sub-segment has been individually normalized. Residual differences in amplitude between experiment and prediction are due to collisional and optical pumping effects inside the buffer gas source, which will be discussed in future work.}
    \label{fig:absorptionspectroscopy}
\end{figure*}

\section{Ab-initio calculations of the odd isotopologue hyperfine structure}
Our measurements provide precise experimental data that can serve as a benchmark for ab initio electronic structure calculations of the molecular hyperfine structure. Accurate calculations of this kind
require a careful treatment of relativistic and electron-correlation effects, which are also essential for accurately modeling a wide range of molecular properties for precision measurements~\cite{Hao2018,Denis2019,Denis2020,Haase2021}. 

In particular, the hyperfine structure constants of the excited \exs\, state of any odd BaF isotoplogues have not been addressed previously. We thus calculate the parallel and perpendicular components $A_\parallel$ and $A_\perp$ of the hyperfine interaction tensor for the \exs\, excited state of \BaFseven, using the relativistic four-component Dirac-Coulomb Hamiltonian combined with the Fock-space coupled cluster method and the finite-field approach. Theoretical predictions of the quadrupole constant require a separate and very different computational investigation and will be presented in future work. The calculated hyperfine interaction tensor is related to the molecular constants by $A_\parallel=2\, h_{1/2}$ and $A_\perp=d$~\cite{Denis2022}. 

In our calculations, we use the $\mathrm{DIRAC}19$ package \cite{DIRAC19,Saue2020}. We correlate all electrons and extrapolate our results to the complete basis set limit, using the dyall.v$n$Z basis sets \cite{Dyall2009,Dyall2012,Dyall2016}. We evaluate the effect of the basis set and electron correlation and estimate the uncertainty on our final values of $\sim5\%$ for the \exs\, excited state. More details of our calculations and uncertainty estimations, including also results for the \exss\, state, are summarized in Ref.~\cite{Chamorro2025}. 

Our final results are presented in Table \ref{table:HFSparameters}, and show excellent agreement with the experimental data. In most of our previous studies, we found that the method used to estimate uncertainties in the calculated molecular parameters tended to produce error bars that were too conservative, that is, the agreement between theoretical predictions and experimental results was consistently better than anticipated \cite{leimbach2020electron,Haase2020}. This is not the case in the present work, where the experimental value lies within the theoretical uncertainty for $h_{1/2}$, and slightly outside it (by 0.05~$\sigma$) for $d$. In both cases, the theoretical values are lower than the experimental ones. A similar pattern---where the calculated values and associated uncertainties slightly underestimated the experimental results---was also observed in our previous study of the $^{19}\mathrm{F}$ hyperfine structure contributions in the excited states of $^{138}\mathrm{BaF}$.

Several effects could be responsible for this pattern. The present study does not account for the Breit interaction or QED effects. However, these contributions were found to be negligible for the ground-state hyperfine parameters of $^{137}$BaF~\cite{Haase2020} and are expected to be even smaller in the excited states. Additional sources of uncertainty likely stem from the limited model space in the Fock-space coupled-cluster approach and the incompleteness of the employed basis sets. The latter has been shown to be particularly significant in calculations involving excited states~\cite{Denis2022}. A more detailed discussion of these missing contributions is provided in Ref.~\cite{Chamorro2025}. 

Overall, our results demonstrate that state-of-the-art relativistic electronic structure methods can yield accurate and reliable predictions of hyperfine structure parameters in excited states of small heavy molecules—a domain that had previously remained largely unexplored computationally.
 
\section{Absorption spectroscopy}
After addressing the previously missing information on the hyperfine structure of \BaFseven\ and \BaFfive, we now explore the combined absorption spectra of all isotopologues from \BaFfour\ to \BaFeight\ to gain a deeper understanding of the isotope shifts in their rovibrational level structure.

To do so, we use a complementary experimental setup, previously used for the spectroscopic analysis~\cite{Albrecht2020,Rockenhaeuser2023} and laser cooling~\cite{Rockenhaeuser2024,Kogel2024isotope} of the most abundant isotopologues \BaFeight\ and \BaFsix. In short, in this setup we create the molecules using laser ablation of a solid $\mathrm{BaF}_2$ precursor in a cryogenic helium buffer gas source at a temperature of $T=3.5\,$ K and subsequently record the in-cell absorption. The probing light is derived from a tunable diode laser, and the intensity of the probing beam is stabilized to around $300\,\mu$W using an electro-optical amplitude modulator to increase the sensitivity to small changes in absorption. Scanning the laser over a range of $20+\,$GHz for various transitions produces spectra containing lines from all naturally occurring isotopologes. The typical linewidths observed are around $70\,$MHz, as given by Doppler, collisional, and residual power broadening in the buffer gas cell. During frequency scans, the probe laser frequency is referenced to wavelength meters that are regularly validated against atomic absorption lines in the same wavelength range. With this, the uncertainty in the absolute frequency calibration is on the order of $50\,$ MHz~\cite{Rockenhaeuser2023,Hiramoto2023}. 

The high signal to noise ratio inside the buffer gas cell allows us to study a large variety of transitions connecting different vibrational levels. Example spectra for the \gs$(\nu=0)$ to \exs$(\nu'=0)$ transition, where $\nu$ and $\nu'$ denote the vibrational quantum numbers of the states, are shown in Fig.~\ref{fig:absorptionspectroscopy}. Building on the precise knowledge of the hyperfine structure obtained in the previous sections, along with the well-established transitions of the more abundant even isotopologues~\cite{Rockenhaeuser2023}, we are  able to identify transitions of \BaFfour, despite its low natural abundance of only $2.4\,$\%. Moreover, using the hyperfine constants determined above, as well as mass scalings where required, we also find excellent agreement for the transition manifolds of the odd isotopologues \BaFfive, and \BaFseven.

We use this data set, as well as similar data sets for transitions between all possible combinations of $\nu=0,1$ and $\nu'=0,1$ to determine the respective absolute energy offsets $T_{\nu',\nu}=T'_\nu-T_\nu$ of the vibrational states for the various isotopologues. We note that the corresponding transitions are highly relevant in the context of laser cooling, as they are used as repumpers for vibrational leaks. Using the relation $T_\nu=T_e +\omega_e(\nu+1/2) -\omega_e\chi_e(\nu+1/2)^2+\omega_e y_e(\nu+1/2)^3$, we fit the values of $T_e$, $\omega'_e$ and $\omega_e$, and summarize the resulting molecular constants for \BaFseven\,,\BaFfive\, and \BaFfour\, in Tab.~\ref{table:constants}.

\begin{figure}[tb]
    \centering
    \includegraphics[width=0.85\columnwidth]{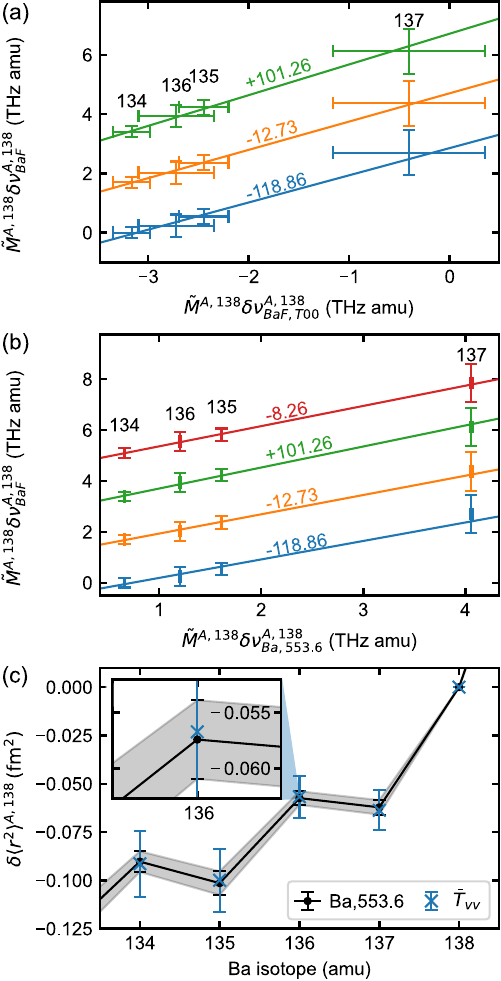} 
    \caption{King plot analysis and odd-even staggering of the barium nuclear charge radius. (a) Isotope shifts of BaF's (from top to bottom) $T_{01}$ (green), $T_{11}$ (orange) and $T_{10}$ (blue) constants relative to the $T_{00}$ constant. (b) Isotope shifts of $T_{00}$ (red), as well as the transitions above, compared to the $553.6\,$nm transition in atomic barium~\cite{FrickeDatabase}. Lines in (a,b) have been vertically offset for visual clarity, with their offset values indicated.
    (c) Relative change in nuclear charge radius $\delta\langle r^2\rangle$, as determined by combining our data with previously known mass and field shift values~\cite{Athanasakis2023}. Data points are averaged over all molecular transitions studied in (a,b), and are in excellent agreement with atomic reference data (solid line). Error bars of the atomic data (shaded area) are derived from uncertainties in the atomic isotope, mass and field shifts~\cite{FrickeDatabase}, and the inset highlights the agreement at the 1–2\% level.}
    \label{fig:kingplot}
\end{figure}

\section{King Plot analysis}
In atoms, frequency shifts such as the ones observed are well known to depend both on the center-of-mass changes when transitioning from one isotope to another --- the so--called mass shift --- and on the overlap of the electronic wave function with the changing nucleus --- the so-called field shift. These shifts can be calculated using electronic structure methods. However, such calculations can be challenging, especially for mass shifts and systems with many open-shell electrons. 

 A well established tool to circumvent this problem for atomic isotope shifts are King plots, where the shifts of two transitions are compared relative to each other, which removes the need for precise knowledge of the nuclear charge radius, and for accurate theoretical predictions of the field and the mass shifts~\cite{King2013}. This results, to first order, in a linear plot, where the slope and the vertical intercept are directly related to the mass and field shift parameters. We note briefly that deviations from this linearity at much higher levels of precision have recently gained interest as probes of physics beyond the Standard Model~\cite{Gebert2015,Ono2022,Hur2022}. 

As in atoms, it has recently been shown that isotope shifts in molecules can also be expressed as the sum of a field shift and a mass shift~\cite{Athanasakis2023}. Several observations of field shifts in molecular spectra have also been previously reported~\cite{Hunter2002,Harms2019,Udrescu2021}. Building on this, our experimental results enable a comprehensive analysis of isotope shifts in BaF molecules, offering two key benefits.

First, the linearity of a King plot provides an additional, robust way to validate line assignments in the observed rovibrational spectra, since an incorrect assignment would immediately stand out in the King plot. This is particularly useful for the analysis of molecular beams containing many different, potentially rare or short-lived isotopologues. 

Second, the argument can be turned around and the King plot can be used to extract information about the nuclear charge radius, and thus about nuclear structure. 
Similar high-precision atomic spectroscopy of isotope shifts has long been a key tool for studying nuclear structure using laser spectroscopy, providing important benchmarks for nuclear theory. As we demonstrate in the following, similar information can be obtained from molecular spectroscopy, which
 is particularly attractive for short-lived nuclei created in on-line facilities, where the formation of molecules, rather than atoms, is often unavoidable~\cite{Athanasakis2023,Blaum2013,Campbell2016}. 

The results of our analysis are summarized in Fig.~\ref{fig:kingplot}. For the King plots, we compare isotope shifts of transitions between different molecular vibrational levels with each other, as encoded by the respective offsets $T_{\nu,\nu'}$. This results in a King plot derived entirely from molecular data, where, unlike in most atomic systems, lasers with vastly different wavelengths are not necessarily required. In addition, we compare the transitions individually with known barium atomic spectroscopy, as previously established in~\cite{Athanasakis2023}. The linearity of the results validates our assignments of the molecular transitions for all isotopologues~\cite{FrickeDatabase}. 
Moreover, using the knowledge of existing mass and field shift constants, we also extract the relative change in the nuclear charge radius of the barium nuclei~\cite{FrickeDatabase}. In this process, we probe nuclei containing $56$ protons, and a varying number of neutrons from $78$ for \BaFfour, to $82$ for \BaFeight.

The resulting nuclear charge radii typically scale as $A^{1/3}$ with the total number of nucleons $A$ in simple nuclear models~\cite{Angeli2013}. However, on top of this well-known increase in size for an increasing number of nucleons, we observe that isotopes with an odd number of neutrons are reduced in size compared to even neutron isotopes. In other words, the addition of an extra neutron can --- counterintuitively --- make the nucleus smaller rather than larger. This odd-even staggering effect arises from nontrivial many-body effects in the nucleus, which form a formidable challenge for nuclear ab initio calculations~\cite{Reinhard2016,Gorges2019,Groote2020}. Our example results highlight that high-resolution molecular spectroscopy, combined with King plot analysis, can provide access to nuclear charge radii—offering a promising approach in cases where molecular spectra are more accessible or experimentally favorable than atomic spectra~\cite{Udrescu2021}. Considering recent progress in molecular spectroscopy, for example, with regard to molecular clocks~\cite{Barontini2021,Leung2023}, it is conceivable that molecules could thus contribute significantly to future measurements of nuclear charge radii and nuclear structure in general.

\begin{table}[tb]
\footnotesize
\centering
\begin{threeparttable}
\begin{tabular}{M{0.22\columnwidth} M{0.23\columnwidth} M{0.23\columnwidth} M{0.22\columnwidth}}
     \toprule
     Parameter & ${}^{137}\mathrm{Ba}^{19}$F& ${}^{135}\mathrm{Ba}^{19}$F & ${}^{134}\mathrm{Ba}^{19}$F\\
     \midrule
\multicolumn{1}{l}{$\mathrm{X}^2\Sigma^+$}\\ %\cmidrule{1-1}
$\omega_e$                & 469.6128(17)\tnote{e} & 470.0381(17)\tnote{e} & 470.2545(17)\tnote{e}\\
$\omega_e \chi_e$         & 1.8362\tnote{b}     & 1.8396\tnote{c}     & 1.8413\tnote{b}    \\
$10^3\, \omega_e y_e$     & 3.0666\tnote{b}     & 3.0749\tnote{c}     & 3.0792\tnote{b}    \\
$10^3\, B_e$              & 216.7210\tnote{a}   & 217.1126\tnote{a}   & 217.3130\tnote{b}  \\
$10^7\, D_e$              & 1.8462\tnote{a}     & 1.8529\tnote{a}     & 1.8563\tnote{b}    \\
$10^3\, \alpha_e$         & 1.1651\tnote{a}     & 1.1683\tnote{a}     & 1.1699\tnote{b}    \\
$10^3\, \gamma$           & 2.7014\tnote{a}     & 2.7063\tnote{a}     & 2.7088\tnote{b}    \\
$10^3\, b_F$ (F)          & 2.1965\tnote{a}     & 2.1965\tnote{a}     & 2.1965\tnote{b}    \\
$10^3\, c$ (F)            & 0.2437\tnote{a}     & 0.2437\tnote{a}     & 0.2436\tnote{b}    \\
$10^3\, b_F$ (Ba)         & 77.6681\tnote{a}    & 69.4887\tnote{a}    &                    \\
$10^3\, c$ (Ba)           & 2.5083\tnote{a}     & 2.2441\tnote{a}     &                    \\
$10^3\, eq_0Q$ (Ba)       & -4.7927\tnote{a}    & -3.1156\tnote{a}    &                    \\
\addlinespace 
\multicolumn{1}{l}{$\mathrm{A}^2\Pi$}\\ %\cmidrule{1-1}
$T_e$                     & 11962.0587(25)\tnote{e} & 11962.0604(25)\tnote{e} & 11962.0578(25)\tnote{e}\\
$\omega_e$                & 438.1472(17)\tnote{e} & 438.5442(17)\tnote{e} & 438.7464(17)\tnote{e}\\
$\omega_e \chi_e$         & 1.8731\tnote{b}     & 1.8765\tnote{c}     & 1.8782\tnote{b}    \\
$10^3\, B_e$              & 212.4585\tnote{b}   & 212.8423\tnote{c}   & 213.0388\tnote{b}  \\
$10^7\, D_e$              & 2.0035\tnote{b}     & 2.0108\tnote{c}     & 2.0145\tnote{b}    \\
$10^3\, \alpha_e$         & 1.1255\tnote{b}     & 1.1286\tnote{c}     & 1.1301\tnote{b}    \\
$A_e$                     & 632.5369\tnote{b}   & 632.5369\tnote{c}   & 632.5369\tnote{b}  \\
$\alpha_A$                & -0.5105\tnote{b}    & -0.5109\tnote{c}    & -0.5112\tnote{b}   \\
$10^3\, A_D$              & 0.0310\tnote{b}     & 0.0311\tnote{c}     & 0.0311\tnote{b}    \\
$p+2q$                    & -0.2578\tnote{b}    & -0.2582\tnote{c}    & -0.2585\tnote{b}   \\
$10^3\, a$ (F)            & 0.8856\tnote{b}     & 0.8856\tnote{c}     & 0.8856\tnote{b}    \\
$10^3\, b_F$ (F)          & -0.0667\tnote{b}    & -0.0667\tnote{c}    & -0.0667\tnote{b}   \\
$10^3\, c$ (F)            & -0.1771\tnote{b}    & -0.1771\tnote{c}    & -0.1771\tnote{b}   \\
$10^3\, d$ (F)            & 0.1194\tnote{b}     & 0.1194\tnote{c}     & 0.1194\tnote{b}    \\
$10^3\, h_{1/2}$ (Ba)     & 6.8948(133)\tnote{d} & 6.1976(200)\tnote{d} &                    \\
$10^3\, d$ (Ba)           & 8.4825(167)\tnote{d} & 7.5953(367)\tnote{d} &                    \\
$10^3\, eq_0Q$ (Ba)       & -2.9721(467)\tnote{d} & -1.8680(901)\tnote{d} &                    \\
     \bottomrule
\end{tabular}

\label{table:constants}
\begin{tablenotes}
\item[a] Reference~\cite{Preston2025}.
\item[b] Value obtained from scaling the value for ${}^{138}\mathrm{Ba}^{19}$F~\cite{Rockenhaeuser2023}.
\item[c] Value obtained from scaling the value for ${}^{137}\mathrm{Ba}^{19}$F.
\item[d] Value obtained from fluorescence spectroscopy fit.
\item[e] Value obtained from absorption spectroscopy fit.

\end{tablenotes}

\caption{Molecular constants in units of cm$^{-1}$ for ${}^{137}\mathrm{Ba}^{19}$F, ${}^{135}\mathrm{Ba}^{19}$F and ${}^{134}\mathrm{Ba}^{19}$F. For unknown values, mass scalings were applied~\cite{Drouin2001,Doppelbauer2022}.}

\end{threeparttable}
\label{tab:molecularconstants}
\end{table}

\section{Conclusion}
We have presented a comprehensive investigation of the energy level structure of most stable isotopologues of the BaF molecule.

Our detailed comparison of ab initio calculations of the excited state hyperfine structure with experiments provides an important benchmark for the former in a so far little explored regime. This, in turn, strengthens similar calculations that directly support the extraction of Standard Model parameters from experiments~\cite{Haase2021}. 

Moreover, the spectroscopic results presented here are an important step towards achieving or improving laser cooling and cycling detection for odd, and by extrapolation, short-lived isotopologues of BaF. While the rovibronic transitions studied provide the necessary frequencies for repumping vibrational losses, the analysis of the excited-state hyperfine structure provides input for the optimization of the cooling forces through engineering of laser sidebands~\cite{Rockenhaeuser2024,Holland2021}. 
Knowledge about this hyperfine structure also allows us to estimate the expected degree of hyperfine mixing in the excited state, which is known to disrupt rotational selection rules in molecules such as YbOH~\cite{Zeng2023}. In the even isotopologues of BaF, the states $J = 1/2^+, F = 1$ and $J = 3/2^+, F = 1$ are mixed via a lambda-doubling-type hyperfine interaction, specifically the $d^{(F)}$ term. In contrast, the odd isotopologues exhibit a greater number of mixed levels due to increased hyperfine interactions from the $d^{(Ba)}$ term, resulting in a more significant and problematic loss mechanism. Although our results indicate slightly less favorable branching ratios than previously assumed~\cite{Steimle2011,Kogel2021}, they show that \BaFseven, and \BaFfive, can scatter at least on the order of $2000$ photons, before branching into the typically unaddressed $N=3$ rotational level occurs. This loss rate is comparable to the one known to occur in BaF through the intermediate electronic $\Delta$ state, which, in contrast, leads to decays to rotational levels $N=0$ and $N=2$ with opposite parity. Losses to all these rotational levels have recently been addressed in the realization of a magneto-optical trap for \BaFeight~\cite{Zeng2024}. We therefore anticipate that similar repumping schemes can be used to realize a magneto-optical trap for odd BaF isotopologues without a significant increase in experimental complexity.

Finally, the precise knowledge of the isotope shifts along a chain of isotopologues provides input to a King plot analysis that reveals the odd-even staggering of the nuclear charge radius of the barium atom. The precision of this approach could be significantly improved in the future using improved molecular spectroscopy techniques~\cite{Aiello2022} or molecular clocks~\cite{Barontini2021,Leung2023} to make it competitive with atomic spectroscopy, where many open questions remain~\cite{Hofsaess2023}. As a similar level of spectroscopic resolution has already been demonstrated with radioactive molecular species, this opens up new approaches for the study of short-lived nuclei, including their higher-order nuclear moments, which provide important enhancements in sensitivity for precision measurements of fundamental nuclear properties~\cite{ArrowsmithKron2024}.

---------
\section*{Acknowledgments}
We thank Andreas Schindewolf, Timothy Steimle, Richard Mawhorter, and Edward Grant for fruitful discussions. D.D. and M.B. thank Sidney B. Cahn for extensive contributions to the fluorescence spectroscopy apparatus.  T.L., F.K., M.R. and T.G. acknowledge Tilman Pfau for generous support, as well as Ralf Albrecht and Einius Pultinevicius for contributions in the early stage of the experiment. The fluorescence spectroscopy work was supported by Laboratory Directed Research and Development (LDRD) funding from Argonne National Laboratory, provided by the Director, Office of Science, of the U.S. DOE under Contract No. DEAC02-06CH11357. A.B. and Y.C. thank the Center for Information Technology at the University of Groningen for their support and for providing access to the Peregrine and Hábrók high-performance computing clusters, and acknowlodge the support from the Dutch Research Council, NWM (VI.Vidi.192.088). The absorption spectroscopy work has received funding from the European Research Council (ERC) under the European Union’s Horizon 2020 research and innovation programme (Grant agreement No. 949431), the RiSC programme of the Ministry of Science, Research and Arts Baden-W\"urttemberg and Carl Zeiss Foundation, and was funded in whole or in part by the Austrian Science Fund (FWF) 10.55776/PAT8306623. 

\bibliography{biblio}

\begin{thebibliography}{100}%
\makeatletter
\providecommand \@ifxundefined [1]{%
 \@ifx{#1\undefined}
}%
\providecommand \@ifnum [1]{%
 \ifnum #1\expandafter \@firstoftwo
 \else \expandafter \@secondoftwo
 \fi
}%
\providecommand \@ifx [1]{%
 \ifx #1\expandafter \@firstoftwo
 \else \expandafter \@secondoftwo
 \fi
}%
\providecommand \natexlab [1]{#1}%
\providecommand \enquote  [1]{``#1''}%
\providecommand \bibnamefont  [1]{#1}%
\providecommand \bibfnamefont [1]{#1}%
\providecommand \citenamefont [1]{#1}%
\providecommand \href@noop [0]{\@secondoftwo}%
\providecommand \href [0]{\begingroup \@sanitize@url \@href}%
\providecommand \@href[1]{\@@startlink{#1}\@@href}%
\providecommand \@@href[1]{\endgroup#1\@@endlink}%
\providecommand \@sanitize@url [0]{\catcode `\\12\catcode `\$12\catcode
  `\&12\catcode `\#12\catcode `\^12\catcode `\_12\catcode `\%12\relax}%
\providecommand \@@startlink[1]{}%
\providecommand \@@endlink[0]{}%
\providecommand \url  [0]{\begingroup\@sanitize@url \@url }%
\providecommand \@url [1]{\endgroup\@href {#1}{\urlprefix }}%
\providecommand \urlprefix  [0]{URL }%
\providecommand \Eprint [0]{\href }%
\providecommand \doibase [0]{https://doi.org/}%
\providecommand \selectlanguage [0]{\@gobble}%
\providecommand \bibinfo  [0]{\@secondoftwo}%
\providecommand \bibfield  [0]{\@secondoftwo}%
\providecommand \translation [1]{[#1]}%
\providecommand \BibitemOpen [0]{}%
\providecommand \bibitemStop [0]{}%
\providecommand \bibitemNoStop [0]{.\EOS\space}%
\providecommand \EOS [0]{\spacefactor3000\relax}%
\providecommand \BibitemShut  [1]{\csname bibitem#1\endcsname}%
\let\auto@bib@innerbib\@empty
%</preamble>
\bibitem [{\citenamefont {Safronova}\ \emph {et~al.}(2018)\citenamefont
  {Safronova}, \citenamefont {Budker}, \citenamefont {DeMille}, \citenamefont
  {Kimball}, \citenamefont {Derevianko},\ and\ \citenamefont
  {Clark}}]{Safronova2018}%
  \BibitemOpen
  \bibfield  {author} {\bibinfo {author} {\bibfnamefont {M.~S.}\ \bibnamefont
  {Safronova}}, \bibinfo {author} {\bibfnamefont {D.}~\bibnamefont {Budker}},
  \bibinfo {author} {\bibfnamefont {D.}~\bibnamefont {DeMille}}, \bibinfo
  {author} {\bibfnamefont {D.~F.~J.}\ \bibnamefont {Kimball}}, \bibinfo
  {author} {\bibfnamefont {A.}~\bibnamefont {Derevianko}},\ and\ \bibinfo
  {author} {\bibfnamefont {C.~W.}\ \bibnamefont {Clark}},\ }\bibfield  {title}
  {\bibinfo {title} {Search for new physics with atoms and molecules},\ }\href
  {https://doi.org/10.1103/RevModPhys.90.025008} {\bibfield  {journal}
  {\bibinfo  {journal} {Rev. Mod. Phys.}\ }\textbf {\bibinfo {volume} {90}},\
  \bibinfo {pages} {025008} (\bibinfo {year} {2018})}\BibitemShut {NoStop}%
\bibitem [{\citenamefont {DeMille}\ \emph {et~al.}(2024)\citenamefont
  {DeMille}, \citenamefont {Hutzler}, \citenamefont {Rey},\ and\ \citenamefont
  {Zelevinsky}}]{DeMille2023}%
  \BibitemOpen
  \bibfield  {author} {\bibinfo {author} {\bibfnamefont {D.}~\bibnamefont
  {DeMille}}, \bibinfo {author} {\bibfnamefont {N.~R.}\ \bibnamefont
  {Hutzler}}, \bibinfo {author} {\bibfnamefont {A.~M.}\ \bibnamefont {Rey}},\
  and\ \bibinfo {author} {\bibfnamefont {T.}~\bibnamefont {Zelevinsky}},\
  }\bibfield  {title} {\bibinfo {title} {Quantum sensing and metrology for
  fundamental physics with molecules},\ }\href
  {https://doi.org/10.1038/s41567-024-02499-9} {\bibfield  {journal} {\bibinfo
  {journal} {Nature Physics}\ }\textbf {\bibinfo {volume} {20}},\ \bibinfo
  {pages} {741} (\bibinfo {year} {2024})}\BibitemShut {NoStop}%
\bibitem [{\citenamefont {Hudson}\ \emph {et~al.}(2002)\citenamefont {Hudson},
  \citenamefont {Sauer}, \citenamefont {Tarbutt},\ and\ \citenamefont
  {Hinds}}]{Hudson2002}%
  \BibitemOpen
  \bibfield  {author} {\bibinfo {author} {\bibfnamefont {J.~J.}\ \bibnamefont
  {Hudson}}, \bibinfo {author} {\bibfnamefont {B.~E.}\ \bibnamefont {Sauer}},
  \bibinfo {author} {\bibfnamefont {M.~R.}\ \bibnamefont {Tarbutt}},\ and\
  \bibinfo {author} {\bibfnamefont {E.~A.}\ \bibnamefont {Hinds}},\ }\bibfield
  {title} {\bibinfo {title} {Measurement of the electron electric dipole moment
  using ybf molecules},\ }\href {https://doi.org/10.1103/PhysRevLett.89.023003}
  {\bibfield  {journal} {\bibinfo  {journal} {Phys. Rev. Lett.}\ }\textbf
  {\bibinfo {volume} {89}},\ \bibinfo {pages} {023003} (\bibinfo {year}
  {2002})}\BibitemShut {NoStop}%
\bibitem [{\citenamefont {Andreev}\ \emph {et~al.}(2018)\citenamefont
  {Andreev}, \citenamefont {Ang}, \citenamefont {DeMille}, \citenamefont
  {Doyle}, \citenamefont {Gabrielse}, \citenamefont {Haefner}, \citenamefont
  {Hutzler}, \citenamefont {Lasner}, \citenamefont {Meisenhelder},
  \citenamefont {O'Leary}, \citenamefont {Panda}, \citenamefont {West},
  \citenamefont {West}, \citenamefont {Wu},\ and\ \citenamefont
  {Collaboration}}]{ACME2018}%
  \BibitemOpen
  \bibfield  {author} {\bibinfo {author} {\bibfnamefont {V.}~\bibnamefont
  {Andreev}}, \bibinfo {author} {\bibfnamefont {D.~G.}\ \bibnamefont {Ang}},
  \bibinfo {author} {\bibfnamefont {D.}~\bibnamefont {DeMille}}, \bibinfo
  {author} {\bibfnamefont {J.~M.}\ \bibnamefont {Doyle}}, \bibinfo {author}
  {\bibfnamefont {G.}~\bibnamefont {Gabrielse}}, \bibinfo {author}
  {\bibfnamefont {J.}~\bibnamefont {Haefner}}, \bibinfo {author} {\bibfnamefont
  {N.~R.}\ \bibnamefont {Hutzler}}, \bibinfo {author} {\bibfnamefont
  {Z.}~\bibnamefont {Lasner}}, \bibinfo {author} {\bibfnamefont
  {C.}~\bibnamefont {Meisenhelder}}, \bibinfo {author} {\bibfnamefont {B.~R.}\
  \bibnamefont {O'Leary}}, \bibinfo {author} {\bibfnamefont {C.~D.}\
  \bibnamefont {Panda}}, \bibinfo {author} {\bibfnamefont {A.~D.}\ \bibnamefont
  {West}}, \bibinfo {author} {\bibfnamefont {E.~P.}\ \bibnamefont {West}},
  \bibinfo {author} {\bibfnamefont {X.}~\bibnamefont {Wu}},\ and\ \bibinfo
  {author} {\bibfnamefont {A.}~\bibnamefont {Collaboration}},\ }\bibfield
  {title} {\bibinfo {title} {Improved limit on the electric dipole moment of
  the electron},\ }\href {https://doi.org/10.1038/s41586-018-0599-8} {\bibfield
   {journal} {\bibinfo  {journal} {Nature}\ }\textbf {\bibinfo {volume}
  {562}},\ \bibinfo {pages} {355} (\bibinfo {year} {2018})}\BibitemShut
  {NoStop}%
\bibitem [{\citenamefont {Roussy}\ \emph {et~al.}(2023)\citenamefont {Roussy},
  \citenamefont {Caldwell}, \citenamefont {Wright}, \citenamefont {Cairncross},
  \citenamefont {Shagam}, \citenamefont {Ng}, \citenamefont {Schlossberger},
  \citenamefont {Park}, \citenamefont {Wang}, \citenamefont {Ye},\ and\
  \citenamefont {Cornell}}]{Roussy2023}%
  \BibitemOpen
  \bibfield  {author} {\bibinfo {author} {\bibfnamefont {T.~S.}\ \bibnamefont
  {Roussy}}, \bibinfo {author} {\bibfnamefont {L.}~\bibnamefont {Caldwell}},
  \bibinfo {author} {\bibfnamefont {T.}~\bibnamefont {Wright}}, \bibinfo
  {author} {\bibfnamefont {W.~B.}\ \bibnamefont {Cairncross}}, \bibinfo
  {author} {\bibfnamefont {Y.}~\bibnamefont {Shagam}}, \bibinfo {author}
  {\bibfnamefont {K.~B.}\ \bibnamefont {Ng}}, \bibinfo {author} {\bibfnamefont
  {N.}~\bibnamefont {Schlossberger}}, \bibinfo {author} {\bibfnamefont {S.~Y.}\
  \bibnamefont {Park}}, \bibinfo {author} {\bibfnamefont {A.}~\bibnamefont
  {Wang}}, \bibinfo {author} {\bibfnamefont {J.}~\bibnamefont {Ye}},\ and\
  \bibinfo {author} {\bibfnamefont {E.~A.}\ \bibnamefont {Cornell}},\
  }\bibfield  {title} {\bibinfo {title} {An improved bound on the electron’s
  electric dipole moment},\ }\href {https://doi.org/10.1126/science.adg4084}
  {\bibfield  {journal} {\bibinfo  {journal} {Science}\ }\textbf {\bibinfo
  {volume} {381}},\ \bibinfo {pages} {46} (\bibinfo {year} {2023})}\BibitemShut
  {NoStop}%
\bibitem [{\citenamefont {Grasdijk}\ \emph {et~al.}(2021)\citenamefont
  {Grasdijk}, \citenamefont {Timgren}, \citenamefont {Kastelic}, \citenamefont
  {Wright}, \citenamefont {Lamoreaux}, \citenamefont {DeMille}, \citenamefont
  {Wenz}, \citenamefont {Aitken}, \citenamefont {Zelevinsky}, \citenamefont
  {Winick},\ and\ \citenamefont {Kawall}}]{Grasdijk2021}%
  \BibitemOpen
  \bibfield  {author} {\bibinfo {author} {\bibfnamefont {O.}~\bibnamefont
  {Grasdijk}}, \bibinfo {author} {\bibfnamefont {O.}~\bibnamefont {Timgren}},
  \bibinfo {author} {\bibfnamefont {J.}~\bibnamefont {Kastelic}}, \bibinfo
  {author} {\bibfnamefont {T.}~\bibnamefont {Wright}}, \bibinfo {author}
  {\bibfnamefont {S.}~\bibnamefont {Lamoreaux}}, \bibinfo {author}
  {\bibfnamefont {D.}~\bibnamefont {DeMille}}, \bibinfo {author} {\bibfnamefont
  {K.}~\bibnamefont {Wenz}}, \bibinfo {author} {\bibfnamefont {M.}~\bibnamefont
  {Aitken}}, \bibinfo {author} {\bibfnamefont {T.}~\bibnamefont {Zelevinsky}},
  \bibinfo {author} {\bibfnamefont {T.}~\bibnamefont {Winick}},\ and\ \bibinfo
  {author} {\bibfnamefont {D.}~\bibnamefont {Kawall}},\ }\bibfield  {title}
  {\bibinfo {title} {Centrex: a new search for time-reversal symmetry violation
  in the 205tl nucleus},\ }\href {https://doi.org/10.1088/2058-9565/abdca3}
  {\bibfield  {journal} {\bibinfo  {journal} {Quantum Science and Technology}\
  }\textbf {\bibinfo {volume} {6}},\ \bibinfo {pages} {044007} (\bibinfo {year}
  {2021})}\BibitemShut {NoStop}%
\bibitem [{\citenamefont {Flambaum}\ \emph {et~al.}(2014)\citenamefont
  {Flambaum}, \citenamefont {DeMille},\ and\ \citenamefont
  {Kozlov}}]{Flambaum2014}%
  \BibitemOpen
  \bibfield  {author} {\bibinfo {author} {\bibfnamefont {V.~V.}\ \bibnamefont
  {Flambaum}}, \bibinfo {author} {\bibfnamefont {D.}~\bibnamefont {DeMille}},\
  and\ \bibinfo {author} {\bibfnamefont {M.~G.}\ \bibnamefont {Kozlov}},\
  }\bibfield  {title} {\bibinfo {title} {Time-reversal symmetry violation in
  molecules induced by nuclear magnetic quadrupole moments},\ }\href
  {https://doi.org/10.1103/PhysRevLett.113.103003} {\bibfield  {journal}
  {\bibinfo  {journal} {Phys. Rev. Lett.}\ }\textbf {\bibinfo {volume} {113}},\
  \bibinfo {pages} {103003} (\bibinfo {year} {2014})}\BibitemShut {NoStop}%
\bibitem [{\citenamefont {Shelkovnikov}\ \emph {et~al.}(2008)\citenamefont
  {Shelkovnikov}, \citenamefont {Butcher}, \citenamefont {Chardonnet},\ and\
  \citenamefont {Amy-Klein}}]{Shelkovnikov2008}%
  \BibitemOpen
  \bibfield  {author} {\bibinfo {author} {\bibfnamefont {A.}~\bibnamefont
  {Shelkovnikov}}, \bibinfo {author} {\bibfnamefont {R.~J.}\ \bibnamefont
  {Butcher}}, \bibinfo {author} {\bibfnamefont {C.}~\bibnamefont
  {Chardonnet}},\ and\ \bibinfo {author} {\bibfnamefont {A.}~\bibnamefont
  {Amy-Klein}},\ }\bibfield  {title} {\bibinfo {title} {Stability of the
  proton-to-electron mass ratio},\ }\href
  {https://doi.org/10.1103/PhysRevLett.100.150801} {\bibfield  {journal}
  {\bibinfo  {journal} {Phys. Rev. Lett.}\ }\textbf {\bibinfo {volume} {100}},\
  \bibinfo {pages} {150801} (\bibinfo {year} {2008})}\BibitemShut {NoStop}%
\bibitem [{\citenamefont {Bethlem}\ and\ \citenamefont
  {Ubachs}(2009)}]{Bethlem2009}%
  \BibitemOpen
  \bibfield  {author} {\bibinfo {author} {\bibfnamefont {H.~L.}\ \bibnamefont
  {Bethlem}}\ and\ \bibinfo {author} {\bibfnamefont {W.}~\bibnamefont
  {Ubachs}},\ }\bibfield  {title} {\bibinfo {title} {Testing the
  time-invariance of fundamental constants using microwave spectroscopy on cold
  diatomic radicals},\ }\href {https://doi.org/10.1039/B819099B} {\bibfield
  {journal} {\bibinfo  {journal} {Faraday Discuss.}\ }\textbf {\bibinfo
  {volume} {142}},\ \bibinfo {pages} {25} (\bibinfo {year} {2009})}\BibitemShut
  {NoStop}%
\bibitem [{\citenamefont {Truppe}\ \emph {et~al.}(2013)\citenamefont {Truppe},
  \citenamefont {Hendricks}, \citenamefont {Tokunaga}, \citenamefont
  {Lewandowski}, \citenamefont {Kozlov}, \citenamefont {Henkel}, \citenamefont
  {Hinds},\ and\ \citenamefont {Tarbutt}}]{Truppe2013}%
  \BibitemOpen
  \bibfield  {author} {\bibinfo {author} {\bibfnamefont {S.}~\bibnamefont
  {Truppe}}, \bibinfo {author} {\bibfnamefont {R.~J.}\ \bibnamefont
  {Hendricks}}, \bibinfo {author} {\bibfnamefont {S.~K.}\ \bibnamefont
  {Tokunaga}}, \bibinfo {author} {\bibfnamefont {H.~J.}\ \bibnamefont
  {Lewandowski}}, \bibinfo {author} {\bibfnamefont {M.~G.}\ \bibnamefont
  {Kozlov}}, \bibinfo {author} {\bibfnamefont {C.}~\bibnamefont {Henkel}},
  \bibinfo {author} {\bibfnamefont {E.~A.}\ \bibnamefont {Hinds}},\ and\
  \bibinfo {author} {\bibfnamefont {M.~R.}\ \bibnamefont {Tarbutt}},\
  }\bibfield  {title} {\bibinfo {title} {A search for varying fundamental
  constants using hertz-level frequency measurements of cold ch molecules},\
  }\href {https://doi.org/10.1038/ncomms3600} {\bibfield  {journal} {\bibinfo
  {journal} {Nature Communications}\ }\textbf {\bibinfo {volume} {4}},\
  \bibinfo {pages} {2600} (\bibinfo {year} {2013})}\BibitemShut {NoStop}%
\bibitem [{\citenamefont {Schiller}\ \emph {et~al.}(2014)\citenamefont
  {Schiller}, \citenamefont {Bakalov},\ and\ \citenamefont
  {Korobov}}]{Schiller2014}%
  \BibitemOpen
  \bibfield  {author} {\bibinfo {author} {\bibfnamefont {S.}~\bibnamefont
  {Schiller}}, \bibinfo {author} {\bibfnamefont {D.}~\bibnamefont {Bakalov}},\
  and\ \bibinfo {author} {\bibfnamefont {V.~I.}\ \bibnamefont {Korobov}},\
  }\bibfield  {title} {\bibinfo {title} {Simplest molecules as candidates for
  precise optical clocks},\ }\href
  {https://doi.org/10.1103/PhysRevLett.113.023004} {\bibfield  {journal}
  {\bibinfo  {journal} {Phys. Rev. Lett.}\ }\textbf {\bibinfo {volume} {113}},\
  \bibinfo {pages} {023004} (\bibinfo {year} {2014})}\BibitemShut {NoStop}%
\bibitem [{\citenamefont {Kobayashi}\ \emph {et~al.}(2019)\citenamefont
  {Kobayashi}, \citenamefont {Ogino},\ and\ \citenamefont
  {Inouye}}]{Kobayashi2019}%
  \BibitemOpen
  \bibfield  {author} {\bibinfo {author} {\bibfnamefont {J.}~\bibnamefont
  {Kobayashi}}, \bibinfo {author} {\bibfnamefont {A.}~\bibnamefont {Ogino}},\
  and\ \bibinfo {author} {\bibfnamefont {S.}~\bibnamefont {Inouye}},\
  }\bibfield  {title} {\bibinfo {title} {Measurement of the variation of
  electron-to-proton mass ratio using ultracold molecules produced from
  laser-cooled atoms},\ }\href {https://doi.org/10.1038/s41467-019-11761-1}
  {\bibfield  {journal} {\bibinfo  {journal} {Nature Communications}\ }\textbf
  {\bibinfo {volume} {10}},\ \bibinfo {pages} {3771} (\bibinfo {year}
  {2019})}\BibitemShut {NoStop}%
\bibitem [{\citenamefont {Leung}\ \emph {et~al.}(2023)\citenamefont {Leung},
  \citenamefont {Iritani}, \citenamefont {Tiberi}, \citenamefont {Majewska},
  \citenamefont {Borkowski}, \citenamefont {Moszynski},\ and\ \citenamefont
  {Zelevinsky}}]{Leung2023}%
  \BibitemOpen
  \bibfield  {author} {\bibinfo {author} {\bibfnamefont {K.~H.}\ \bibnamefont
  {Leung}}, \bibinfo {author} {\bibfnamefont {B.}~\bibnamefont {Iritani}},
  \bibinfo {author} {\bibfnamefont {E.}~\bibnamefont {Tiberi}}, \bibinfo
  {author} {\bibfnamefont {I.}~\bibnamefont {Majewska}}, \bibinfo {author}
  {\bibfnamefont {M.}~\bibnamefont {Borkowski}}, \bibinfo {author}
  {\bibfnamefont {R.}~\bibnamefont {Moszynski}},\ and\ \bibinfo {author}
  {\bibfnamefont {T.}~\bibnamefont {Zelevinsky}},\ }\bibfield  {title}
  {\bibinfo {title} {Terahertz vibrational molecular clock with systematic
  uncertainty at the ${10}^{\ensuremath{-}14}$ level},\ }\href
  {https://doi.org/10.1103/PhysRevX.13.011047} {\bibfield  {journal} {\bibinfo
  {journal} {Phys. Rev. X}\ }\textbf {\bibinfo {volume} {13}},\ \bibinfo
  {pages} {011047} (\bibinfo {year} {2023})}\BibitemShut {NoStop}%
\bibitem [{\citenamefont {Athanasakis-Kaklamanakis}\ \emph
  {et~al.}(2023)\citenamefont {Athanasakis-Kaklamanakis}, \citenamefont
  {Wilkins}, \citenamefont {Breier},\ and\ \citenamefont
  {Neyens}}]{Athanasakis2023}%
  \BibitemOpen
  \bibfield  {author} {\bibinfo {author} {\bibfnamefont {M.}~\bibnamefont
  {Athanasakis-Kaklamanakis}}, \bibinfo {author} {\bibfnamefont {S.~G.}\
  \bibnamefont {Wilkins}}, \bibinfo {author} {\bibfnamefont {A.~A.}\
  \bibnamefont {Breier}},\ and\ \bibinfo {author} {\bibfnamefont
  {G.}~\bibnamefont {Neyens}},\ }\bibfield  {title} {\bibinfo {title}
  {King-plot analysis of isotope shifts in simple diatomic molecules},\ }\href
  {https://doi.org/10.1103/PhysRevX.13.011015} {\bibfield  {journal} {\bibinfo
  {journal} {Phys. Rev. X}\ }\textbf {\bibinfo {volume} {13}},\ \bibinfo
  {pages} {011015} (\bibinfo {year} {2023})}\BibitemShut {NoStop}%
\bibitem [{\citenamefont {Wilkins}\ \emph {et~al.}(2023)\citenamefont
  {Wilkins}, \citenamefont {Udrescu}, \citenamefont {Athanasakis-Kaklamanakis},
  \citenamefont {Ruiz}, \citenamefont {Au}, \citenamefont {Belosevic},
  \citenamefont {Berger}, \citenamefont {Bissell}, \citenamefont {Breier},
  \citenamefont {Brinson}, \citenamefont {Chrysalidis}, \citenamefont
  {Cocolios}, \citenamefont {de~Groote}, \citenamefont {Dorne}, \citenamefont
  {Flanagan}, \citenamefont {Franchoo}, \citenamefont {Gaul}, \citenamefont
  {Geldhof}, \citenamefont {Giesen}, \citenamefont {Hanstorp}, \citenamefont
  {Heinke}, \citenamefont {Isaev}, \citenamefont {Koszorus}, \citenamefont
  {Kujanp\"a\"a}, \citenamefont {Lalanne}, \citenamefont {Neyens},
  \citenamefont {Nichols}, \citenamefont {Perrett}, \citenamefont {Reilly},
  \citenamefont {Skripnikov}, \citenamefont {Rothe}, \citenamefont {van~den
  Borne}, \citenamefont {Wang}, \citenamefont {Wessolek}, \citenamefont
  {Yang},\ and\ \citenamefont {Zülch}}]{Wilkins2023}%
  \BibitemOpen
  \bibfield  {author} {\bibinfo {author} {\bibfnamefont {S.~G.}\ \bibnamefont
  {Wilkins}}, \bibinfo {author} {\bibfnamefont {S.~M.}\ \bibnamefont
  {Udrescu}}, \bibinfo {author} {\bibfnamefont {M.}~\bibnamefont
  {Athanasakis-Kaklamanakis}}, \bibinfo {author} {\bibfnamefont {R.~F.~G.}\
  \bibnamefont {Ruiz}}, \bibinfo {author} {\bibfnamefont {M.}~\bibnamefont
  {Au}}, \bibinfo {author} {\bibfnamefont {I.}~\bibnamefont {Belosevic}},
  \bibinfo {author} {\bibfnamefont {R.}~\bibnamefont {Berger}}, \bibinfo
  {author} {\bibfnamefont {M.~L.}\ \bibnamefont {Bissell}}, \bibinfo {author}
  {\bibfnamefont {A.~A.}\ \bibnamefont {Breier}}, \bibinfo {author}
  {\bibfnamefont {A.~J.}\ \bibnamefont {Brinson}}, \bibinfo {author}
  {\bibfnamefont {K.}~\bibnamefont {Chrysalidis}}, \bibinfo {author}
  {\bibfnamefont {T.~E.}\ \bibnamefont {Cocolios}}, \bibinfo {author}
  {\bibfnamefont {R.~P.}\ \bibnamefont {de~Groote}}, \bibinfo {author}
  {\bibfnamefont {A.}~\bibnamefont {Dorne}}, \bibinfo {author} {\bibfnamefont
  {K.~T.}\ \bibnamefont {Flanagan}}, \bibinfo {author} {\bibfnamefont
  {S.}~\bibnamefont {Franchoo}}, \bibinfo {author} {\bibfnamefont
  {K.}~\bibnamefont {Gaul}}, \bibinfo {author} {\bibfnamefont {S.}~\bibnamefont
  {Geldhof}}, \bibinfo {author} {\bibfnamefont {T.~F.}\ \bibnamefont {Giesen}},
  \bibinfo {author} {\bibfnamefont {D.}~\bibnamefont {Hanstorp}}, \bibinfo
  {author} {\bibfnamefont {R.}~\bibnamefont {Heinke}}, \bibinfo {author}
  {\bibfnamefont {T.}~\bibnamefont {Isaev}}, \bibinfo {author} {\bibfnamefont
  {A.}~\bibnamefont {Koszorus}}, \bibinfo {author} {\bibfnamefont
  {S.}~\bibnamefont {Kujanp\"a\"a}}, \bibinfo {author} {\bibfnamefont
  {L.}~\bibnamefont {Lalanne}}, \bibinfo {author} {\bibfnamefont
  {G.}~\bibnamefont {Neyens}}, \bibinfo {author} {\bibfnamefont
  {M.}~\bibnamefont {Nichols}}, \bibinfo {author} {\bibfnamefont {H.~A.}\
  \bibnamefont {Perrett}}, \bibinfo {author} {\bibfnamefont {J.~R.}\
  \bibnamefont {Reilly}}, \bibinfo {author} {\bibfnamefont {L.~V.}\
  \bibnamefont {Skripnikov}}, \bibinfo {author} {\bibfnamefont
  {S.}~\bibnamefont {Rothe}}, \bibinfo {author} {\bibfnamefont
  {B.}~\bibnamefont {van~den Borne}}, \bibinfo {author} {\bibfnamefont
  {Q.}~\bibnamefont {Wang}}, \bibinfo {author} {\bibfnamefont {J.}~\bibnamefont
  {Wessolek}}, \bibinfo {author} {\bibfnamefont {X.~F.}\ \bibnamefont {Yang}},\
  and\ \bibinfo {author} {\bibfnamefont {C.}~\bibnamefont {Zülch}},\
  }\href@noop {} {\bibinfo {title} {Observation of the distribution of nuclear
  magnetization in a molecule}} (\bibinfo {year} {2023}),\ \Eprint
  {https://arxiv.org/abs/2311.04121} {arXiv:2311.04121} \BibitemShut {NoStop}%
\bibitem [{\citenamefont {Udrescu}\ \emph {et~al.}(2024)\citenamefont
  {Udrescu}, \citenamefont {Wilkins}, \citenamefont {Breier}, \citenamefont
  {Athanasakis-Kaklamanakis}, \citenamefont {{Garcia Ruiz}}, \citenamefont
  {Au}, \citenamefont {Belo{\v{s}}evi{\'{c}}}, \citenamefont {Berger},
  \citenamefont {Bissell}, \citenamefont {Binnersley}, \citenamefont {Brinson},
  \citenamefont {Chrysalidis}, \citenamefont {Cocolios}, \citenamefont
  {de~Groote}, \citenamefont {Dorne}, \citenamefont {Flanagan}, \citenamefont
  {Franchoo}, \citenamefont {Gaul}, \citenamefont {Geldhof}, \citenamefont
  {Giesen}, \citenamefont {Hanstorp}, \citenamefont {Heinke}, \citenamefont
  {Koszor{\'{u}}s}, \citenamefont {Kujanp{\"{a}}{\"{a}}}, \citenamefont
  {Lalanne}, \citenamefont {Neyens}, \citenamefont {Nichols}, \citenamefont
  {Perrett}, \citenamefont {Reilly}, \citenamefont {Rothe}, \citenamefont
  {van~den Borne}, \citenamefont {Vernon}, \citenamefont {Wang}, \citenamefont
  {Wessolek}, \citenamefont {Yang},\ and\ \citenamefont
  {Z{\"{u}}lch}}]{Udrescu2024}%
  \BibitemOpen
  \bibfield  {author} {\bibinfo {author} {\bibfnamefont {S.~M.}\ \bibnamefont
  {Udrescu}}, \bibinfo {author} {\bibfnamefont {S.~G.}\ \bibnamefont
  {Wilkins}}, \bibinfo {author} {\bibfnamefont {A.~A.}\ \bibnamefont {Breier}},
  \bibinfo {author} {\bibfnamefont {M.}~\bibnamefont
  {Athanasakis-Kaklamanakis}}, \bibinfo {author} {\bibfnamefont {R.~F.}\
  \bibnamefont {{Garcia Ruiz}}}, \bibinfo {author} {\bibfnamefont
  {M.}~\bibnamefont {Au}}, \bibinfo {author} {\bibfnamefont {I.}~\bibnamefont
  {Belo{\v{s}}evi{\'{c}}}}, \bibinfo {author} {\bibfnamefont {R.}~\bibnamefont
  {Berger}}, \bibinfo {author} {\bibfnamefont {M.~L.}\ \bibnamefont {Bissell}},
  \bibinfo {author} {\bibfnamefont {C.~L.}\ \bibnamefont {Binnersley}},
  \bibinfo {author} {\bibfnamefont {A.~J.}\ \bibnamefont {Brinson}}, \bibinfo
  {author} {\bibfnamefont {K.}~\bibnamefont {Chrysalidis}}, \bibinfo {author}
  {\bibfnamefont {T.~E.}\ \bibnamefont {Cocolios}}, \bibinfo {author}
  {\bibfnamefont {R.~P.}\ \bibnamefont {de~Groote}}, \bibinfo {author}
  {\bibfnamefont {A.}~\bibnamefont {Dorne}}, \bibinfo {author} {\bibfnamefont
  {K.~T.}\ \bibnamefont {Flanagan}}, \bibinfo {author} {\bibfnamefont
  {S.}~\bibnamefont {Franchoo}}, \bibinfo {author} {\bibfnamefont
  {K.}~\bibnamefont {Gaul}}, \bibinfo {author} {\bibfnamefont {S.}~\bibnamefont
  {Geldhof}}, \bibinfo {author} {\bibfnamefont {T.~F.}\ \bibnamefont {Giesen}},
  \bibinfo {author} {\bibfnamefont {D.}~\bibnamefont {Hanstorp}}, \bibinfo
  {author} {\bibfnamefont {R.}~\bibnamefont {Heinke}}, \bibinfo {author}
  {\bibfnamefont {{\'{A}}.}~\bibnamefont {Koszor{\'{u}}s}}, \bibinfo {author}
  {\bibfnamefont {S.}~\bibnamefont {Kujanp{\"{a}}{\"{a}}}}, \bibinfo {author}
  {\bibfnamefont {L.}~\bibnamefont {Lalanne}}, \bibinfo {author} {\bibfnamefont
  {G.}~\bibnamefont {Neyens}}, \bibinfo {author} {\bibfnamefont
  {M.}~\bibnamefont {Nichols}}, \bibinfo {author} {\bibfnamefont {H.~A.}\
  \bibnamefont {Perrett}}, \bibinfo {author} {\bibfnamefont {J.~R.}\
  \bibnamefont {Reilly}}, \bibinfo {author} {\bibfnamefont {S.}~\bibnamefont
  {Rothe}}, \bibinfo {author} {\bibfnamefont {B.}~\bibnamefont {van~den
  Borne}}, \bibinfo {author} {\bibfnamefont {A.~R.}\ \bibnamefont {Vernon}},
  \bibinfo {author} {\bibfnamefont {Q.}~\bibnamefont {Wang}}, \bibinfo {author}
  {\bibfnamefont {J.}~\bibnamefont {Wessolek}}, \bibinfo {author}
  {\bibfnamefont {X.~F.}\ \bibnamefont {Yang}},\ and\ \bibinfo {author}
  {\bibfnamefont {C.}~\bibnamefont {Z{\"{u}}lch}},\ }\bibfield  {title}
  {\bibinfo {title} {{Precision spectroscopy and laser-cooling scheme of a
  radium-containing molecule}},\ }\href
  {https://doi.org/10.1038/s41567-023-02296-w} {\bibfield  {journal} {\bibinfo
  {journal} {Nature Physics}\ }\textbf {\bibinfo {volume} {20}},\ \bibinfo
  {pages} {202} (\bibinfo {year} {2024})}\BibitemShut {NoStop}%
\bibitem [{\citenamefont {DeMille}\ \emph {et~al.}(2008)\citenamefont
  {DeMille}, \citenamefont {Cahn}, \citenamefont {Murphree}, \citenamefont
  {Rahmlow},\ and\ \citenamefont {Kozlov}}]{Demille2008}%
  \BibitemOpen
  \bibfield  {author} {\bibinfo {author} {\bibfnamefont {D.}~\bibnamefont
  {DeMille}}, \bibinfo {author} {\bibfnamefont {S.~B.}\ \bibnamefont {Cahn}},
  \bibinfo {author} {\bibfnamefont {D.}~\bibnamefont {Murphree}}, \bibinfo
  {author} {\bibfnamefont {D.~A.}\ \bibnamefont {Rahmlow}},\ and\ \bibinfo
  {author} {\bibfnamefont {M.~G.}\ \bibnamefont {Kozlov}},\ }\bibfield  {title}
  {\bibinfo {title} {{Using Molecules to Measure Nuclear Spin-Dependent Parity
  Violation}},\ }\href {https://doi.org/10.1103/PhysRevLett.100.023003}
  {\bibfield  {journal} {\bibinfo  {journal} {Phys. Rev. Lett.}\ }\textbf
  {\bibinfo {volume} {100}},\ \bibinfo {pages} {23003} (\bibinfo {year}
  {2008})}\BibitemShut {NoStop}%
\bibitem [{\citenamefont {Cahn}\ \emph {et~al.}(2014)\citenamefont {Cahn},
  \citenamefont {Ammon}, \citenamefont {Kirilov}, \citenamefont {Gurevich},
  \citenamefont {Murphree}, \citenamefont {Paolino}, \citenamefont {Rahmlow},
  \citenamefont {Kozlov},\ and\ \citenamefont {DeMille}}]{Cahn2014}%
  \BibitemOpen
  \bibfield  {author} {\bibinfo {author} {\bibfnamefont {S.~B.}\ \bibnamefont
  {Cahn}}, \bibinfo {author} {\bibfnamefont {J.}~\bibnamefont {Ammon}},
  \bibinfo {author} {\bibfnamefont {E.}~\bibnamefont {Kirilov}}, \bibinfo
  {author} {\bibfnamefont {Y.~V.}\ \bibnamefont {Gurevich}}, \bibinfo {author}
  {\bibfnamefont {D.}~\bibnamefont {Murphree}}, \bibinfo {author}
  {\bibfnamefont {R.}~\bibnamefont {Paolino}}, \bibinfo {author} {\bibfnamefont
  {D.~A.}\ \bibnamefont {Rahmlow}}, \bibinfo {author} {\bibfnamefont {M.~G.}\
  \bibnamefont {Kozlov}},\ and\ \bibinfo {author} {\bibfnamefont
  {D.}~\bibnamefont {DeMille}},\ }\bibfield  {title} {\bibinfo {title}
  {Zeeman-tuned rotational level-crossing spectroscopy in a diatomic free
  radical},\ }\href {https://doi.org/10.1103/PhysRevLett.112.163002} {\bibfield
   {journal} {\bibinfo  {journal} {Phys. Rev. Lett.}\ }\textbf {\bibinfo
  {volume} {112}},\ \bibinfo {pages} {163002} (\bibinfo {year}
  {2014})}\BibitemShut {NoStop}%
\bibitem [{\citenamefont {Norrgard}\ \emph {et~al.}(2019)\citenamefont
  {Norrgard}, \citenamefont {Barker}, \citenamefont {Eckel}, \citenamefont
  {Fedchak}, \citenamefont {Klimov},\ and\ \citenamefont
  {Scherschligt}}]{Norrgard2019}%
  \BibitemOpen
  \bibfield  {author} {\bibinfo {author} {\bibfnamefont {E.~B.}\ \bibnamefont
  {Norrgard}}, \bibinfo {author} {\bibfnamefont {D.~S.}\ \bibnamefont
  {Barker}}, \bibinfo {author} {\bibfnamefont {S.}~\bibnamefont {Eckel}},
  \bibinfo {author} {\bibfnamefont {J.~A.}\ \bibnamefont {Fedchak}}, \bibinfo
  {author} {\bibfnamefont {N.~N.}\ \bibnamefont {Klimov}},\ and\ \bibinfo
  {author} {\bibfnamefont {J.}~\bibnamefont {Scherschligt}},\ }\bibfield
  {title} {\bibinfo {title} {Nuclear-spin dependent parity violation in
  optically trapped polyatomic molecules},\ }\href
  {https://doi.org/10.1038/s42005-019-0181-1} {\bibfield  {journal} {\bibinfo
  {journal} {Communications Physics}\ }\textbf {\bibinfo {volume} {2}},\
  \bibinfo {pages} {77} (\bibinfo {year} {2019})}\BibitemShut {NoStop}%
\bibitem [{\citenamefont {Lim}\ \emph {et~al.}(2018)\citenamefont {Lim},
  \citenamefont {Almond}, \citenamefont {Trigatzis}, \citenamefont {Devlin},
  \citenamefont {Fitch}, \citenamefont {Sauer}, \citenamefont {Tarbutt},\ and\
  \citenamefont {Hinds}}]{Lim2018}%
  \BibitemOpen
  \bibfield  {author} {\bibinfo {author} {\bibfnamefont {J.}~\bibnamefont
  {Lim}}, \bibinfo {author} {\bibfnamefont {J.~R.}\ \bibnamefont {Almond}},
  \bibinfo {author} {\bibfnamefont {M.~A.}\ \bibnamefont {Trigatzis}}, \bibinfo
  {author} {\bibfnamefont {J.~A.}\ \bibnamefont {Devlin}}, \bibinfo {author}
  {\bibfnamefont {N.~J.}\ \bibnamefont {Fitch}}, \bibinfo {author}
  {\bibfnamefont {B.~E.}\ \bibnamefont {Sauer}}, \bibinfo {author}
  {\bibfnamefont {M.~R.}\ \bibnamefont {Tarbutt}},\ and\ \bibinfo {author}
  {\bibfnamefont {E.~A.}\ \bibnamefont {Hinds}},\ }\bibfield  {title} {\bibinfo
  {title} {Laser cooled {YbF} molecules for measuring the electron's electric
  dipole moment},\ }\href {https://doi.org/10.1103/PhysRevLett.120.123201}
  {\bibfield  {journal} {\bibinfo  {journal} {Phys. Rev. Lett.}\ }\textbf
  {\bibinfo {volume} {120}},\ \bibinfo {pages} {123201} (\bibinfo {year}
  {2018})}\BibitemShut {NoStop}%
\bibitem [{\citenamefont {Fitch}\ \emph {et~al.}(2020)\citenamefont {Fitch},
  \citenamefont {Lim}, \citenamefont {Hinds}, \citenamefont {Sauer},\ and\
  \citenamefont {Tarbutt}}]{Fitch2020methods}%
  \BibitemOpen
  \bibfield  {author} {\bibinfo {author} {\bibfnamefont {N.~J.}\ \bibnamefont
  {Fitch}}, \bibinfo {author} {\bibfnamefont {J.}~\bibnamefont {Lim}}, \bibinfo
  {author} {\bibfnamefont {E.~A.}\ \bibnamefont {Hinds}}, \bibinfo {author}
  {\bibfnamefont {B.~E.}\ \bibnamefont {Sauer}},\ and\ \bibinfo {author}
  {\bibfnamefont {M.~R.}\ \bibnamefont {Tarbutt}},\ }\bibfield  {title}
  {\bibinfo {title} {Methods for measuring the electron’s electric dipole
  moment using ultracold ybf molecules},\ }\href
  {https://doi.org/10.1088/2058-9565/abc931} {\bibfield  {journal} {\bibinfo
  {journal} {Quantum Science and Technology}\ }\textbf {\bibinfo {volume}
  {6}},\ \bibinfo {pages} {014006} (\bibinfo {year} {2020})}\BibitemShut
  {NoStop}%
\bibitem [{\citenamefont {Kozyryev}\ and\ \citenamefont
  {Hutzler}(2017)}]{Kozyryev2017}%
  \BibitemOpen
  \bibfield  {author} {\bibinfo {author} {\bibfnamefont {I.}~\bibnamefont
  {Kozyryev}}\ and\ \bibinfo {author} {\bibfnamefont {N.~R.}\ \bibnamefont
  {Hutzler}},\ }\bibfield  {title} {\bibinfo {title} {Precision measurement of
  time-reversal symmetry violation with laser-cooled polyatomic molecules},\
  }\href {https://doi.org/10.1103/PhysRevLett.119.133002} {\bibfield  {journal}
  {\bibinfo  {journal} {Phys. Rev. Lett.}\ }\textbf {\bibinfo {volume} {119}},\
  \bibinfo {pages} {133002} (\bibinfo {year} {2017})}\BibitemShut {NoStop}%
\bibitem [{\citenamefont {Aggarwal}\ \emph {et~al.}(2018)\citenamefont
  {Aggarwal}, \citenamefont {Bethlem}, \citenamefont {Borschevsky},
  \citenamefont {Denis}, \citenamefont {Esajas}, \citenamefont {Haase},
  \citenamefont {Hao}, \citenamefont {Hoekstra}, \citenamefont {Jungmann},
  \citenamefont {Meijknecht}, \citenamefont {Mooij}, \citenamefont
  {Timmermans}, \citenamefont {Ubachs}, \citenamefont {Willmann},\ and\
  \citenamefont {Zapara}}]{Aggarwal2018}%
  \BibitemOpen
  \bibfield  {author} {\bibinfo {author} {\bibfnamefont {P.}~\bibnamefont
  {Aggarwal}}, \bibinfo {author} {\bibfnamefont {H.~L.}\ \bibnamefont
  {Bethlem}}, \bibinfo {author} {\bibfnamefont {A.}~\bibnamefont
  {Borschevsky}}, \bibinfo {author} {\bibfnamefont {M.}~\bibnamefont {Denis}},
  \bibinfo {author} {\bibfnamefont {K.}~\bibnamefont {Esajas}}, \bibinfo
  {author} {\bibfnamefont {P.~A.~B.}\ \bibnamefont {Haase}}, \bibinfo {author}
  {\bibfnamefont {Y.}~\bibnamefont {Hao}}, \bibinfo {author} {\bibfnamefont
  {S.}~\bibnamefont {Hoekstra}}, \bibinfo {author} {\bibfnamefont
  {K.}~\bibnamefont {Jungmann}}, \bibinfo {author} {\bibfnamefont {T.~B.}\
  \bibnamefont {Meijknecht}}, \bibinfo {author} {\bibfnamefont {M.~C.}\
  \bibnamefont {Mooij}}, \bibinfo {author} {\bibfnamefont {R.~G.~E.}\
  \bibnamefont {Timmermans}}, \bibinfo {author} {\bibfnamefont
  {W.}~\bibnamefont {Ubachs}}, \bibinfo {author} {\bibfnamefont
  {L.}~\bibnamefont {Willmann}},\ and\ \bibinfo {author} {\bibfnamefont
  {A.}~\bibnamefont {Zapara}},\ }\bibfield  {title} {\bibinfo {title}
  {{Measuring the electric dipole moment of the electron in BaF}},\ }\href
  {https://doi.org/10.1140/epjd/e2018-90192-9} {\bibfield  {journal} {\bibinfo
  {journal} {Eur. Phys. J. D}\ }\textbf {\bibinfo {volume} {72}},\ \bibinfo
  {pages} {197} (\bibinfo {year} {2018})}\BibitemShut {NoStop}%
\bibitem [{\citenamefont {Fitch}\ and\ \citenamefont
  {Tarbutt}(2021)}]{Fitch2021}%
  \BibitemOpen
  \bibfield  {author} {\bibinfo {author} {\bibfnamefont {N.}~\bibnamefont
  {Fitch}}\ and\ \bibinfo {author} {\bibfnamefont {M.}~\bibnamefont
  {Tarbutt}},\ }\bibfield  {title} {\bibinfo {title} {Laser-cooled molecules},\
  }\href {https://doi.org/https://doi.org/10.1016/bs.aamop.2021.04.003}
  {\bibfield  {journal} {\bibinfo  {journal} {Adv. At. Mol. Opt.}\ }\textbf
  {\bibinfo {volume} {70}},\ \bibinfo {pages} {157} (\bibinfo {year}
  {2021})}\BibitemShut {NoStop}%
\bibitem [{\citenamefont {Langen}\ \emph {et~al.}(2024)\citenamefont {Langen},
  \citenamefont {Valtolina}, \citenamefont {Wang},\ and\ \citenamefont
  {Ye}}]{Langen2023}%
  \BibitemOpen
  \bibfield  {author} {\bibinfo {author} {\bibfnamefont {T.}~\bibnamefont
  {Langen}}, \bibinfo {author} {\bibfnamefont {G.}~\bibnamefont {Valtolina}},
  \bibinfo {author} {\bibfnamefont {D.}~\bibnamefont {Wang}},\ and\ \bibinfo
  {author} {\bibfnamefont {J.}~\bibnamefont {Ye}},\ }\bibfield  {title}
  {\bibinfo {title} {Quantum state manipulation and cooling of ultracold
  molecules},\ }\href {https://doi.org/10.1038/s41567-024-02423-1} {\bibfield
  {journal} {\bibinfo  {journal} {Nature Physics}\ }\textbf {\bibinfo {volume}
  {20}},\ \bibinfo {pages} {702} (\bibinfo {year} {2024})}\BibitemShut
  {NoStop}%
\bibitem [{\citenamefont {Rockenh\"auser}\ \emph {et~al.}(2023)\citenamefont
  {Rockenh\"auser}, \citenamefont {Kogel}, \citenamefont {Pultinevicius},\ and\
  \citenamefont {Langen}}]{Rockenhaeuser2023}%
  \BibitemOpen
  \bibfield  {author} {\bibinfo {author} {\bibfnamefont {M.}~\bibnamefont
  {Rockenh\"auser}}, \bibinfo {author} {\bibfnamefont {F.}~\bibnamefont
  {Kogel}}, \bibinfo {author} {\bibfnamefont {E.}~\bibnamefont
  {Pultinevicius}},\ and\ \bibinfo {author} {\bibfnamefont {T.}~\bibnamefont
  {Langen}},\ }\bibfield  {title} {\bibinfo {title} {Absorption spectroscopy
  for laser cooling and high-fidelity detection of barium monofluoride
  molecules},\ }\href {https://doi.org/10.1103/PhysRevA.108.062812} {\bibfield
  {journal} {\bibinfo  {journal} {Phys. Rev. A}\ }\textbf {\bibinfo {volume}
  {108}},\ \bibinfo {pages} {062812} (\bibinfo {year} {2023})}\BibitemShut
  {NoStop}%
\bibitem [{\citenamefont {Albrecht}\ \emph {et~al.}(2020)\citenamefont
  {Albrecht}, \citenamefont {Scharwaechter}, \citenamefont {Sixt},
  \citenamefont {Hofer},\ and\ \citenamefont {Langen}}]{Albrecht2020}%
  \BibitemOpen
  \bibfield  {author} {\bibinfo {author} {\bibfnamefont {R.}~\bibnamefont
  {Albrecht}}, \bibinfo {author} {\bibfnamefont {M.}~\bibnamefont
  {Scharwaechter}}, \bibinfo {author} {\bibfnamefont {T.}~\bibnamefont {Sixt}},
  \bibinfo {author} {\bibfnamefont {L.}~\bibnamefont {Hofer}},\ and\ \bibinfo
  {author} {\bibfnamefont {T.}~\bibnamefont {Langen}},\ }\bibfield  {title}
  {\bibinfo {title} {Buffer-gas cooling, high-resolution spectroscopy, and
  optical cycling of barium monofluoride molecules},\ }\href
  {https://doi.org/10.1103/PhysRevA.101.013413} {\bibfield  {journal} {\bibinfo
   {journal} {Phys. Rev. A}\ }\textbf {\bibinfo {volume} {101}},\ \bibinfo
  {pages} {013413} (\bibinfo {year} {2020})}\BibitemShut {NoStop}%
\bibitem [{\citenamefont {Chen}\ \emph {et~al.}(2017)\citenamefont {Chen},
  \citenamefont {Bu},\ and\ \citenamefont {Yan}}]{Chen2017}%
  \BibitemOpen
  \bibfield  {author} {\bibinfo {author} {\bibfnamefont {T.}~\bibnamefont
  {Chen}}, \bibinfo {author} {\bibfnamefont {W.}~\bibnamefont {Bu}},\ and\
  \bibinfo {author} {\bibfnamefont {B.}~\bibnamefont {Yan}},\ }\bibfield
  {title} {\bibinfo {title} {Radiative deflection of a {BaF} molecular beam via
  optical cycling},\ }\href {https://doi.org/10.1103/PhysRevA.96.053401}
  {\bibfield  {journal} {\bibinfo  {journal} {Phys. Rev. A}\ }\textbf {\bibinfo
  {volume} {96}},\ \bibinfo {pages} {053401} (\bibinfo {year}
  {2017})}\BibitemShut {NoStop}%
\bibitem [{\citenamefont {Rockenh\"auser}\ \emph {et~al.}(2024)\citenamefont
  {Rockenh\"auser}, \citenamefont {Kogel}, \citenamefont {Garg}, \citenamefont
  {Morales-Ram\'{\i}rez},\ and\ \citenamefont {Langen}}]{Rockenhaeuser2024}%
  \BibitemOpen
  \bibfield  {author} {\bibinfo {author} {\bibfnamefont {M.}~\bibnamefont
  {Rockenh\"auser}}, \bibinfo {author} {\bibfnamefont {F.}~\bibnamefont
  {Kogel}}, \bibinfo {author} {\bibfnamefont {T.}~\bibnamefont {Garg}},
  \bibinfo {author} {\bibfnamefont {S.~A.}\ \bibnamefont
  {Morales-Ram\'{\i}rez}},\ and\ \bibinfo {author} {\bibfnamefont
  {T.}~\bibnamefont {Langen}},\ }\bibfield  {title} {\bibinfo {title} {Laser
  cooling of barium monofluoride molecules using synthesized optical spectra},\
  }\href {https://doi.org/10.1103/PhysRevResearch.6.043161} {\bibfield
  {journal} {\bibinfo  {journal} {Phys. Rev. Res.}\ }\textbf {\bibinfo {volume}
  {6}},\ \bibinfo {pages} {043161} (\bibinfo {year} {2024})}\BibitemShut
  {NoStop}%
\bibitem [{\citenamefont {Kogel}\ \emph
  {et~al.}(2025{\natexlab{a}})\citenamefont {Kogel}, \citenamefont {Garg},
  \citenamefont {Rockenh{\"a}user}, \citenamefont {Morales-Ram{\'\i}rez},\ and\
  \citenamefont {Langen}}]{Kogel2024serrodynes}%
  \BibitemOpen
  \bibfield  {author} {\bibinfo {author} {\bibfnamefont {F.}~\bibnamefont
  {Kogel}}, \bibinfo {author} {\bibfnamefont {T.}~\bibnamefont {Garg}},
  \bibinfo {author} {\bibfnamefont {M.}~\bibnamefont {Rockenh{\"a}user}},
  \bibinfo {author} {\bibfnamefont {S.~A.}\ \bibnamefont
  {Morales-Ram{\'\i}rez}},\ and\ \bibinfo {author} {\bibfnamefont
  {T.}~\bibnamefont {Langen}},\ }\bibfield  {title} {\bibinfo {title}
  {Molecular laser cooling using serrodynes: implementation, characterization
  and prospects},\ }\href {https://doi.org/10.1088/1367-2630/add0d4} {\bibfield
   {journal} {\bibinfo  {journal} {New Journal of Physics}\ }\textbf {\bibinfo
  {volume} {27}},\ \bibinfo {pages} {055001} (\bibinfo {year}
  {2025}{\natexlab{a}})}\BibitemShut {NoStop}%
\bibitem [{\citenamefont {Zeng}\ \emph {et~al.}(2024)\citenamefont {Zeng},
  \citenamefont {Deng}, \citenamefont {Yang},\ and\ \citenamefont
  {Yan}}]{Zeng2024}%
  \BibitemOpen
  \bibfield  {author} {\bibinfo {author} {\bibfnamefont {Z.}~\bibnamefont
  {Zeng}}, \bibinfo {author} {\bibfnamefont {S.}~\bibnamefont {Deng}}, \bibinfo
  {author} {\bibfnamefont {S.}~\bibnamefont {Yang}},\ and\ \bibinfo {author}
  {\bibfnamefont {B.}~\bibnamefont {Yan}},\ }\bibfield  {title} {\bibinfo
  {title} {Three-dimensional magneto-optical trapping of barium monofluoride},\
  }\href {https://doi.org/10.1103/PhysRevLett.133.143404} {\bibfield  {journal}
  {\bibinfo  {journal} {Phys. Rev. Lett.}\ }\textbf {\bibinfo {volume} {133}},\
  \bibinfo {pages} {143404} (\bibinfo {year} {2024})}\BibitemShut {NoStop}%
\bibitem [{\citenamefont {Kogel}\ \emph
  {et~al.}(2025{\natexlab{b}})\citenamefont {Kogel}, \citenamefont {Garg},
  \citenamefont {Rockenh{\"{a}}user}, \citenamefont {Morales-Ram{\'{i}}rez},\
  and\ \citenamefont {Langen}}]{Kogel2024isotope}%
  \BibitemOpen
  \bibfield  {author} {\bibinfo {author} {\bibfnamefont {F.}~\bibnamefont
  {Kogel}}, \bibinfo {author} {\bibfnamefont {T.}~\bibnamefont {Garg}},
  \bibinfo {author} {\bibfnamefont {M.}~\bibnamefont {Rockenh{\"{a}}user}},
  \bibinfo {author} {\bibfnamefont {S.~A.}\ \bibnamefont
  {Morales-Ram{\'{i}}rez}},\ and\ \bibinfo {author} {\bibfnamefont
  {T.}~\bibnamefont {Langen}},\ }\bibfield  {title} {\bibinfo {title}
  {{Isotopologue-selective laser cooling of molecules}},\ }\href
  {https://doi.org/10.1088/1367-2630/ada3f0} {\bibfield  {journal} {\bibinfo
  {journal} {New Journal of Physics}\ }\textbf {\bibinfo {volume} {27}},\
  \bibinfo {pages} {13001} (\bibinfo {year} {2025}{\natexlab{b}})}\BibitemShut
  {NoStop}%
\bibitem [{\citenamefont {Kogel}\ \emph
  {et~al.}(2025{\natexlab{c}})\citenamefont {Kogel}, \citenamefont {Garg},
  \citenamefont {Rockenh\"auser},\ and\ \citenamefont
  {Langen}}]{Kogel2025lasercooling137}%
  \BibitemOpen
  \bibfield  {author} {\bibinfo {author} {\bibfnamefont {F.}~\bibnamefont
  {Kogel}}, \bibinfo {author} {\bibfnamefont {T.}~\bibnamefont {Garg}},
  \bibinfo {author} {\bibfnamefont {M.}~\bibnamefont {Rockenh\"auser}},\ and\
  \bibinfo {author} {\bibfnamefont {T.}~\bibnamefont {Langen}},\ }\bibfield
  {title} {\bibinfo {title} {Laser-cooled $^{137}\mathrm{BaF}$ molecules for
  measuring nuclear-spin-dependent parity violation},\ }\href
  {https://doi.org/10.1103/PhysRevResearch.7.L022041} {\bibfield  {journal}
  {\bibinfo  {journal} {Phys. Rev. Res.}\ }\textbf {\bibinfo {volume} {7}},\
  \bibinfo {pages} {L022041} (\bibinfo {year}
  {2025}{\natexlab{c}})}\BibitemShut {NoStop}%
\bibitem [{\citenamefont {Boeschoten}\ \emph {et~al.}(2024)\citenamefont
  {Boeschoten}, \citenamefont {Marshall}, \citenamefont {Meijknecht},
  \citenamefont {Touwen}, \citenamefont {Bethlem}, \citenamefont {Borschevsky},
  \citenamefont {Hoekstra}, \citenamefont {van Hofslot}, \citenamefont
  {Jungmann}, \citenamefont {Mooij}, \citenamefont {Timmermans}, \citenamefont
  {Ubachs},\ and\ \citenamefont {Willmann}}]{Boeschoten2024}%
  \BibitemOpen
  \bibfield  {author} {\bibinfo {author} {\bibfnamefont {A.}~\bibnamefont
  {Boeschoten}}, \bibinfo {author} {\bibfnamefont {V.~R.}\ \bibnamefont
  {Marshall}}, \bibinfo {author} {\bibfnamefont {T.~B.}\ \bibnamefont
  {Meijknecht}}, \bibinfo {author} {\bibfnamefont {A.}~\bibnamefont {Touwen}},
  \bibinfo {author} {\bibfnamefont {H.~L.}\ \bibnamefont {Bethlem}}, \bibinfo
  {author} {\bibfnamefont {A.}~\bibnamefont {Borschevsky}}, \bibinfo {author}
  {\bibfnamefont {S.}~\bibnamefont {Hoekstra}}, \bibinfo {author}
  {\bibfnamefont {J.~W.~F.}\ \bibnamefont {van Hofslot}}, \bibinfo {author}
  {\bibfnamefont {K.}~\bibnamefont {Jungmann}}, \bibinfo {author}
  {\bibfnamefont {M.~C.}\ \bibnamefont {Mooij}}, \bibinfo {author}
  {\bibfnamefont {R.~G.~E.}\ \bibnamefont {Timmermans}}, \bibinfo {author}
  {\bibfnamefont {W.}~\bibnamefont {Ubachs}},\ and\ \bibinfo {author}
  {\bibfnamefont {L.}~\bibnamefont {Willmann}} (\bibinfo {collaboration}
  {NL-$e$EDM Collaboration}),\ }\bibfield  {title} {\bibinfo {title}
  {Spin-precession method for sensitive electric dipole moment searches},\
  }\href {https://doi.org/10.1103/PhysRevA.110.L010801} {\bibfield  {journal}
  {\bibinfo  {journal} {Phys. Rev. A}\ }\textbf {\bibinfo {volume} {110}},\
  \bibinfo {pages} {L010801} (\bibinfo {year} {2024})}\BibitemShut {NoStop}%
\bibitem [{\citenamefont {Altuntas}\ \emph {et~al.}(2018)\citenamefont
  {Altuntas}, \citenamefont {Ammon}, \citenamefont {Cahn},\ and\ \citenamefont
  {DeMille}}]{Altuntas2018}%
  \BibitemOpen
  \bibfield  {author} {\bibinfo {author} {\bibfnamefont {E.}~\bibnamefont
  {Altuntas}}, \bibinfo {author} {\bibfnamefont {J.}~\bibnamefont {Ammon}},
  \bibinfo {author} {\bibfnamefont {S.~B.}\ \bibnamefont {Cahn}},\ and\
  \bibinfo {author} {\bibfnamefont {D.}~\bibnamefont {DeMille}},\ }\bibfield
  {title} {\bibinfo {title} {{Demonstration of a Sensitive Method to Measure
  Nuclear-Spin-Dependent Parity Violation}},\ }\href
  {https://doi.org/10.1103/PhysRevLett.120.142501} {\bibfield  {journal}
  {\bibinfo  {journal} {Phys. Rev. Lett.}\ }\textbf {\bibinfo {volume} {120}},\
  \bibinfo {pages} {142501} (\bibinfo {year} {2018})}\BibitemShut {NoStop}%
\bibitem [{\citenamefont {Phillips}\ \emph {et~al.}(1986)\citenamefont
  {Phillips}, \citenamefont {Ahmad}, \citenamefont {Emling}, \citenamefont
  {Holzmann}, \citenamefont {Janssens}, \citenamefont {Khoo},\ and\
  \citenamefont {Drigert}}]{Phillips1986}%
  \BibitemOpen
  \bibfield  {author} {\bibinfo {author} {\bibfnamefont {W.~R.}\ \bibnamefont
  {Phillips}}, \bibinfo {author} {\bibfnamefont {I.}~\bibnamefont {Ahmad}},
  \bibinfo {author} {\bibfnamefont {H.}~\bibnamefont {Emling}}, \bibinfo
  {author} {\bibfnamefont {R.}~\bibnamefont {Holzmann}}, \bibinfo {author}
  {\bibfnamefont {R.~V.~F.}\ \bibnamefont {Janssens}}, \bibinfo {author}
  {\bibfnamefont {T.~L.}\ \bibnamefont {Khoo}},\ and\ \bibinfo {author}
  {\bibfnamefont {M.~W.}\ \bibnamefont {Drigert}},\ }\bibfield  {title}
  {\bibinfo {title} {Octupole deformation in neutron-rich barium isotopes},\
  }\href {https://doi.org/10.1103/PhysRevLett.57.3257} {\bibfield  {journal}
  {\bibinfo  {journal} {Phys. Rev. Lett.}\ }\textbf {\bibinfo {volume} {57}},\
  \bibinfo {pages} {3257} (\bibinfo {year} {1986})}\BibitemShut {NoStop}%
\bibitem [{\citenamefont {Arrowsmith-Kron}\ \emph {et~al.}(2024)\citenamefont
  {Arrowsmith-Kron}, \citenamefont {Athanasakis-Kaklamanakis}, \citenamefont
  {Au}, \citenamefont {Ballof}, \citenamefont {Berger}, \citenamefont
  {Borschevsky}, \citenamefont {Breier}, \citenamefont {Buchinger},
  \citenamefont {Budker}, \citenamefont {Caldwell}, \citenamefont {Charles},
  \citenamefont {Dattani}, \citenamefont {de~Groote}, \citenamefont {DeMille},
  \citenamefont {Dickel}, \citenamefont {Dobaczewski}, \citenamefont
  {Düllmann}, \citenamefont {Eliav}, \citenamefont {Engel}, \citenamefont
  {Fan}, \citenamefont {Flambaum}, \citenamefont {Flanagan}, \citenamefont
  {Gaiser}, \citenamefont {Ruiz}, \citenamefont {Gaul}, \citenamefont {Giesen},
  \citenamefont {Ginges}, \citenamefont {Gottberg}, \citenamefont {Gwinner},
  \citenamefont {Heinke}, \citenamefont {Hoekstra}, \citenamefont {Holt},
  \citenamefont {Hutzler}, \citenamefont {Jayich}, \citenamefont {Karthein},
  \citenamefont {Leach}, \citenamefont {Madison}, \citenamefont
  {Malbrunot-Ettenauer}, \citenamefont {Miyagi}, \citenamefont {Moore},
  \citenamefont {Moroch}, \citenamefont {Navratil}, \citenamefont {Nazarewicz},
  \citenamefont {Neyens}, \citenamefont {Norrgard}, \citenamefont {Nusgart},
  \citenamefont {Pašteka}, \citenamefont {Petrov}, \citenamefont {Plaß},
  \citenamefont {Ready}, \citenamefont {Reiter}, \citenamefont {Reponen},
  \citenamefont {Rothe}, \citenamefont {Safronova}, \citenamefont
  {Scheidenerger}, \citenamefont {Shindler}, \citenamefont {Singh},
  \citenamefont {Skripnikov}, \citenamefont {Titov}, \citenamefont {Udrescu},
  \citenamefont {Wilkins},\ and\ \citenamefont {Yang}}]{ArrowsmithKron2024}%
  \BibitemOpen
  \bibfield  {author} {\bibinfo {author} {\bibfnamefont {G.}~\bibnamefont
  {Arrowsmith-Kron}}, \bibinfo {author} {\bibfnamefont {M.}~\bibnamefont
  {Athanasakis-Kaklamanakis}}, \bibinfo {author} {\bibfnamefont
  {M.}~\bibnamefont {Au}}, \bibinfo {author} {\bibfnamefont {J.}~\bibnamefont
  {Ballof}}, \bibinfo {author} {\bibfnamefont {R.}~\bibnamefont {Berger}},
  \bibinfo {author} {\bibfnamefont {A.}~\bibnamefont {Borschevsky}}, \bibinfo
  {author} {\bibfnamefont {A.~A.}\ \bibnamefont {Breier}}, \bibinfo {author}
  {\bibfnamefont {F.}~\bibnamefont {Buchinger}}, \bibinfo {author}
  {\bibfnamefont {D.}~\bibnamefont {Budker}}, \bibinfo {author} {\bibfnamefont
  {L.}~\bibnamefont {Caldwell}}, \bibinfo {author} {\bibfnamefont
  {C.}~\bibnamefont {Charles}}, \bibinfo {author} {\bibfnamefont
  {N.}~\bibnamefont {Dattani}}, \bibinfo {author} {\bibfnamefont {R.~P.}\
  \bibnamefont {de~Groote}}, \bibinfo {author} {\bibfnamefont {D.}~\bibnamefont
  {DeMille}}, \bibinfo {author} {\bibfnamefont {T.}~\bibnamefont {Dickel}},
  \bibinfo {author} {\bibfnamefont {J.}~\bibnamefont {Dobaczewski}}, \bibinfo
  {author} {\bibfnamefont {C.~E.}\ \bibnamefont {Düllmann}}, \bibinfo {author}
  {\bibfnamefont {E.}~\bibnamefont {Eliav}}, \bibinfo {author} {\bibfnamefont
  {J.}~\bibnamefont {Engel}}, \bibinfo {author} {\bibfnamefont
  {M.}~\bibnamefont {Fan}}, \bibinfo {author} {\bibfnamefont {V.}~\bibnamefont
  {Flambaum}}, \bibinfo {author} {\bibfnamefont {K.~T.}\ \bibnamefont
  {Flanagan}}, \bibinfo {author} {\bibfnamefont {A.~N.}\ \bibnamefont
  {Gaiser}}, \bibinfo {author} {\bibfnamefont {R.~F.~G.}\ \bibnamefont {Ruiz}},
  \bibinfo {author} {\bibfnamefont {K.}~\bibnamefont {Gaul}}, \bibinfo {author}
  {\bibfnamefont {T.~F.}\ \bibnamefont {Giesen}}, \bibinfo {author}
  {\bibfnamefont {J.~S.~M.}\ \bibnamefont {Ginges}}, \bibinfo {author}
  {\bibfnamefont {A.}~\bibnamefont {Gottberg}}, \bibinfo {author}
  {\bibfnamefont {G.}~\bibnamefont {Gwinner}}, \bibinfo {author} {\bibfnamefont
  {R.}~\bibnamefont {Heinke}}, \bibinfo {author} {\bibfnamefont
  {S.}~\bibnamefont {Hoekstra}}, \bibinfo {author} {\bibfnamefont {J.~D.}\
  \bibnamefont {Holt}}, \bibinfo {author} {\bibfnamefont {N.~R.}\ \bibnamefont
  {Hutzler}}, \bibinfo {author} {\bibfnamefont {A.}~\bibnamefont {Jayich}},
  \bibinfo {author} {\bibfnamefont {J.}~\bibnamefont {Karthein}}, \bibinfo
  {author} {\bibfnamefont {K.~G.}\ \bibnamefont {Leach}}, \bibinfo {author}
  {\bibfnamefont {K.~W.}\ \bibnamefont {Madison}}, \bibinfo {author}
  {\bibfnamefont {S.}~\bibnamefont {Malbrunot-Ettenauer}}, \bibinfo {author}
  {\bibfnamefont {T.}~\bibnamefont {Miyagi}}, \bibinfo {author} {\bibfnamefont
  {I.~D.}\ \bibnamefont {Moore}}, \bibinfo {author} {\bibfnamefont
  {S.}~\bibnamefont {Moroch}}, \bibinfo {author} {\bibfnamefont
  {P.}~\bibnamefont {Navratil}}, \bibinfo {author} {\bibfnamefont
  {W.}~\bibnamefont {Nazarewicz}}, \bibinfo {author} {\bibfnamefont
  {G.}~\bibnamefont {Neyens}}, \bibinfo {author} {\bibfnamefont {E.~B.}\
  \bibnamefont {Norrgard}}, \bibinfo {author} {\bibfnamefont {N.}~\bibnamefont
  {Nusgart}}, \bibinfo {author} {\bibfnamefont {L.~F.}\ \bibnamefont
  {Pašteka}}, \bibinfo {author} {\bibfnamefont {A.~N.}\ \bibnamefont
  {Petrov}}, \bibinfo {author} {\bibfnamefont {W.~R.}\ \bibnamefont {Plaß}},
  \bibinfo {author} {\bibfnamefont {R.~A.}\ \bibnamefont {Ready}}, \bibinfo
  {author} {\bibfnamefont {M.~P.}\ \bibnamefont {Reiter}}, \bibinfo {author}
  {\bibfnamefont {M.}~\bibnamefont {Reponen}}, \bibinfo {author} {\bibfnamefont
  {S.}~\bibnamefont {Rothe}}, \bibinfo {author} {\bibfnamefont {M.~S.}\
  \bibnamefont {Safronova}}, \bibinfo {author} {\bibfnamefont {C.}~\bibnamefont
  {Scheidenerger}}, \bibinfo {author} {\bibfnamefont {A.}~\bibnamefont
  {Shindler}}, \bibinfo {author} {\bibfnamefont {J.~T.}\ \bibnamefont {Singh}},
  \bibinfo {author} {\bibfnamefont {L.~V.}\ \bibnamefont {Skripnikov}},
  \bibinfo {author} {\bibfnamefont {A.~V.}\ \bibnamefont {Titov}}, \bibinfo
  {author} {\bibfnamefont {S.-M.}\ \bibnamefont {Udrescu}}, \bibinfo {author}
  {\bibfnamefont {S.~G.}\ \bibnamefont {Wilkins}},\ and\ \bibinfo {author}
  {\bibfnamefont {X.}~\bibnamefont {Yang}},\ }\bibfield  {title} {\bibinfo
  {title} {Opportunities for fundamental physics research with radioactive
  molecules},\ }\href {https://doi.org/10.1088/1361-6633/ad1e39} {\bibfield
  {journal} {\bibinfo  {journal} {Reports on Progress in Physics}\ }\textbf
  {\bibinfo {volume} {87}},\ \bibinfo {pages} {084301} (\bibinfo {year}
  {2024})}\BibitemShut {NoStop}%
\bibitem [{\citenamefont {Bu}\ \emph {et~al.}(2022)\citenamefont {Bu},
  \citenamefont {Zhang}, \citenamefont {Liang}, \citenamefont {Chen},\ and\
  \citenamefont {Yan}}]{Bu2022}%
  \BibitemOpen
  \bibfield  {author} {\bibinfo {author} {\bibfnamefont {W.}~\bibnamefont
  {Bu}}, \bibinfo {author} {\bibfnamefont {Y.}~\bibnamefont {Zhang}}, \bibinfo
  {author} {\bibfnamefont {Q.}~\bibnamefont {Liang}}, \bibinfo {author}
  {\bibfnamefont {T.}~\bibnamefont {Chen}},\ and\ \bibinfo {author}
  {\bibfnamefont {B.}~\bibnamefont {Yan}},\ }\bibfield  {title} {\bibinfo
  {title} {{Saturated absorption spectroscopy of buffer-gas-cooled Barium
  monofluoride molecules}},\ }\href {https://doi.org/10.1007/s11467-022-1194-x}
  {\bibfield  {journal} {\bibinfo  {journal} {Frontiers of Physics}\ }\textbf
  {\bibinfo {volume} {17}},\ \bibinfo {pages} {62502} (\bibinfo {year}
  {2022})}\BibitemShut {NoStop}%
\bibitem [{\citenamefont {Ryzlewicz}\ and\ \citenamefont
  {Törring}(1980)}]{Ryzlewicz1980}%
  \BibitemOpen
  \bibfield  {author} {\bibinfo {author} {\bibfnamefont {C.}~\bibnamefont
  {Ryzlewicz}}\ and\ \bibinfo {author} {\bibfnamefont {T.}~\bibnamefont
  {Törring}},\ }\bibfield  {title} {\bibinfo {title} {Formation and microwave
  spectrum of the {${}^2\Sigma$} radical barium monofluoride},\ }\href
  {https://doi.org/https://doi.org/10.1016/0301-0104(80)80107-8} {\bibfield
  {journal} {\bibinfo  {journal} {Chemical Physics}\ }\textbf {\bibinfo
  {volume} {51}},\ \bibinfo {pages} {329} (\bibinfo {year} {1980})}\BibitemShut
  {NoStop}%
\bibitem [{\citenamefont {Effantin}\ \emph {et~al.}(1990)\citenamefont
  {Effantin}, \citenamefont {Bernard}, \citenamefont {d'Incan}, \citenamefont
  {Wannous}, \citenamefont {Vergès},\ and\ \citenamefont
  {Barrow}}]{Effantin1990}%
  \BibitemOpen
  \bibfield  {author} {\bibinfo {author} {\bibfnamefont {C.}~\bibnamefont
  {Effantin}}, \bibinfo {author} {\bibfnamefont {A.}~\bibnamefont {Bernard}},
  \bibinfo {author} {\bibfnamefont {J.}~\bibnamefont {d'Incan}}, \bibinfo
  {author} {\bibfnamefont {G.}~\bibnamefont {Wannous}}, \bibinfo {author}
  {\bibfnamefont {J.}~\bibnamefont {Vergès}},\ and\ \bibinfo {author}
  {\bibfnamefont {R.}~\bibnamefont {Barrow}},\ }\bibfield  {title} {\bibinfo
  {title} {Studies of the electronic states of the {BaF} molecule},\ }\href
  {https://doi.org/10.1080/00268979000101311} {\bibfield  {journal} {\bibinfo
  {journal} {Molecular Physics}\ }\textbf {\bibinfo {volume} {70}},\ \bibinfo
  {pages} {735} (\bibinfo {year} {1990})}\BibitemShut {NoStop}%
\bibitem [{\citenamefont {Ernst}\ \emph {et~al.}(1986)\citenamefont {Ernst},
  \citenamefont {K{\"{a}}ndler},\ and\ \citenamefont
  {T{\"{o}}rring}}]{Ernst1986}%
  \BibitemOpen
  \bibfield  {author} {\bibinfo {author} {\bibfnamefont {W.~E.}\ \bibnamefont
  {Ernst}}, \bibinfo {author} {\bibfnamefont {J.}~\bibnamefont
  {K{\"{a}}ndler}},\ and\ \bibinfo {author} {\bibfnamefont {T.}~\bibnamefont
  {T{\"{o}}rring}},\ }\bibfield  {title} {\bibinfo {title} {{Hyperfine
  structure and electric dipole moment of BaF X2$\Sigma$+}},\ }\href
  {https://doi.org/10.1063/1.449961} {\bibfield  {journal} {\bibinfo  {journal}
  {The Journal of Chemical Physics}\ }\textbf {\bibinfo {volume} {84}},\
  \bibinfo {pages} {4769} (\bibinfo {year} {1986})}\BibitemShut {NoStop}%
\bibitem [{\citenamefont {Steimle}\ \emph {et~al.}(2011)\citenamefont
  {Steimle}, \citenamefont {Frey}, \citenamefont {Le}, \citenamefont {DeMille},
  \citenamefont {Rahmlow},\ and\ \citenamefont {Linton}}]{Steimle2011}%
  \BibitemOpen
  \bibfield  {author} {\bibinfo {author} {\bibfnamefont {T.~C.}\ \bibnamefont
  {Steimle}}, \bibinfo {author} {\bibfnamefont {S.}~\bibnamefont {Frey}},
  \bibinfo {author} {\bibfnamefont {A.}~\bibnamefont {Le}}, \bibinfo {author}
  {\bibfnamefont {D.}~\bibnamefont {DeMille}}, \bibinfo {author} {\bibfnamefont
  {D.~A.}\ \bibnamefont {Rahmlow}},\ and\ \bibinfo {author} {\bibfnamefont
  {C.}~\bibnamefont {Linton}},\ }\bibfield  {title} {\bibinfo {title}
  {Molecular-beam optical {Stark} and {Zeeman} study of the {$A$
  ${}^{2}\ensuremath{\Pi}$--$X$ ${}^{2}{\ensuremath{\Sigma}}^{+}$} (0,0) band
  system of {BaF}},\ }\href {https://doi.org/10.1103/PhysRevA.84.012508}
  {\bibfield  {journal} {\bibinfo  {journal} {Phys. Rev. A}\ }\textbf {\bibinfo
  {volume} {84}},\ \bibinfo {pages} {012508} (\bibinfo {year}
  {2011})}\BibitemShut {NoStop}%
\bibitem [{\citenamefont {Preston}\ \emph {et~al.}(2025)\citenamefont
  {Preston}, \citenamefont {Aufderheide}, \citenamefont {Ballard},
  \citenamefont {Mawhorter},\ and\ \citenamefont {Grabow}}]{Preston2025}%
  \BibitemOpen
  \bibfield  {author} {\bibinfo {author} {\bibfnamefont {A.}~\bibnamefont
  {Preston}}, \bibinfo {author} {\bibfnamefont {G.}~\bibnamefont
  {Aufderheide}}, \bibinfo {author} {\bibfnamefont {W.}~\bibnamefont
  {Ballard}}, \bibinfo {author} {\bibfnamefont {R.}~\bibnamefont {Mawhorter}},\
  and\ \bibinfo {author} {\bibfnamefont {J.-U.}\ \bibnamefont {Grabow}},\
  }\href {https://arxiv.org/abs/2503.21688} {\bibinfo {title} {{Global isotopic
  analysis of rotational spectroscopic data for barium monofluoride, BaF}}}
  (\bibinfo {year} {2025}),\ \Eprint {https://arxiv.org/abs/2503.21688}
  {arXiv:2503.21688 [physics.atom-ph]} \BibitemShut {NoStop}%
\bibitem [{\citenamefont {Kogel}\ \emph {et~al.}(2021)\citenamefont {Kogel},
  \citenamefont {Rockenhäuser}, \citenamefont {Albrecht},\ and\ \citenamefont
  {Langen}}]{Kogel2021}%
  \BibitemOpen
  \bibfield  {author} {\bibinfo {author} {\bibfnamefont {F.}~\bibnamefont
  {Kogel}}, \bibinfo {author} {\bibfnamefont {M.}~\bibnamefont
  {Rockenhäuser}}, \bibinfo {author} {\bibfnamefont {R.}~\bibnamefont
  {Albrecht}},\ and\ \bibinfo {author} {\bibfnamefont {T.}~\bibnamefont
  {Langen}},\ }\bibfield  {title} {\bibinfo {title} {A laser cooling scheme for
  precision measurements using fermionic barium monofluoride ({137Ba19F})
  molecules},\ }\href {https://doi.org/10.1088/1367-2630/ac1df2} {\bibfield
  {journal} {\bibinfo  {journal} {New Journal of Physics}\ }\textbf {\bibinfo
  {volume} {23}},\ \bibinfo {pages} {095003} (\bibinfo {year}
  {2021})}\BibitemShut {NoStop}%
\bibitem [{\citenamefont {Altunta\ifmmode~\mbox{\c{s}}\else \c{s}\fi{}}\ \emph
  {et~al.}(2018)\citenamefont {Altunta\ifmmode~\mbox{\c{s}}\else \c{s}\fi{}},
  \citenamefont {Ammon}, \citenamefont {Cahn},\ and\ \citenamefont
  {DeMille}}]{AltuntasPRA}%
  \BibitemOpen
  \bibfield  {author} {\bibinfo {author} {\bibfnamefont {E.}~\bibnamefont
  {Altunta\ifmmode~\mbox{\c{s}}\else \c{s}\fi{}}}, \bibinfo {author}
  {\bibfnamefont {J.}~\bibnamefont {Ammon}}, \bibinfo {author} {\bibfnamefont
  {S.~B.}\ \bibnamefont {Cahn}},\ and\ \bibinfo {author} {\bibfnamefont
  {D.}~\bibnamefont {DeMille}},\ }\bibfield  {title} {\bibinfo {title}
  {Measuring nuclear-spin-dependent parity violation with molecules:
  Experimental methods and analysis of systematic errors},\ }\href
  {https://doi.org/10.1103/PhysRevA.97.042101} {\bibfield  {journal} {\bibinfo
  {journal} {Phys. Rev. A}\ }\textbf {\bibinfo {volume} {97}},\ \bibinfo
  {pages} {042101} (\bibinfo {year} {2018})}\BibitemShut {NoStop}%
\bibitem [{\citenamefont {Denis}\ \emph {et~al.}(2022)\citenamefont {Denis},
  \citenamefont {Haase}, \citenamefont {Mooij}, \citenamefont {Chamorro},
  \citenamefont {Aggarwal}, \citenamefont {Bethlem}, \citenamefont
  {Boeschoten}, \citenamefont {Borschevsky}, \citenamefont {Esajas},
  \citenamefont {Hao} \emph {et~al.}}]{Denis2022}%
  \BibitemOpen
  \bibfield  {author} {\bibinfo {author} {\bibfnamefont {M.}~\bibnamefont
  {Denis}}, \bibinfo {author} {\bibfnamefont {P.~A.}\ \bibnamefont {Haase}},
  \bibinfo {author} {\bibfnamefont {M.~C.}\ \bibnamefont {Mooij}}, \bibinfo
  {author} {\bibfnamefont {Y.}~\bibnamefont {Chamorro}}, \bibinfo {author}
  {\bibfnamefont {P.}~\bibnamefont {Aggarwal}}, \bibinfo {author}
  {\bibfnamefont {H.~L.}\ \bibnamefont {Bethlem}}, \bibinfo {author}
  {\bibfnamefont {A.}~\bibnamefont {Boeschoten}}, \bibinfo {author}
  {\bibfnamefont {A.}~\bibnamefont {Borschevsky}}, \bibinfo {author}
  {\bibfnamefont {K.}~\bibnamefont {Esajas}}, \bibinfo {author} {\bibfnamefont
  {Y.}~\bibnamefont {Hao}}, \emph {et~al.},\ }\bibfield  {title} {\bibinfo
  {title} {Benchmarking of the fock-space coupled-cluster method and
  uncertainty estimation: Magnetic hyperfine interaction in the excited state
  of baf},\ }\href@noop {} {\bibfield  {journal} {\bibinfo  {journal} {Physical
  Review A}\ }\textbf {\bibinfo {volume} {105}},\ \bibinfo {pages} {052811}
  (\bibinfo {year} {2022})}\BibitemShut {NoStop}%
\bibitem [{\citenamefont {Hutzler}\ \emph {et~al.}(2011)\citenamefont
  {Hutzler}, \citenamefont {Parsons}, \citenamefont {Gurevich}, \citenamefont
  {Hess}, \citenamefont {Petrik}, \citenamefont {Spaun}, \citenamefont {Vutha},
  \citenamefont {DeMille}, \citenamefont {Gabrielse},\ and\ \citenamefont
  {Doyle}}]{Hutzler2011}%
  \BibitemOpen
  \bibfield  {author} {\bibinfo {author} {\bibfnamefont {N.~R.}\ \bibnamefont
  {Hutzler}}, \bibinfo {author} {\bibfnamefont {M.~F.}\ \bibnamefont
  {Parsons}}, \bibinfo {author} {\bibfnamefont {Y.~V.}\ \bibnamefont
  {Gurevich}}, \bibinfo {author} {\bibfnamefont {P.~W.}\ \bibnamefont {Hess}},
  \bibinfo {author} {\bibfnamefont {E.}~\bibnamefont {Petrik}}, \bibinfo
  {author} {\bibfnamefont {B.}~\bibnamefont {Spaun}}, \bibinfo {author}
  {\bibfnamefont {A.~C.}\ \bibnamefont {Vutha}}, \bibinfo {author}
  {\bibfnamefont {D.}~\bibnamefont {DeMille}}, \bibinfo {author} {\bibfnamefont
  {G.}~\bibnamefont {Gabrielse}},\ and\ \bibinfo {author} {\bibfnamefont
  {J.~M.}\ \bibnamefont {Doyle}},\ }\bibfield  {title} {\bibinfo {title} {A
  cryogenic beam of refractory{,} chemically reactive molecules with expansion
  cooling},\ }\href {https://doi.org/10.1039/C1CP20901A} {\bibfield  {journal}
  {\bibinfo  {journal} {Phys. Chem. Chem. Phys.}\ }\textbf {\bibinfo {volume}
  {13}},\ \bibinfo {pages} {18976} (\bibinfo {year} {2011})}\BibitemShut
  {NoStop}%
\bibitem [{\citenamefont {Hutzler}\ \emph {et~al.}(2012)\citenamefont
  {Hutzler}, \citenamefont {Lu},\ and\ \citenamefont {Doyle}}]{Hutzler2012}%
  \BibitemOpen
  \bibfield  {author} {\bibinfo {author} {\bibfnamefont {N.~R.}\ \bibnamefont
  {Hutzler}}, \bibinfo {author} {\bibfnamefont {H.-I.}\ \bibnamefont {Lu}},\
  and\ \bibinfo {author} {\bibfnamefont {J.~M.}\ \bibnamefont {Doyle}},\
  }\bibfield  {title} {\bibinfo {title} {The buffer gas beam: An intense, cold,
  and slow source for atoms and molecules},\ }\href
  {https://doi.org/10.1021/cr200362u} {\bibfield  {journal} {\bibinfo
  {journal} {Chemical Reviews}\ }\textbf {\bibinfo {volume} {112}},\ \bibinfo
  {pages} {4803} (\bibinfo {year} {2012})},\ \bibinfo {note} {pMID:
  22571401}\BibitemShut {NoStop}%
\bibitem [{\citenamefont {Bu}\ \emph {et~al.}(2017)\citenamefont {Bu},
  \citenamefont {Chen}, \citenamefont {Lv},\ and\ \citenamefont
  {Yan}}]{Bu2017}%
  \BibitemOpen
  \bibfield  {author} {\bibinfo {author} {\bibfnamefont {W.}~\bibnamefont
  {Bu}}, \bibinfo {author} {\bibfnamefont {T.}~\bibnamefont {Chen}}, \bibinfo
  {author} {\bibfnamefont {G.}~\bibnamefont {Lv}},\ and\ \bibinfo {author}
  {\bibfnamefont {B.}~\bibnamefont {Yan}},\ }\bibfield  {title} {\bibinfo
  {title} {Cold collision and high-resolution spectroscopy of buffer-gas-cooled
  baf molecules},\ }\href {https://doi.org/10.1103/PhysRevA.95.032701}
  {\bibfield  {journal} {\bibinfo  {journal} {Phys. Rev. A}\ }\textbf {\bibinfo
  {volume} {95}},\ \bibinfo {pages} {032701} (\bibinfo {year}
  {2017})}\BibitemShut {NoStop}%
\bibitem [{\citenamefont {Zhao}\ \emph {et~al.}(1998)\citenamefont {Zhao},
  \citenamefont {Simsarian}, \citenamefont {Orozco},\ and\ \citenamefont
  {Sprouse}}]{Zhao1998}%
  \BibitemOpen
  \bibfield  {author} {\bibinfo {author} {\bibfnamefont {W.~Z.}\ \bibnamefont
  {Zhao}}, \bibinfo {author} {\bibfnamefont {J.~E.}\ \bibnamefont {Simsarian}},
  \bibinfo {author} {\bibfnamefont {L.~A.}\ \bibnamefont {Orozco}},\ and\
  \bibinfo {author} {\bibfnamefont {G.~D.}\ \bibnamefont {Sprouse}},\
  }\bibfield  {title} {\bibinfo {title} {A computer-based digital feedback
  control of frequency drift of multiple lasers},\ }\href
  {https://doi.org/10.1063/1.1149171} {\bibfield  {journal} {\bibinfo
  {journal} {Review of Scientific Instruments}\ }\textbf {\bibinfo {volume}
  {69}},\ \bibinfo {pages} {3737} (\bibinfo {year} {1998})}\BibitemShut
  {NoStop}%
\bibitem [{\citenamefont {Gomez}\ \emph {et~al.}(2004)\citenamefont {Gomez},
  \citenamefont {Aubin}, \citenamefont {Orozco},\ and\ \citenamefont
  {Sprouse}}]{Gomez2004}%
  \BibitemOpen
  \bibfield  {author} {\bibinfo {author} {\bibfnamefont {E.}~\bibnamefont
  {Gomez}}, \bibinfo {author} {\bibfnamefont {S.}~\bibnamefont {Aubin}},
  \bibinfo {author} {\bibfnamefont {L.~A.}\ \bibnamefont {Orozco}},\ and\
  \bibinfo {author} {\bibfnamefont {G.~D.}\ \bibnamefont {Sprouse}},\
  }\bibfield  {title} {\bibinfo {title} {Lifetime and hyperfine splitting
  measurements on the 7s and 6p levels in rubidium},\ }\href
  {https://doi.org/10.1364/JOSAB.21.002058} {\bibfield  {journal} {\bibinfo
  {journal} {J. Opt. Soc. Am. B}\ }\textbf {\bibinfo {volume} {21}},\ \bibinfo
  {pages} {2058} (\bibinfo {year} {2004})}\BibitemShut {NoStop}%
\bibitem [{\citenamefont {Rahmlow}(2010)}]{DaveRahmlow}%
  \BibitemOpen
  \bibfield  {author} {\bibinfo {author} {\bibfnamefont {D.}~\bibnamefont
  {Rahmlow}},\ }\emph {\bibinfo {title} {Towards a measurement of parity
  nonconservation in diatomic molecules}},\ \href@noop {} {Ph.D. thesis},\
  \bibinfo  {school} {Yale University} (\bibinfo {year} {2010})\BibitemShut
  {NoStop}%
\bibitem [{\citenamefont {Shimizu}\ and\ \citenamefont
  {Shimizu}(1983)}]{Shimizu1983}%
  \BibitemOpen
  \bibfield  {author} {\bibinfo {author} {\bibfnamefont {K.}~\bibnamefont
  {Shimizu}}\ and\ \bibinfo {author} {\bibfnamefont {F.}~\bibnamefont
  {Shimizu}},\ }\bibfield  {title} {\bibinfo {title} {Laser induced
  fluorescence spectra of the a {$^3\Pi$}u-{X$^1\Sigma^+$} g band of {$Na_2$}
  by molecular beam},\ }\href@noop {} {\bibfield  {journal} {\bibinfo
  {journal} {Journal of Chemical Physics}\ }\textbf {\bibinfo {volume} {78}},\
  \bibinfo {pages} {1126} (\bibinfo {year} {1983})}\BibitemShut {NoStop}%
\bibitem [{\citenamefont {Ryzlewicz}\ \emph {et~al.}(1982)\citenamefont
  {Ryzlewicz}, \citenamefont {Schütze-Pahlmann}, \citenamefont {Hoeft},\ and\
  \citenamefont {Törring}}]{Ryzlewicz1982}%
  \BibitemOpen
  \bibfield  {author} {\bibinfo {author} {\bibfnamefont {C.}~\bibnamefont
  {Ryzlewicz}}, \bibinfo {author} {\bibfnamefont {H.-U.}\ \bibnamefont
  {Schütze-Pahlmann}}, \bibinfo {author} {\bibfnamefont {J.}~\bibnamefont
  {Hoeft}},\ and\ \bibinfo {author} {\bibfnamefont {T.}~\bibnamefont
  {Törring}},\ }\bibfield  {title} {\bibinfo {title} {{Rotational spectrum and
  hyperfine structure of the $^2\Sigma$ radicals BaF and BaCl}},\ }\href
  {https://doi.org/https://doi.org/10.1016/0301-0104(82)85045-3} {\bibfield
  {journal} {\bibinfo  {journal} {Chemical Physics}\ }\textbf {\bibinfo
  {volume} {71}},\ \bibinfo {pages} {389} (\bibinfo {year} {1982})}\BibitemShut
  {NoStop}%
\bibitem [{\citenamefont {Hay}(1941)}]{Hay1941}%
  \BibitemOpen
  \bibfield  {author} {\bibinfo {author} {\bibfnamefont {R.~H.}\ \bibnamefont
  {Hay}},\ }\bibfield  {title} {\bibinfo {title} {{The Nuclear Magnetic Moments
  of ${\mathrm{C}}^{13}$, ${\mathrm{Ba}}^{135}$ and ${\mathrm{Ba}}^{137}$}},\
  }\href {https://doi.org/10.1103/PhysRev.60.75} {\bibfield  {journal}
  {\bibinfo  {journal} {Phys. Rev.}\ }\textbf {\bibinfo {volume} {60}},\
  \bibinfo {pages} {75} (\bibinfo {year} {1941})}\BibitemShut {NoStop}%
\bibitem [{\citenamefont {Pyykkö}(2008)}]{Pekka2008}%
  \BibitemOpen
  \bibfield  {author} {\bibinfo {author} {\bibfnamefont {P.}~\bibnamefont
  {Pyykkö}},\ }\bibfield  {title} {\bibinfo {title} {Year-2008 nuclear
  quadrupole moments},\ }\href {https://doi.org/10.1080/00268970802018367}
  {\bibfield  {journal} {\bibinfo  {journal} {Molecular Physics}\ }\textbf
  {\bibinfo {volume} {106}},\ \bibinfo {pages} {1965} (\bibinfo {year}
  {2008})}\BibitemShut {NoStop}%
\bibitem [{\citenamefont {Hao}\ \emph {et~al.}(2018)\citenamefont {Hao},
  \citenamefont {Ilias}, \citenamefont {Eliav}, \citenamefont {Schwerdtfeger},
  \citenamefont {Flambaum},\ and\ \citenamefont {Borschevsky}}]{Hao2018}%
  \BibitemOpen
  \bibfield  {author} {\bibinfo {author} {\bibfnamefont {Y.}~\bibnamefont
  {Hao}}, \bibinfo {author} {\bibfnamefont {M.}~\bibnamefont {Ilias}}, \bibinfo
  {author} {\bibfnamefont {E.}~\bibnamefont {Eliav}}, \bibinfo {author}
  {\bibfnamefont {P.}~\bibnamefont {Schwerdtfeger}}, \bibinfo {author}
  {\bibfnamefont {V.~V.}\ \bibnamefont {Flambaum}},\ and\ \bibinfo {author}
  {\bibfnamefont {A.}~\bibnamefont {Borschevsky}},\ }\bibfield  {title}
  {\bibinfo {title} {Nuclear anapole moment interaction in baf from
  relativistic coupled-cluster theory},\ }\href
  {https://doi.org/10.1103/PhysRevA.98.032510} {\bibfield  {journal} {\bibinfo
  {journal} {Phys. Rev. A}\ }\textbf {\bibinfo {volume} {98}},\ \bibinfo
  {pages} {032510} (\bibinfo {year} {2018})}\BibitemShut {NoStop}%
\bibitem [{\citenamefont {Denis}\ \emph {et~al.}(2019)\citenamefont {Denis},
  \citenamefont {Haase}, \citenamefont {Timmermans}, \citenamefont {Eliav},
  \citenamefont {Hutzler},\ and\ \citenamefont {Borschevsky}}]{Denis2019}%
  \BibitemOpen
  \bibfield  {author} {\bibinfo {author} {\bibfnamefont {M.}~\bibnamefont
  {Denis}}, \bibinfo {author} {\bibfnamefont {P.~A.~B.}\ \bibnamefont {Haase}},
  \bibinfo {author} {\bibfnamefont {R.~G.~E.}\ \bibnamefont {Timmermans}},
  \bibinfo {author} {\bibfnamefont {E.}~\bibnamefont {Eliav}}, \bibinfo
  {author} {\bibfnamefont {N.~R.}\ \bibnamefont {Hutzler}},\ and\ \bibinfo
  {author} {\bibfnamefont {A.}~\bibnamefont {Borschevsky}},\ }\bibfield
  {title} {\bibinfo {title} {Enhancement factor for the electric dipole moment
  of the electron in the baoh and yboh molecules},\ }\href
  {https://doi.org/10.1103/PhysRevA.99.042512} {\bibfield  {journal} {\bibinfo
  {journal} {Phys. Rev. A}\ }\textbf {\bibinfo {volume} {99}},\ \bibinfo
  {pages} {042512} (\bibinfo {year} {2019})}\BibitemShut {NoStop}%
\bibitem [{\citenamefont {Denis}\ \emph {et~al.}(2020)\citenamefont {Denis},
  \citenamefont {Hao}, \citenamefont {Eliav}, \citenamefont {Hutzler},
  \citenamefont {Nayak}, \citenamefont {Timmermans},\ and\ \citenamefont
  {Borschesvky}}]{Denis2020}%
  \BibitemOpen
  \bibfield  {author} {\bibinfo {author} {\bibfnamefont {M.}~\bibnamefont
  {Denis}}, \bibinfo {author} {\bibfnamefont {Y.}~\bibnamefont {Hao}}, \bibinfo
  {author} {\bibfnamefont {E.}~\bibnamefont {Eliav}}, \bibinfo {author}
  {\bibfnamefont {N.~R.}\ \bibnamefont {Hutzler}}, \bibinfo {author}
  {\bibfnamefont {M.~K.}\ \bibnamefont {Nayak}}, \bibinfo {author}
  {\bibfnamefont {R.~G.~E.}\ \bibnamefont {Timmermans}},\ and\ \bibinfo
  {author} {\bibfnamefont {A.}~\bibnamefont {Borschesvky}},\ }\bibfield
  {title} {\bibinfo {title} {Enhanced p,t-violating nuclear magnetic quadrupole
  moment effects in laser-coolable molecules},\ }\href
  {https://doi.org/10.1063/1.5141065} {\bibfield  {journal} {\bibinfo
  {journal} {The Journal of Chemical Physics}\ }\textbf {\bibinfo {volume}
  {152}},\ \bibinfo {pages} {084303} (\bibinfo {year} {2020})}\BibitemShut
  {NoStop}%
\bibitem [{\citenamefont {Haase}\ \emph {et~al.}(2021)\citenamefont {Haase},
  \citenamefont {Doeglas}, \citenamefont {Boeschoten}, \citenamefont {Eliav},
  \citenamefont {Ilias}, \citenamefont {Aggarwal}, \citenamefont {Bethlem},
  \citenamefont {Borschevsky}, \citenamefont {Esajas}, \citenamefont {Hao},
  \citenamefont {Hoekstra}, \citenamefont {Marshall}, \citenamefont
  {Meijknecht}, \citenamefont {Mooij}, \citenamefont {Steinebach},
  \citenamefont {Timmermans}, \citenamefont {Touwen}, \citenamefont {Ubachs},
  \citenamefont {Willmann}, \citenamefont {Yin},\ and\ \citenamefont
  {Collaboration)}}]{Haase2021}%
  \BibitemOpen
  \bibfield  {author} {\bibinfo {author} {\bibfnamefont {P.~A.~B.}\
  \bibnamefont {Haase}}, \bibinfo {author} {\bibfnamefont {D.~J.}\ \bibnamefont
  {Doeglas}}, \bibinfo {author} {\bibfnamefont {A.}~\bibnamefont {Boeschoten}},
  \bibinfo {author} {\bibfnamefont {E.}~\bibnamefont {Eliav}}, \bibinfo
  {author} {\bibfnamefont {M.}~\bibnamefont {Ilias}}, \bibinfo {author}
  {\bibfnamefont {P.}~\bibnamefont {Aggarwal}}, \bibinfo {author}
  {\bibfnamefont {H.~L.}\ \bibnamefont {Bethlem}}, \bibinfo {author}
  {\bibfnamefont {A.}~\bibnamefont {Borschevsky}}, \bibinfo {author}
  {\bibfnamefont {K.}~\bibnamefont {Esajas}}, \bibinfo {author} {\bibfnamefont
  {Y.}~\bibnamefont {Hao}}, \bibinfo {author} {\bibfnamefont {S.}~\bibnamefont
  {Hoekstra}}, \bibinfo {author} {\bibfnamefont {V.~R.}\ \bibnamefont
  {Marshall}}, \bibinfo {author} {\bibfnamefont {T.~B.}\ \bibnamefont
  {Meijknecht}}, \bibinfo {author} {\bibfnamefont {M.~C.}\ \bibnamefont
  {Mooij}}, \bibinfo {author} {\bibfnamefont {K.}~\bibnamefont {Steinebach}},
  \bibinfo {author} {\bibfnamefont {R.~G.~E.}\ \bibnamefont {Timmermans}},
  \bibinfo {author} {\bibfnamefont {A.~P.}\ \bibnamefont {Touwen}}, \bibinfo
  {author} {\bibfnamefont {W.}~\bibnamefont {Ubachs}}, \bibinfo {author}
  {\bibfnamefont {L.}~\bibnamefont {Willmann}}, \bibinfo {author}
  {\bibfnamefont {Y.}~\bibnamefont {Yin}},\ and\ \bibinfo {author}
  {\bibfnamefont {N.-e.}\ \bibnamefont {Collaboration)}},\ }\bibfield  {title}
  {\bibinfo {title} {{Systematic study and uncertainty evaluation of P, T-odd
  molecular enhancement factors in BaF}},\ }\href
  {https://doi.org/10.1063/5.0047344} {\bibfield  {journal} {\bibinfo
  {journal} {The Journal of Chemical Physics}\ }\textbf {\bibinfo {volume}
  {155}},\ \bibinfo {pages} {34309} (\bibinfo {year} {2021})}\BibitemShut
  {NoStop}%
\bibitem [{DIR()}]{DIRAC19}%
  \BibitemOpen
  \href@noop {} {}\bibinfo {note} {{DIRAC}, a relativistic ab initio electronic
  structure program, Release {DIRAC19} (2019), written by A.~S.~P.~Gomes,
  T.~Saue, L.~Visscher, H.~J.~{\relax Aa}.~Jensen, and R.~Bast, with
  contributions from I.~A.~Aucar, V.~Bakken, K.~G.~Dyall, S.~Dubillard,
  U.~Ekstr{\"o}m, E.~Eliav, T.~Enevoldsen, E.~Fa{\ss}hauer, T.~Fleig,
  O.~Fossgaard, L.~Halbert, E.~D.~Hedeg{\aa}rd, B.~Heimlich--Paris,
  T.~Helgaker, J.~Henriksson, M.~Ilias, Ch.~R.~Jacob, S.~Knecht,
  S.~Komorovsk{\'y}, O.~Kullie, J.~K.~L{\ae}rdahl, C.~V.~Larsen, Y.~S.~Lee,
  H.~S.~Nataraj, M.~K.~Nayak, P.~Norman, G.~Olejniczak, J.~Olsen,
  J.~M.~H.~Olsen, Y.~C.~Park, J.~K.~Pedersen, M.~Pernpointner, R.~di~Remigio,
  K.~Ruud, P.~Sa{\l}ek, B.~Schimmelpfennig, B.~Senjean, A.~Shee, J.~Sikkema,
  A.~J.~Thorvaldsen, J.~Thyssen, J.~van~Stralen, M.~L.~Vidal, S.~Villaume,
  O.~Visser, T.~Winther, and S.~Yamamoto (available at
  \url{http://dx.doi.org/10.5281/zenodo.3572669}, see also
  \url{http://www.diracprogram.org})}\BibitemShut {NoStop}%
\bibitem [{\citenamefont {Saue}\ \emph {et~al.}(2020)\citenamefont {Saue},
  \citenamefont {Bast}, \citenamefont {Gomes}, \citenamefont {Jensen},
  \citenamefont {Visscher}, \citenamefont {Aucar}, \citenamefont {Di~Remigio},
  \citenamefont {Dyall}, \citenamefont {Eliav}, \citenamefont {Fasshauer} \emph
  {et~al.}}]{Saue2020}%
  \BibitemOpen
  \bibfield  {author} {\bibinfo {author} {\bibfnamefont {T.}~\bibnamefont
  {Saue}}, \bibinfo {author} {\bibfnamefont {R.}~\bibnamefont {Bast}}, \bibinfo
  {author} {\bibfnamefont {A.~S.~P.}\ \bibnamefont {Gomes}}, \bibinfo {author}
  {\bibfnamefont {H.~J.~A.}\ \bibnamefont {Jensen}}, \bibinfo {author}
  {\bibfnamefont {L.}~\bibnamefont {Visscher}}, \bibinfo {author}
  {\bibfnamefont {I.~A.}\ \bibnamefont {Aucar}}, \bibinfo {author}
  {\bibfnamefont {R.}~\bibnamefont {Di~Remigio}}, \bibinfo {author}
  {\bibfnamefont {K.~G.}\ \bibnamefont {Dyall}}, \bibinfo {author}
  {\bibfnamefont {E.}~\bibnamefont {Eliav}}, \bibinfo {author} {\bibfnamefont
  {E.}~\bibnamefont {Fasshauer}}, \emph {et~al.},\ }\bibfield  {title}
  {\bibinfo {title} {The dirac code for relativistic molecular calculations},\
  }\href@noop {} {\bibfield  {journal} {\bibinfo  {journal} {The Journal of
  chemical physics}\ }\textbf {\bibinfo {volume} {152}},\ \bibinfo {pages}
  {204104} (\bibinfo {year} {2020})}\BibitemShut {NoStop}%
\bibitem [{\citenamefont {Dyall}(2009)}]{Dyall2009}%
  \BibitemOpen
  \bibfield  {author} {\bibinfo {author} {\bibfnamefont {K.~G.}\ \bibnamefont
  {Dyall}},\ }\bibfield  {title} {\bibinfo {title} {Relativistic double-zeta,
  triple-zeta, and quadruple-zeta basis sets for the 4s, 5s, 6s, and 7s
  elements},\ }\href@noop {} {\bibfield  {journal} {\bibinfo  {journal} {The
  Journal of Physical Chemistry A}\ }\textbf {\bibinfo {volume} {113}},\
  \bibinfo {pages} {12638} (\bibinfo {year} {2009})}\BibitemShut {NoStop}%
\bibitem [{\citenamefont {Dyall}(2012)}]{Dyall2012}%
  \BibitemOpen
  \bibfield  {author} {\bibinfo {author} {\bibfnamefont {K.~G.}\ \bibnamefont
  {Dyall}},\ }\bibfield  {title} {\bibinfo {title} {Core correlating basis
  functions for elements 31--118},\ }\href@noop {} {\bibfield  {journal}
  {\bibinfo  {journal} {Theoretical Chemistry Accounts}\ }\textbf {\bibinfo
  {volume} {131}},\ \bibinfo {pages} {1} (\bibinfo {year} {2012})}\BibitemShut
  {NoStop}%
\bibitem [{\citenamefont {Dyall}(2016)}]{Dyall2016}%
  \BibitemOpen
  \bibfield  {author} {\bibinfo {author} {\bibfnamefont {K.~G.}\ \bibnamefont
  {Dyall}},\ }\bibfield  {title} {\bibinfo {title} {Relativistic double-zeta,
  triple-zeta, and quadruple-zeta basis sets for the light elements
  {H}--{Ar}},\ }\href@noop {} {\bibfield  {journal} {\bibinfo  {journal}
  {Theoretical Chemistry Accounts}\ }\textbf {\bibinfo {volume} {135}},\
  \bibinfo {pages} {128} (\bibinfo {year} {2016})}\BibitemShut {NoStop}%
\bibitem [{\citenamefont {Chamorro}\ \emph {et~al.}(2025)\citenamefont
  {Chamorro}, \citenamefont {Kogel}, \citenamefont {Langen},\ and\
  \citenamefont {Borschevsky}}]{Chamorro2025}%
  \BibitemOpen
  \bibfield  {author} {\bibinfo {author} {\bibfnamefont {Y.}~\bibnamefont
  {Chamorro}}, \bibinfo {author} {\bibfnamefont {F.}~\bibnamefont {Kogel}},
  \bibinfo {author} {\bibfnamefont {T.}~\bibnamefont {Langen}},\ and\ \bibinfo
  {author} {\bibfnamefont {A.}~\bibnamefont {Borschevsky}},\ }\bibfield
  {title} {\bibinfo {title} {Magnetic hyperfine structure constants of
  {$^{137}$BaF} in the {$^2\Pi_{1/2}$} and {$^2\Pi_{3/2}$} excited states},\
  }\href {https://arxiv.org/abs/2506.12972} {\bibfield  {journal} {\bibinfo
  {journal} {arXiv:2506.12972}\ } (\bibinfo {year} {2025})}\BibitemShut
  {NoStop}%
\bibitem [{\citenamefont {Leimbach}\ \emph {et~al.}(2020)\citenamefont
  {Leimbach}, \citenamefont {Karls}, \citenamefont {Guo}, \citenamefont
  {Ahmed}, \citenamefont {Ballof}, \citenamefont {Bengtsson}, \citenamefont
  {Boix~Pamies}, \citenamefont {Borschevsky}, \citenamefont {Chrysalidis},
  \citenamefont {Eliav} \emph {et~al.}}]{leimbach2020electron}%
  \BibitemOpen
  \bibfield  {author} {\bibinfo {author} {\bibfnamefont {D.}~\bibnamefont
  {Leimbach}}, \bibinfo {author} {\bibfnamefont {J.}~\bibnamefont {Karls}},
  \bibinfo {author} {\bibfnamefont {Y.}~\bibnamefont {Guo}}, \bibinfo {author}
  {\bibfnamefont {R.}~\bibnamefont {Ahmed}}, \bibinfo {author} {\bibfnamefont
  {J.}~\bibnamefont {Ballof}}, \bibinfo {author} {\bibfnamefont
  {L.}~\bibnamefont {Bengtsson}}, \bibinfo {author} {\bibfnamefont
  {F.}~\bibnamefont {Boix~Pamies}}, \bibinfo {author} {\bibfnamefont
  {A.}~\bibnamefont {Borschevsky}}, \bibinfo {author} {\bibfnamefont
  {K.}~\bibnamefont {Chrysalidis}}, \bibinfo {author} {\bibfnamefont
  {E.}~\bibnamefont {Eliav}}, \emph {et~al.},\ }\bibfield  {title} {\bibinfo
  {title} {The electron affinity of astatine},\ }\href@noop {} {\bibfield
  {journal} {\bibinfo  {journal} {Nature communications}\ }\textbf {\bibinfo
  {volume} {11}},\ \bibinfo {pages} {3824} (\bibinfo {year}
  {2020})}\BibitemShut {NoStop}%
\bibitem [{\citenamefont {Haase}\ \emph {et~al.}(2020)\citenamefont {Haase},
  \citenamefont {Eliav}, \citenamefont {Ilias},\ and\ \citenamefont
  {Borschevsky}}]{Haase2020}%
  \BibitemOpen
  \bibfield  {author} {\bibinfo {author} {\bibfnamefont {P.~A.}\ \bibnamefont
  {Haase}}, \bibinfo {author} {\bibfnamefont {E.}~\bibnamefont {Eliav}},
  \bibinfo {author} {\bibfnamefont {M.}~\bibnamefont {Ilias}},\ and\ \bibinfo
  {author} {\bibfnamefont {A.}~\bibnamefont {Borschevsky}},\ }\bibfield
  {title} {\bibinfo {title} {Hyperfine structure constants on the relativistic
  coupled cluster level with associated uncertainties},\ }\href@noop {}
  {\bibfield  {journal} {\bibinfo  {journal} {The Journal of Physical Chemistry
  A}\ }\textbf {\bibinfo {volume} {124}},\ \bibinfo {pages} {3157} (\bibinfo
  {year} {2020})}\BibitemShut {NoStop}%
\bibitem [{\citenamefont {Hiramoto}\ \emph {et~al.}(2023)\citenamefont
  {Hiramoto}, \citenamefont {Baba}, \citenamefont {Enomoto}, \citenamefont
  {Iwakuni}, \citenamefont {Kuma}, \citenamefont {Takahashi}, \citenamefont
  {Tobaru},\ and\ \citenamefont {Miyamoto}}]{Hiramoto2023}%
  \BibitemOpen
  \bibfield  {author} {\bibinfo {author} {\bibfnamefont {A.}~\bibnamefont
  {Hiramoto}}, \bibinfo {author} {\bibfnamefont {M.}~\bibnamefont {Baba}},
  \bibinfo {author} {\bibfnamefont {K.}~\bibnamefont {Enomoto}}, \bibinfo
  {author} {\bibfnamefont {K.}~\bibnamefont {Iwakuni}}, \bibinfo {author}
  {\bibfnamefont {S.}~\bibnamefont {Kuma}}, \bibinfo {author} {\bibfnamefont
  {Y.}~\bibnamefont {Takahashi}}, \bibinfo {author} {\bibfnamefont
  {R.}~\bibnamefont {Tobaru}},\ and\ \bibinfo {author} {\bibfnamefont
  {Y.}~\bibnamefont {Miyamoto}},\ }\bibfield  {title} {\bibinfo {title}
  {Measurement of doppler effects in a cryogenic buffer-gas cell},\ }\href
  {https://doi.org/10.1103/PhysRevA.107.043114} {\bibfield  {journal} {\bibinfo
   {journal} {Phys. Rev. A}\ }\textbf {\bibinfo {volume} {107}},\ \bibinfo
  {pages} {043114} (\bibinfo {year} {2023})}\BibitemShut {NoStop}%
\bibitem [{\citenamefont {Fricke}\ and\ \citenamefont
  {Heilig}()}]{FrickeDatabase}%
  \BibitemOpen
  \bibfield  {author} {\bibinfo {author} {\bibfnamefont {G.}~\bibnamefont
  {Fricke}}\ and\ \bibinfo {author} {\bibfnamefont {K.}~\bibnamefont
  {Heilig}},\ }\href {https://doi.org/10.1007/10856314{\_}58} {\bibinfo {title}
  {Nuclear charge radii 56-ba barium: Datasheet from {Landolt-B{\"o}rnstein} -
  {Group I} elementary particles, nuclei and atoms. vol. 20: "{Nuclear} charge
  radii" in {Springer Materials}}}\BibitemShut {NoStop}%
\bibitem [{\citenamefont {King}(2013)}]{King2013}%
  \BibitemOpen
  \bibfield  {author} {\bibinfo {author} {\bibfnamefont {W.~H.}\ \bibnamefont
  {King}},\ }\href@noop {} {\emph {\bibinfo {title} {Isotope shifts in atomic
  spectra}}}\ (\bibinfo  {publisher} {Springer Science \& Business Media},\
  \bibinfo {year} {2013})\BibitemShut {NoStop}%
\bibitem [{\citenamefont {Gebert}\ \emph {et~al.}(2015)\citenamefont {Gebert},
  \citenamefont {Wan}, \citenamefont {Wolf}, \citenamefont {Angstmann},
  \citenamefont {Berengut},\ and\ \citenamefont {Schmidt}}]{Gebert2015}%
  \BibitemOpen
  \bibfield  {author} {\bibinfo {author} {\bibfnamefont {F.}~\bibnamefont
  {Gebert}}, \bibinfo {author} {\bibfnamefont {Y.}~\bibnamefont {Wan}},
  \bibinfo {author} {\bibfnamefont {F.}~\bibnamefont {Wolf}}, \bibinfo {author}
  {\bibfnamefont {C.~N.}\ \bibnamefont {Angstmann}}, \bibinfo {author}
  {\bibfnamefont {J.~C.}\ \bibnamefont {Berengut}},\ and\ \bibinfo {author}
  {\bibfnamefont {P.~O.}\ \bibnamefont {Schmidt}},\ }\bibfield  {title}
  {\bibinfo {title} {Precision isotope shift measurements in calcium ions using
  quantum logic detection schemes},\ }\href
  {https://doi.org/10.1103/PhysRevLett.115.053003} {\bibfield  {journal}
  {\bibinfo  {journal} {Phys. Rev. Lett.}\ }\textbf {\bibinfo {volume} {115}},\
  \bibinfo {pages} {053003} (\bibinfo {year} {2015})}\BibitemShut {NoStop}%
\bibitem [{\citenamefont {Ono}\ \emph {et~al.}(2022)\citenamefont {Ono},
  \citenamefont {Saito}, \citenamefont {Ishiyama}, \citenamefont {Higomoto},
  \citenamefont {Takano}, \citenamefont {Takasu}, \citenamefont {Yamamoto},
  \citenamefont {Tanaka},\ and\ \citenamefont {Takahashi}}]{Ono2022}%
  \BibitemOpen
  \bibfield  {author} {\bibinfo {author} {\bibfnamefont {K.}~\bibnamefont
  {Ono}}, \bibinfo {author} {\bibfnamefont {Y.}~\bibnamefont {Saito}}, \bibinfo
  {author} {\bibfnamefont {T.}~\bibnamefont {Ishiyama}}, \bibinfo {author}
  {\bibfnamefont {T.}~\bibnamefont {Higomoto}}, \bibinfo {author}
  {\bibfnamefont {T.}~\bibnamefont {Takano}}, \bibinfo {author} {\bibfnamefont
  {Y.}~\bibnamefont {Takasu}}, \bibinfo {author} {\bibfnamefont
  {Y.}~\bibnamefont {Yamamoto}}, \bibinfo {author} {\bibfnamefont
  {M.}~\bibnamefont {Tanaka}},\ and\ \bibinfo {author} {\bibfnamefont
  {Y.}~\bibnamefont {Takahashi}},\ }\bibfield  {title} {\bibinfo {title}
  {Observation of nonlinearity of generalized king plot in the search for new
  boson},\ }\href {https://doi.org/10.1103/PhysRevX.12.021033} {\bibfield
  {journal} {\bibinfo  {journal} {Phys. Rev. X}\ }\textbf {\bibinfo {volume}
  {12}},\ \bibinfo {pages} {021033} (\bibinfo {year} {2022})}\BibitemShut
  {NoStop}%
\bibitem [{\citenamefont {Hur}\ \emph {et~al.}(2022)\citenamefont {Hur},
  \citenamefont {Aude~Craik}, \citenamefont {Counts}, \citenamefont {Knyazev},
  \citenamefont {Caldwell}, \citenamefont {Leung}, \citenamefont {Pandey},
  \citenamefont {Berengut}, \citenamefont {Geddes}, \citenamefont {Nazarewicz},
  \citenamefont {Reinhard}, \citenamefont {Kawasaki}, \citenamefont {Jeon},
  \citenamefont {Jhe},\ and\ \citenamefont {Vuleti\ifmmode~\acute{c}\else
  \'{c}\fi{}}}]{Hur2022}%
  \BibitemOpen
  \bibfield  {author} {\bibinfo {author} {\bibfnamefont {J.}~\bibnamefont
  {Hur}}, \bibinfo {author} {\bibfnamefont {D.~P.~L.}\ \bibnamefont
  {Aude~Craik}}, \bibinfo {author} {\bibfnamefont {I.}~\bibnamefont {Counts}},
  \bibinfo {author} {\bibfnamefont {E.}~\bibnamefont {Knyazev}}, \bibinfo
  {author} {\bibfnamefont {L.}~\bibnamefont {Caldwell}}, \bibinfo {author}
  {\bibfnamefont {C.}~\bibnamefont {Leung}}, \bibinfo {author} {\bibfnamefont
  {S.}~\bibnamefont {Pandey}}, \bibinfo {author} {\bibfnamefont {J.~C.}\
  \bibnamefont {Berengut}}, \bibinfo {author} {\bibfnamefont {A.}~\bibnamefont
  {Geddes}}, \bibinfo {author} {\bibfnamefont {W.}~\bibnamefont {Nazarewicz}},
  \bibinfo {author} {\bibfnamefont {P.-G.}\ \bibnamefont {Reinhard}}, \bibinfo
  {author} {\bibfnamefont {A.}~\bibnamefont {Kawasaki}}, \bibinfo {author}
  {\bibfnamefont {H.}~\bibnamefont {Jeon}}, \bibinfo {author} {\bibfnamefont
  {W.}~\bibnamefont {Jhe}},\ and\ \bibinfo {author} {\bibfnamefont
  {V.}~\bibnamefont {Vuleti\ifmmode~\acute{c}\else \'{c}\fi{}}},\ }\bibfield
  {title} {\bibinfo {title} {Evidence of two-source king plot nonlinearity in
  spectroscopic search for new boson},\ }\href
  {https://doi.org/10.1103/PhysRevLett.128.163201} {\bibfield  {journal}
  {\bibinfo  {journal} {Phys. Rev. Lett.}\ }\textbf {\bibinfo {volume} {128}},\
  \bibinfo {pages} {163201} (\bibinfo {year} {2022})}\BibitemShut {NoStop}%
\bibitem [{\citenamefont {Hunter}\ \emph {et~al.}(2002)\citenamefont {Hunter},
  \citenamefont {Maxwell}, \citenamefont {Ulmer}, \citenamefont {Charney},
  \citenamefont {Peck}, \citenamefont {Krause}, \citenamefont {Ter-Avetisyan},\
  and\ \citenamefont {DeMille}}]{Hunter2002}%
  \BibitemOpen
  \bibfield  {author} {\bibinfo {author} {\bibfnamefont {L.~R.}\ \bibnamefont
  {Hunter}}, \bibinfo {author} {\bibfnamefont {S.~E.}\ \bibnamefont {Maxwell}},
  \bibinfo {author} {\bibfnamefont {K.~A.}\ \bibnamefont {Ulmer}}, \bibinfo
  {author} {\bibfnamefont {N.~D.}\ \bibnamefont {Charney}}, \bibinfo {author}
  {\bibfnamefont {S.~K.}\ \bibnamefont {Peck}}, \bibinfo {author}
  {\bibfnamefont {D.}~\bibnamefont {Krause}}, \bibinfo {author} {\bibfnamefont
  {S.}~\bibnamefont {Ter-Avetisyan}},\ and\ \bibinfo {author} {\bibfnamefont
  {D.}~\bibnamefont {DeMille}},\ }\bibfield  {title} {\bibinfo {title}
  {Detailed spectroscopy of the $a(1)$ ${[}^{3}{\ensuremath{\Sigma}}^{+}]$
  state of pbo},\ }\href {https://doi.org/10.1103/PhysRevA.65.030501}
  {\bibfield  {journal} {\bibinfo  {journal} {Phys. Rev. A}\ }\textbf {\bibinfo
  {volume} {65}},\ \bibinfo {pages} {030501} (\bibinfo {year}
  {2002})}\BibitemShut {NoStop}%
\bibitem [{\citenamefont {Harms}\ \emph {et~al.}(2019)\citenamefont {Harms},
  \citenamefont {O’Brien},\ and\ \citenamefont {O’Brien}}]{Harms2019}%
  \BibitemOpen
  \bibfield  {author} {\bibinfo {author} {\bibfnamefont {J.~C.}\ \bibnamefont
  {Harms}}, \bibinfo {author} {\bibfnamefont {L.~C.}\ \bibnamefont
  {O’Brien}},\ and\ \bibinfo {author} {\bibfnamefont {J.~J.}\ \bibnamefont
  {O’Brien}},\ }\bibfield  {title} {\bibinfo {title} {Mass-independent dunham
  analysis of the known electronic states of platinum sulfide, pts, and
  analysis of the electronic field-shift effect},\ }\href
  {https://doi.org/10.1063/1.5113510} {\bibfield  {journal} {\bibinfo
  {journal} {The Journal of Chemical Physics}\ }\textbf {\bibinfo {volume}
  {151}},\ \bibinfo {pages} {094303} (\bibinfo {year} {2019})}\BibitemShut
  {NoStop}%
\bibitem [{\citenamefont {Udrescu}\ \emph {et~al.}(2021)\citenamefont
  {Udrescu}, \citenamefont {Brinson}, \citenamefont {Ruiz}, \citenamefont
  {Gaul}, \citenamefont {Berger}, \citenamefont {Billowes}, \citenamefont
  {Binnersley}, \citenamefont {Bissell}, \citenamefont {Breier}, \citenamefont
  {Chrysalidis}, \citenamefont {Cocolios}, \citenamefont {Cooper},
  \citenamefont {Flanagan}, \citenamefont {Giesen}, \citenamefont {de~Groote},
  \citenamefont {Franchoo}, \citenamefont {Gustafsson}, \citenamefont {Isaev},
  \citenamefont {Koszor\'us}, \citenamefont {Neyens}, \citenamefont {Perrett},
  \citenamefont {Ricketts}, \citenamefont {Rothe}, \citenamefont {Vernon},
  \citenamefont {Wendt}, \citenamefont {Wienholtz}, \citenamefont {Wilkins},\
  and\ \citenamefont {Yang}}]{Udrescu2021}%
  \BibitemOpen
  \bibfield  {author} {\bibinfo {author} {\bibfnamefont {S.~M.}\ \bibnamefont
  {Udrescu}}, \bibinfo {author} {\bibfnamefont {A.~J.}\ \bibnamefont
  {Brinson}}, \bibinfo {author} {\bibfnamefont {R.~F.~G.}\ \bibnamefont
  {Ruiz}}, \bibinfo {author} {\bibfnamefont {K.}~\bibnamefont {Gaul}}, \bibinfo
  {author} {\bibfnamefont {R.}~\bibnamefont {Berger}}, \bibinfo {author}
  {\bibfnamefont {J.}~\bibnamefont {Billowes}}, \bibinfo {author}
  {\bibfnamefont {C.~L.}\ \bibnamefont {Binnersley}}, \bibinfo {author}
  {\bibfnamefont {M.~L.}\ \bibnamefont {Bissell}}, \bibinfo {author}
  {\bibfnamefont {A.~A.}\ \bibnamefont {Breier}}, \bibinfo {author}
  {\bibfnamefont {K.}~\bibnamefont {Chrysalidis}}, \bibinfo {author}
  {\bibfnamefont {T.~E.}\ \bibnamefont {Cocolios}}, \bibinfo {author}
  {\bibfnamefont {B.~S.}\ \bibnamefont {Cooper}}, \bibinfo {author}
  {\bibfnamefont {K.~T.}\ \bibnamefont {Flanagan}}, \bibinfo {author}
  {\bibfnamefont {T.~F.}\ \bibnamefont {Giesen}}, \bibinfo {author}
  {\bibfnamefont {R.~P.}\ \bibnamefont {de~Groote}}, \bibinfo {author}
  {\bibfnamefont {S.}~\bibnamefont {Franchoo}}, \bibinfo {author}
  {\bibfnamefont {F.~P.}\ \bibnamefont {Gustafsson}}, \bibinfo {author}
  {\bibfnamefont {T.~A.}\ \bibnamefont {Isaev}}, \bibinfo {author}
  {\bibfnamefont {A.}~\bibnamefont {Koszor\'us}}, \bibinfo {author}
  {\bibfnamefont {G.}~\bibnamefont {Neyens}}, \bibinfo {author} {\bibfnamefont
  {H.~A.}\ \bibnamefont {Perrett}}, \bibinfo {author} {\bibfnamefont {C.~M.}\
  \bibnamefont {Ricketts}}, \bibinfo {author} {\bibfnamefont {S.}~\bibnamefont
  {Rothe}}, \bibinfo {author} {\bibfnamefont {A.~R.}\ \bibnamefont {Vernon}},
  \bibinfo {author} {\bibfnamefont {K.~D.~A.}\ \bibnamefont {Wendt}}, \bibinfo
  {author} {\bibfnamefont {F.}~\bibnamefont {Wienholtz}}, \bibinfo {author}
  {\bibfnamefont {S.~G.}\ \bibnamefont {Wilkins}},\ and\ \bibinfo {author}
  {\bibfnamefont {X.~F.}\ \bibnamefont {Yang}},\ }\bibfield  {title} {\bibinfo
  {title} {Isotope shifts of radium monofluoride molecules},\ }\href
  {https://doi.org/10.1103/PhysRevLett.127.033001} {\bibfield  {journal}
  {\bibinfo  {journal} {Phys. Rev. Lett.}\ }\textbf {\bibinfo {volume} {127}},\
  \bibinfo {pages} {033001} (\bibinfo {year} {2021})}\BibitemShut {NoStop}%
\bibitem [{\citenamefont {Blaum}\ \emph {et~al.}(2013)\citenamefont {Blaum},
  \citenamefont {Dilling},\ and\ \citenamefont
  {N{\"o}rtersh{\"a}user}}]{Blaum2013}%
  \BibitemOpen
  \bibfield  {author} {\bibinfo {author} {\bibfnamefont {K.}~\bibnamefont
  {Blaum}}, \bibinfo {author} {\bibfnamefont {J.}~\bibnamefont {Dilling}},\
  and\ \bibinfo {author} {\bibfnamefont {W.}~\bibnamefont
  {N{\"o}rtersh{\"a}user}},\ }\bibfield  {title} {\bibinfo {title} {Precision
  atomic physics techniques for nuclear physics with radioactive beams},\
  }\href {https://doi.org/10.1088/0031-8949/2013/T152/014017} {\bibfield
  {journal} {\bibinfo  {journal} {Physica Scripta}\ }\textbf {\bibinfo {volume}
  {2013}},\ \bibinfo {pages} {014017} (\bibinfo {year} {2013})}\BibitemShut
  {NoStop}%
\bibitem [{\citenamefont {Campbell}\ \emph {et~al.}(2016)\citenamefont
  {Campbell}, \citenamefont {Moore},\ and\ \citenamefont
  {Pearson}}]{Campbell2016}%
  \BibitemOpen
  \bibfield  {author} {\bibinfo {author} {\bibfnamefont {P.}~\bibnamefont
  {Campbell}}, \bibinfo {author} {\bibfnamefont {I.~D.}\ \bibnamefont
  {Moore}},\ and\ \bibinfo {author} {\bibfnamefont {M.~R.}\ \bibnamefont
  {Pearson}},\ }\bibfield  {title} {\bibinfo {title} {Laser spectroscopy for
  nuclear structure physics},\ }\href
  {https://doi.org/https://doi.org/10.1016/j.ppnp.2015.09.003} {\bibfield
  {journal} {\bibinfo  {journal} {Progress in Particle and Nuclear Physics}\
  }\textbf {\bibinfo {volume} {86}},\ \bibinfo {pages} {127} (\bibinfo {year}
  {2016})}\BibitemShut {NoStop}%
\bibitem [{\citenamefont {Angeli}\ and\ \citenamefont
  {Marinova}(2013)}]{Angeli2013}%
  \BibitemOpen
  \bibfield  {author} {\bibinfo {author} {\bibfnamefont {I.}~\bibnamefont
  {Angeli}}\ and\ \bibinfo {author} {\bibfnamefont {K.}~\bibnamefont
  {Marinova}},\ }\bibfield  {title} {\bibinfo {title} {Table of experimental
  nuclear ground state charge radii: An update},\ }\href
  {https://doi.org/https://doi.org/10.1016/j.adt.2011.12.006} {\bibfield
  {journal} {\bibinfo  {journal} {Atomic Data and Nuclear Data Tables}\
  }\textbf {\bibinfo {volume} {99}},\ \bibinfo {pages} {69} (\bibinfo {year}
  {2013})}\BibitemShut {NoStop}%
\bibitem [{\citenamefont {Reinhard}\ and\ \citenamefont
  {Nazarewicz}(2016)}]{Reinhard2016}%
  \BibitemOpen
  \bibfield  {author} {\bibinfo {author} {\bibfnamefont {P.-G.}\ \bibnamefont
  {Reinhard}}\ and\ \bibinfo {author} {\bibfnamefont {W.}~\bibnamefont
  {Nazarewicz}},\ }\bibfield  {title} {\bibinfo {title} {Nuclear charge and
  neutron radii and nuclear matter: Trend analysis in skyrme
  density-functional-theory approach},\ }\href
  {https://doi.org/10.1103/PhysRevC.93.051303} {\bibfield  {journal} {\bibinfo
  {journal} {Phys. Rev. C}\ }\textbf {\bibinfo {volume} {93}},\ \bibinfo
  {pages} {051303} (\bibinfo {year} {2016})}\BibitemShut {NoStop}%
\bibitem [{\citenamefont {Gorges}\ \emph {et~al.}(2019)\citenamefont {Gorges},
  \citenamefont {Rodr\'{\i}guez}, \citenamefont {Balabanski}, \citenamefont
  {Bissell}, \citenamefont {Blaum}, \citenamefont {Cheal}, \citenamefont
  {Garcia~Ruiz}, \citenamefont {Georgiev}, \citenamefont {Gins}, \citenamefont
  {Heylen}, \citenamefont {Kanellakopoulos}, \citenamefont {Kaufmann},
  \citenamefont {Kowalska}, \citenamefont {Lagaki}, \citenamefont {Lechner},
  \citenamefont {Maa\ss{}}, \citenamefont {Malbrunot-Ettenauer}, \citenamefont
  {Nazarewicz}, \citenamefont {Neugart}, \citenamefont {Neyens}, \citenamefont
  {N\"ortersh\"auser}, \citenamefont {Reinhard}, \citenamefont {Sailer},
  \citenamefont {S\'anchez}, \citenamefont {Schmidt}, \citenamefont {Wehner},
  \citenamefont {Wraith}, \citenamefont {Xie}, \citenamefont {Xu},
  \citenamefont {Yang},\ and\ \citenamefont {Yordanov}}]{Gorges2019}%
  \BibitemOpen
  \bibfield  {author} {\bibinfo {author} {\bibfnamefont {C.}~\bibnamefont
  {Gorges}}, \bibinfo {author} {\bibfnamefont {L.~V.}\ \bibnamefont
  {Rodr\'{\i}guez}}, \bibinfo {author} {\bibfnamefont {D.~L.}\ \bibnamefont
  {Balabanski}}, \bibinfo {author} {\bibfnamefont {M.~L.}\ \bibnamefont
  {Bissell}}, \bibinfo {author} {\bibfnamefont {K.}~\bibnamefont {Blaum}},
  \bibinfo {author} {\bibfnamefont {B.}~\bibnamefont {Cheal}}, \bibinfo
  {author} {\bibfnamefont {R.~F.}\ \bibnamefont {Garcia~Ruiz}}, \bibinfo
  {author} {\bibfnamefont {G.}~\bibnamefont {Georgiev}}, \bibinfo {author}
  {\bibfnamefont {W.}~\bibnamefont {Gins}}, \bibinfo {author} {\bibfnamefont
  {H.}~\bibnamefont {Heylen}}, \bibinfo {author} {\bibfnamefont
  {A.}~\bibnamefont {Kanellakopoulos}}, \bibinfo {author} {\bibfnamefont
  {S.}~\bibnamefont {Kaufmann}}, \bibinfo {author} {\bibfnamefont
  {M.}~\bibnamefont {Kowalska}}, \bibinfo {author} {\bibfnamefont
  {V.}~\bibnamefont {Lagaki}}, \bibinfo {author} {\bibfnamefont
  {S.}~\bibnamefont {Lechner}}, \bibinfo {author} {\bibfnamefont
  {B.}~\bibnamefont {Maa\ss{}}}, \bibinfo {author} {\bibfnamefont
  {S.}~\bibnamefont {Malbrunot-Ettenauer}}, \bibinfo {author} {\bibfnamefont
  {W.}~\bibnamefont {Nazarewicz}}, \bibinfo {author} {\bibfnamefont
  {R.}~\bibnamefont {Neugart}}, \bibinfo {author} {\bibfnamefont
  {G.}~\bibnamefont {Neyens}}, \bibinfo {author} {\bibfnamefont
  {W.}~\bibnamefont {N\"ortersh\"auser}}, \bibinfo {author} {\bibfnamefont
  {P.-G.}\ \bibnamefont {Reinhard}}, \bibinfo {author} {\bibfnamefont
  {S.}~\bibnamefont {Sailer}}, \bibinfo {author} {\bibfnamefont
  {R.}~\bibnamefont {S\'anchez}}, \bibinfo {author} {\bibfnamefont
  {S.}~\bibnamefont {Schmidt}}, \bibinfo {author} {\bibfnamefont
  {L.}~\bibnamefont {Wehner}}, \bibinfo {author} {\bibfnamefont
  {C.}~\bibnamefont {Wraith}}, \bibinfo {author} {\bibfnamefont
  {L.}~\bibnamefont {Xie}}, \bibinfo {author} {\bibfnamefont {Z.~Y.}\
  \bibnamefont {Xu}}, \bibinfo {author} {\bibfnamefont {X.~F.}\ \bibnamefont
  {Yang}},\ and\ \bibinfo {author} {\bibfnamefont {D.~T.}\ \bibnamefont
  {Yordanov}},\ }\bibfield  {title} {\bibinfo {title} {Laser spectroscopy of
  neutron-rich tin isotopes: A discontinuity in charge radii across the $n=82$
  shell closure},\ }\href {https://doi.org/10.1103/PhysRevLett.122.192502}
  {\bibfield  {journal} {\bibinfo  {journal} {Phys. Rev. Lett.}\ }\textbf
  {\bibinfo {volume} {122}},\ \bibinfo {pages} {192502} (\bibinfo {year}
  {2019})}\BibitemShut {NoStop}%
\bibitem [{\citenamefont {de~Groote}\ \emph {et~al.}(2020)\citenamefont
  {de~Groote}, \citenamefont {Billowes}, \citenamefont {Binnersley},
  \citenamefont {Bissell}, \citenamefont {Cocolios}, \citenamefont
  {Day~Goodacre}, \citenamefont {Farooq-Smith}, \citenamefont {Fedorov},
  \citenamefont {Flanagan}, \citenamefont {Franchoo}, \citenamefont
  {Garcia~Ruiz}, \citenamefont {Gins}, \citenamefont {Holt}, \citenamefont
  {Koszor{\'u}s}, \citenamefont {Lynch}, \citenamefont {Miyagi}, \citenamefont
  {Nazarewicz}, \citenamefont {Neyens}, \citenamefont {Reinhard}, \citenamefont
  {Rothe}, \citenamefont {Stroke}, \citenamefont {Vernon}, \citenamefont
  {Wendt}, \citenamefont {Wilkins}, \citenamefont {Xu},\ and\ \citenamefont
  {Yang}}]{Groote2020}%
  \BibitemOpen
  \bibfield  {author} {\bibinfo {author} {\bibfnamefont {R.~P.}\ \bibnamefont
  {de~Groote}}, \bibinfo {author} {\bibfnamefont {J.}~\bibnamefont {Billowes}},
  \bibinfo {author} {\bibfnamefont {C.~L.}\ \bibnamefont {Binnersley}},
  \bibinfo {author} {\bibfnamefont {M.~L.}\ \bibnamefont {Bissell}}, \bibinfo
  {author} {\bibfnamefont {T.~E.}\ \bibnamefont {Cocolios}}, \bibinfo {author}
  {\bibfnamefont {T.}~\bibnamefont {Day~Goodacre}}, \bibinfo {author}
  {\bibfnamefont {G.~J.}\ \bibnamefont {Farooq-Smith}}, \bibinfo {author}
  {\bibfnamefont {D.~V.}\ \bibnamefont {Fedorov}}, \bibinfo {author}
  {\bibfnamefont {K.~T.}\ \bibnamefont {Flanagan}}, \bibinfo {author}
  {\bibfnamefont {S.}~\bibnamefont {Franchoo}}, \bibinfo {author}
  {\bibfnamefont {R.~F.}\ \bibnamefont {Garcia~Ruiz}}, \bibinfo {author}
  {\bibfnamefont {W.}~\bibnamefont {Gins}}, \bibinfo {author} {\bibfnamefont
  {J.~D.}\ \bibnamefont {Holt}}, \bibinfo {author} {\bibfnamefont
  {{\'A}.}~\bibnamefont {Koszor{\'u}s}}, \bibinfo {author} {\bibfnamefont
  {K.~M.}\ \bibnamefont {Lynch}}, \bibinfo {author} {\bibfnamefont
  {T.}~\bibnamefont {Miyagi}}, \bibinfo {author} {\bibfnamefont
  {W.}~\bibnamefont {Nazarewicz}}, \bibinfo {author} {\bibfnamefont
  {G.}~\bibnamefont {Neyens}}, \bibinfo {author} {\bibfnamefont {P.~G.}\
  \bibnamefont {Reinhard}}, \bibinfo {author} {\bibfnamefont {S.}~\bibnamefont
  {Rothe}}, \bibinfo {author} {\bibfnamefont {H.~H.}\ \bibnamefont {Stroke}},
  \bibinfo {author} {\bibfnamefont {A.~R.}\ \bibnamefont {Vernon}}, \bibinfo
  {author} {\bibfnamefont {K.~D.~A.}\ \bibnamefont {Wendt}}, \bibinfo {author}
  {\bibfnamefont {S.~G.}\ \bibnamefont {Wilkins}}, \bibinfo {author}
  {\bibfnamefont {Z.~Y.}\ \bibnamefont {Xu}},\ and\ \bibinfo {author}
  {\bibfnamefont {X.~F.}\ \bibnamefont {Yang}},\ }\bibfield  {title} {\bibinfo
  {title} {Measurement and microscopic description of odd--even staggering of
  charge radii of exotic copper isotopes},\ }\href
  {https://doi.org/10.1038/s41567-020-0868-y} {\bibfield  {journal} {\bibinfo
  {journal} {Nature Physics}\ }\textbf {\bibinfo {volume} {16}},\ \bibinfo
  {pages} {620} (\bibinfo {year} {2020})}\BibitemShut {NoStop}%
\bibitem [{\citenamefont {Barontini}\ \emph {et~al.}(2021)\citenamefont
  {Barontini}, \citenamefont {Boyer}, \citenamefont {Calmet}, \citenamefont
  {Fitch}, \citenamefont {Forgan}, \citenamefont {Godun}, \citenamefont
  {Goldwin}, \citenamefont {Guarrera}, \citenamefont {Hill}, \citenamefont
  {Jeong}, \citenamefont {Keller}, \citenamefont {Kuipers}, \citenamefont
  {Margolis}, \citenamefont {Newman}, \citenamefont {Prokhorov}, \citenamefont
  {Rodewald}, \citenamefont {Sauer}, \citenamefont {Schioppo}, \citenamefont
  {Sherrill}, \citenamefont {Tarbutt}, \citenamefont {Vecchio},\ and\
  \citenamefont {Worm}}]{Barontini2021}%
  \BibitemOpen
  \bibfield  {author} {\bibinfo {author} {\bibfnamefont {G.}~\bibnamefont
  {Barontini}}, \bibinfo {author} {\bibfnamefont {V.}~\bibnamefont {Boyer}},
  \bibinfo {author} {\bibfnamefont {X.}~\bibnamefont {Calmet}}, \bibinfo
  {author} {\bibfnamefont {N.~J.}\ \bibnamefont {Fitch}}, \bibinfo {author}
  {\bibfnamefont {E.~M.}\ \bibnamefont {Forgan}}, \bibinfo {author}
  {\bibfnamefont {R.~M.}\ \bibnamefont {Godun}}, \bibinfo {author}
  {\bibfnamefont {J.}~\bibnamefont {Goldwin}}, \bibinfo {author} {\bibfnamefont
  {V.}~\bibnamefont {Guarrera}}, \bibinfo {author} {\bibfnamefont {I.~R.}\
  \bibnamefont {Hill}}, \bibinfo {author} {\bibfnamefont {M.}~\bibnamefont
  {Jeong}}, \bibinfo {author} {\bibfnamefont {M.}~\bibnamefont {Keller}},
  \bibinfo {author} {\bibfnamefont {F.}~\bibnamefont {Kuipers}}, \bibinfo
  {author} {\bibfnamefont {H.~S.}\ \bibnamefont {Margolis}}, \bibinfo {author}
  {\bibfnamefont {P.}~\bibnamefont {Newman}}, \bibinfo {author} {\bibfnamefont
  {L.}~\bibnamefont {Prokhorov}}, \bibinfo {author} {\bibfnamefont
  {J.}~\bibnamefont {Rodewald}}, \bibinfo {author} {\bibfnamefont {B.~E.}\
  \bibnamefont {Sauer}}, \bibinfo {author} {\bibfnamefont {M.}~\bibnamefont
  {Schioppo}}, \bibinfo {author} {\bibfnamefont {N.}~\bibnamefont {Sherrill}},
  \bibinfo {author} {\bibfnamefont {M.~R.}\ \bibnamefont {Tarbutt}}, \bibinfo
  {author} {\bibfnamefont {A.}~\bibnamefont {Vecchio}},\ and\ \bibinfo {author}
  {\bibfnamefont {S.}~\bibnamefont {Worm}},\ }\bibfield  {title} {\bibinfo
  {title} {{QSNET, a network of clocks for measuring the stability of
  fundamental constants}},\ }in\ \href {https://doi.org/10.1117/12.2600493}
  {\emph {\bibinfo {booktitle} {Quantum Technology: Driving Commercialisation
  of an Enabling Science II}}},\ Vol.\ \bibinfo {volume} {11881},\ \bibinfo
  {editor} {edited by\ \bibinfo {editor} {\bibfnamefont {M.~J.}\ \bibnamefont
  {Padgett}}, \bibinfo {editor} {\bibfnamefont {K.}~\bibnamefont {Bongs}},
  \bibinfo {editor} {\bibfnamefont {A.}~\bibnamefont {Fedrizzi}},\ and\
  \bibinfo {editor} {\bibfnamefont {A.}~\bibnamefont {Politi}}},\ \bibinfo
  {organization} {International Society for Optics and Photonics}\ (\bibinfo
  {publisher} {SPIE},\ \bibinfo {year} {2021})\ pp.\ \bibinfo {pages} {63 --
  66}\BibitemShut {NoStop}%
\bibitem [{\citenamefont {Drouin}\ \emph {et~al.}(2001)\citenamefont {Drouin},
  \citenamefont {Miller}, \citenamefont {Müller},\ and\ \citenamefont
  {Cohen}}]{Drouin2001}%
  \BibitemOpen
  \bibfield  {author} {\bibinfo {author} {\bibfnamefont {B.~J.}\ \bibnamefont
  {Drouin}}, \bibinfo {author} {\bibfnamefont {C.~E.}\ \bibnamefont {Miller}},
  \bibinfo {author} {\bibfnamefont {H.~S.}\ \bibnamefont {Müller}},\ and\
  \bibinfo {author} {\bibfnamefont {E.~A.}\ \bibnamefont {Cohen}},\ }\bibfield
  {title} {\bibinfo {title} {The rotational spectra, isotopically independent
  parameters, and interatomic potentials for the {$X\Pi_{3/2}$} and
  {$X\Pi_{1/2}$} states of {BrO}},\ }\href
  {https://doi.org/https://doi.org/10.1006/jmsp.2000.8252} {\bibfield
  {journal} {\bibinfo  {journal} {Journal of Molecular Spectroscopy}\ }\textbf
  {\bibinfo {volume} {205}},\ \bibinfo {pages} {128} (\bibinfo {year}
  {2001})}\BibitemShut {NoStop}%
\bibitem [{\citenamefont {Doppelbauer}\ \emph {et~al.}(2022)\citenamefont
  {Doppelbauer}, \citenamefont {Wright}, \citenamefont {Hofsäss},
  \citenamefont {Sartakov}, \citenamefont {Meijer},\ and\ \citenamefont
  {Truppe}}]{Doppelbauer2022}%
  \BibitemOpen
  \bibfield  {author} {\bibinfo {author} {\bibfnamefont {M.}~\bibnamefont
  {Doppelbauer}}, \bibinfo {author} {\bibfnamefont {S.~C.}\ \bibnamefont
  {Wright}}, \bibinfo {author} {\bibfnamefont {S.}~\bibnamefont {Hofsäss}},
  \bibinfo {author} {\bibfnamefont {B.~G.}\ \bibnamefont {Sartakov}}, \bibinfo
  {author} {\bibfnamefont {G.}~\bibnamefont {Meijer}},\ and\ \bibinfo {author}
  {\bibfnamefont {S.}~\bibnamefont {Truppe}},\ }\bibfield  {title} {\bibinfo
  {title} {Hyperfine-resolved optical spectroscopy of the {$A2\Pi$}
  $\leftarrow$ {$X2\Sigma+$} transition in {MgF}},\ }\href@noop {} {\bibfield
  {journal} {\bibinfo  {journal} {The Journal of Chemical Physics}\ }\textbf
  {\bibinfo {volume} {156}},\ \bibinfo {pages} {134301} (\bibinfo {year}
  {2022})}\BibitemShut {NoStop}%
\bibitem [{\citenamefont {Holland}\ \emph {et~al.}(2021)\citenamefont
  {Holland}, \citenamefont {Lu},\ and\ \citenamefont {Cheuk}}]{Holland2021}%
  \BibitemOpen
  \bibfield  {author} {\bibinfo {author} {\bibfnamefont {C.~M.}\ \bibnamefont
  {Holland}}, \bibinfo {author} {\bibfnamefont {Y.}~\bibnamefont {Lu}},\ and\
  \bibinfo {author} {\bibfnamefont {L.~W.}\ \bibnamefont {Cheuk}},\ }\bibfield
  {title} {\bibinfo {title} {Synthesizing optical spectra using
  computer-generated holography techniques},\ }\href
  {https://doi.org/10.1088/1367-2630/abe973} {\bibfield  {journal} {\bibinfo
  {journal} {New Journal of Physics}\ }\textbf {\bibinfo {volume} {23}},\
  \bibinfo {pages} {033028} (\bibinfo {year} {2021})}\BibitemShut {NoStop}%
\bibitem [{\citenamefont {Zeng}\ \emph {et~al.}(2023)\citenamefont {Zeng},
  \citenamefont {Jadbabaie}, \citenamefont {Patel}, \citenamefont {Yu},
  \citenamefont {Steimle},\ and\ \citenamefont {Hutzler}}]{Zeng2023}%
  \BibitemOpen
  \bibfield  {author} {\bibinfo {author} {\bibfnamefont {Y.}~\bibnamefont
  {Zeng}}, \bibinfo {author} {\bibfnamefont {A.}~\bibnamefont {Jadbabaie}},
  \bibinfo {author} {\bibfnamefont {A.~N.}\ \bibnamefont {Patel}}, \bibinfo
  {author} {\bibfnamefont {P.}~\bibnamefont {Yu}}, \bibinfo {author}
  {\bibfnamefont {T.~C.}\ \bibnamefont {Steimle}},\ and\ \bibinfo {author}
  {\bibfnamefont {N.~R.}\ \bibnamefont {Hutzler}},\ }\bibfield  {title}
  {\bibinfo {title} {Optical cycling in polyatomic molecules with complex
  hyperfine structure},\ }\href {https://doi.org/10.1103/PhysRevA.108.012813}
  {\bibfield  {journal} {\bibinfo  {journal} {Phys. Rev. A}\ }\textbf {\bibinfo
  {volume} {108}},\ \bibinfo {pages} {012813} (\bibinfo {year}
  {2023})}\BibitemShut {NoStop}%
\bibitem [{\citenamefont {Aiello}\ \emph {et~al.}(2022)\citenamefont {Aiello},
  \citenamefont {Di~Sarno}, \citenamefont {Delli~Santi}, \citenamefont
  {De~Rosa}, \citenamefont {Ricciardi}, \citenamefont {De~Natale},
  \citenamefont {Santamaria}, \citenamefont {Giusfredi},\ and\ \citenamefont
  {Maddaloni}}]{Aiello2022}%
  \BibitemOpen
  \bibfield  {author} {\bibinfo {author} {\bibfnamefont {R.}~\bibnamefont
  {Aiello}}, \bibinfo {author} {\bibfnamefont {V.}~\bibnamefont {Di~Sarno}},
  \bibinfo {author} {\bibfnamefont {M.~G.}\ \bibnamefont {Delli~Santi}},
  \bibinfo {author} {\bibfnamefont {M.}~\bibnamefont {De~Rosa}}, \bibinfo
  {author} {\bibfnamefont {I.}~\bibnamefont {Ricciardi}}, \bibinfo {author}
  {\bibfnamefont {P.}~\bibnamefont {De~Natale}}, \bibinfo {author}
  {\bibfnamefont {L.}~\bibnamefont {Santamaria}}, \bibinfo {author}
  {\bibfnamefont {G.}~\bibnamefont {Giusfredi}},\ and\ \bibinfo {author}
  {\bibfnamefont {P.}~\bibnamefont {Maddaloni}},\ }\bibfield  {title} {\bibinfo
  {title} {Absolute frequency metrology of buffer-gas-cooled molecular spectra
  at 1 khz accuracy level},\ }\href
  {https://doi.org/10.1038/s41467-022-34758-9} {\bibfield  {journal} {\bibinfo
  {journal} {Nature Communications}\ }\textbf {\bibinfo {volume} {13}},\
  \bibinfo {pages} {7016} (\bibinfo {year} {2022})}\BibitemShut {NoStop}%
\bibitem [{\citenamefont {Hofs\"ass}\ \emph {et~al.}(2023)\citenamefont
  {Hofs\"ass}, \citenamefont {Padilla-Castillo}, \citenamefont {Wright},
  \citenamefont {Kray}, \citenamefont {Thomas}, \citenamefont {Sartakov},
  \citenamefont {Ohayon}, \citenamefont {Meijer},\ and\ \citenamefont
  {Truppe}}]{Hofsaess2023}%
  \BibitemOpen
  \bibfield  {author} {\bibinfo {author} {\bibfnamefont {S.}~\bibnamefont
  {Hofs\"ass}}, \bibinfo {author} {\bibfnamefont {J.~E.}\ \bibnamefont
  {Padilla-Castillo}}, \bibinfo {author} {\bibfnamefont {S.~C.}\ \bibnamefont
  {Wright}}, \bibinfo {author} {\bibfnamefont {S.}~\bibnamefont {Kray}},
  \bibinfo {author} {\bibfnamefont {R.}~\bibnamefont {Thomas}}, \bibinfo
  {author} {\bibfnamefont {B.~G.}\ \bibnamefont {Sartakov}}, \bibinfo {author}
  {\bibfnamefont {B.}~\bibnamefont {Ohayon}}, \bibinfo {author} {\bibfnamefont
  {G.}~\bibnamefont {Meijer}},\ and\ \bibinfo {author} {\bibfnamefont
  {S.}~\bibnamefont {Truppe}},\ }\bibfield  {title} {\bibinfo {title}
  {High-resolution isotope-shift spectroscopy of cd i},\ }\href
  {https://doi.org/10.1103/PhysRevResearch.5.013043} {\bibfield  {journal}
  {\bibinfo  {journal} {Phys. Rev. Res.}\ }\textbf {\bibinfo {volume} {5}},\
  \bibinfo {pages} {013043} (\bibinfo {year} {2023})}\BibitemShut {NoStop}%
\bibitem [{\citenamefont {Kogel}(2025)}]{FelixKogel}%
  \BibitemOpen
  \bibfield  {author} {\bibinfo {author} {\bibfnamefont {F.}~\bibnamefont
  {Kogel}},\ }\emph {\bibinfo {title} {Laser cooling of molecules for precision
  measurements of parity violation}},\ \href@noop {} {Ph.D. thesis},\ \bibinfo
  {school} {University of Stuttgart} (\bibinfo {year} {2025})\BibitemShut
  {NoStop}%
\bibitem [{\citenamefont {Brown}\ and\ \citenamefont
  {Carrington}(2003)}]{Brown2003}%
  \BibitemOpen
  \bibfield  {author} {\bibinfo {author} {\bibfnamefont {J.}~\bibnamefont
  {Brown}}\ and\ \bibinfo {author} {\bibfnamefont {A.}~\bibnamefont
  {Carrington}},\ }\href {https://books.google.de/books?id=BcZHngEACAAJ} {\emph
  {\bibinfo {title} {Rotational Spectroscopy of Diatomic Molecules}}},\
  Cambridge molecular science series\ (\bibinfo  {publisher} {Cambridge
  University Press},\ \bibinfo {year} {2003})\BibitemShut {NoStop}%
\bibitem [{\citenamefont {Frosch}\ and\ \citenamefont
  {Foley}(1952)}]{Frosch1952}%
  \BibitemOpen
  \bibfield  {author} {\bibinfo {author} {\bibfnamefont {R.~A.}\ \bibnamefont
  {Frosch}}\ and\ \bibinfo {author} {\bibfnamefont {H.~M.}\ \bibnamefont
  {Foley}},\ }\bibfield  {title} {\bibinfo {title} {Magnetic hyperfine
  structure in diatomic molecules},\ }\href
  {https://doi.org/10.1103/PhysRev.88.1337} {\bibfield  {journal} {\bibinfo
  {journal} {Phys. Rev.}\ }\textbf {\bibinfo {volume} {88}},\ \bibinfo {pages}
  {1337} (\bibinfo {year} {1952})}\BibitemShut {NoStop}%
\bibitem [{\citenamefont {Chen}\ \emph {et~al.}(2016)\citenamefont {Chen},
  \citenamefont {Bu},\ and\ \citenamefont {Yan}}]{Chen2016}%
  \BibitemOpen
  \bibfield  {author} {\bibinfo {author} {\bibfnamefont {T.}~\bibnamefont
  {Chen}}, \bibinfo {author} {\bibfnamefont {W.}~\bibnamefont {Bu}},\ and\
  \bibinfo {author} {\bibfnamefont {B.}~\bibnamefont {Yan}},\ }\bibfield
  {title} {\bibinfo {title} {Structure, branching ratios, and a laser-cooling
  scheme for the $^{138}\mathrm{BaF}$ molecule},\ }\href
  {https://doi.org/10.1103/PhysRevA.94.063415} {\bibfield  {journal} {\bibinfo
  {journal} {Phys. Rev. A}\ }\textbf {\bibinfo {volume} {94}},\ \bibinfo
  {pages} {063415} (\bibinfo {year} {2016})}\BibitemShut {NoStop}%
\bibitem [{\citenamefont {Hao}\ \emph {et~al.}(2019)\citenamefont {Hao},
  \citenamefont {Pasteka}, \citenamefont {Visscher}, \citenamefont {Aggarwal},
  \citenamefont {Bethlem}, \citenamefont {Boeschoten}, \citenamefont
  {Borschevsky}, \citenamefont {Denis}, \citenamefont {Esajas}, \citenamefont
  {Hoekstra}, \citenamefont {Jungmann}, \citenamefont {Marshall}, \citenamefont
  {Meijknecht}, \citenamefont {Mooij}, \citenamefont {Timmermans},
  \citenamefont {Touwen}, \citenamefont {Ubachs}, \citenamefont {Willmann},
  \citenamefont {Yin},\ and\ \citenamefont {Zapara}}]{Hao2019}%
  \BibitemOpen
  \bibfield  {author} {\bibinfo {author} {\bibfnamefont {Y.}~\bibnamefont
  {Hao}}, \bibinfo {author} {\bibfnamefont {L.~F.}\ \bibnamefont {Pasteka}},
  \bibinfo {author} {\bibfnamefont {L.}~\bibnamefont {Visscher}}, \bibinfo
  {author} {\bibfnamefont {P.}~\bibnamefont {Aggarwal}}, \bibinfo {author}
  {\bibfnamefont {H.~L.}\ \bibnamefont {Bethlem}}, \bibinfo {author}
  {\bibfnamefont {A.}~\bibnamefont {Boeschoten}}, \bibinfo {author}
  {\bibfnamefont {A.}~\bibnamefont {Borschevsky}}, \bibinfo {author}
  {\bibfnamefont {M.}~\bibnamefont {Denis}}, \bibinfo {author} {\bibfnamefont
  {K.}~\bibnamefont {Esajas}}, \bibinfo {author} {\bibfnamefont
  {S.}~\bibnamefont {Hoekstra}}, \bibinfo {author} {\bibfnamefont
  {K.}~\bibnamefont {Jungmann}}, \bibinfo {author} {\bibfnamefont {V.~R.}\
  \bibnamefont {Marshall}}, \bibinfo {author} {\bibfnamefont {T.~B.}\
  \bibnamefont {Meijknecht}}, \bibinfo {author} {\bibfnamefont {M.~C.}\
  \bibnamefont {Mooij}}, \bibinfo {author} {\bibfnamefont {R.~G.~E.}\
  \bibnamefont {Timmermans}}, \bibinfo {author} {\bibfnamefont
  {A.}~\bibnamefont {Touwen}}, \bibinfo {author} {\bibfnamefont
  {W.}~\bibnamefont {Ubachs}}, \bibinfo {author} {\bibfnamefont
  {L.}~\bibnamefont {Willmann}}, \bibinfo {author} {\bibfnamefont
  {Y.}~\bibnamefont {Yin}},\ and\ \bibinfo {author} {\bibfnamefont
  {A.}~\bibnamefont {Zapara}},\ }\bibfield  {title} {\bibinfo {title} {High
  accuracy theoretical investigations of {CaF}, {SrF}, and {BaF} and
  implications for laser-cooling},\ }\href@noop {} {\bibfield  {journal}
  {\bibinfo  {journal} {The Journal of Chemical Physics}\ }\textbf {\bibinfo
  {volume} {151}},\ \bibinfo {pages} {034302} (\bibinfo {year}
  {2019})}\BibitemShut {NoStop}%
\bibitem [{\citenamefont {Yeo}\ \emph {et~al.}(2015)\citenamefont {Yeo},
  \citenamefont {Hummon}, \citenamefont {Collopy}, \citenamefont {Yan},
  \citenamefont {Hemmerling}, \citenamefont {Chae}, \citenamefont {Doyle},\
  and\ \citenamefont {Ye}}]{Yeo2015}%
  \BibitemOpen
  \bibfield  {author} {\bibinfo {author} {\bibfnamefont {M.}~\bibnamefont
  {Yeo}}, \bibinfo {author} {\bibfnamefont {M.~T.}\ \bibnamefont {Hummon}},
  \bibinfo {author} {\bibfnamefont {A.~L.}\ \bibnamefont {Collopy}}, \bibinfo
  {author} {\bibfnamefont {B.}~\bibnamefont {Yan}}, \bibinfo {author}
  {\bibfnamefont {B.}~\bibnamefont {Hemmerling}}, \bibinfo {author}
  {\bibfnamefont {E.}~\bibnamefont {Chae}}, \bibinfo {author} {\bibfnamefont
  {J.~M.}\ \bibnamefont {Doyle}},\ and\ \bibinfo {author} {\bibfnamefont
  {J.}~\bibnamefont {Ye}},\ }\bibfield  {title} {\bibinfo {title} {Rotational
  state microwave mixing for laser cooling of complex diatomic molecules},\
  }\href {https://doi.org/10.1103/PhysRevLett.114.223003} {\bibfield  {journal}
  {\bibinfo  {journal} {Phys. Rev. Lett.}\ }\textbf {\bibinfo {volume} {114}},\
  \bibinfo {pages} {223003} (\bibinfo {year} {2015})}\BibitemShut {NoStop}%
\bibitem [{\citenamefont {Collopy}\ \emph {et~al.}(2018)\citenamefont
  {Collopy}, \citenamefont {Ding}, \citenamefont {Wu}, \citenamefont
  {Finneran}, \citenamefont {Anderegg}, \citenamefont {Augenbraun},
  \citenamefont {Doyle},\ and\ \citenamefont {Ye}}]{Collopy2018}%
  \BibitemOpen
  \bibfield  {author} {\bibinfo {author} {\bibfnamefont {A.~L.}\ \bibnamefont
  {Collopy}}, \bibinfo {author} {\bibfnamefont {S.}~\bibnamefont {Ding}},
  \bibinfo {author} {\bibfnamefont {Y.}~\bibnamefont {Wu}}, \bibinfo {author}
  {\bibfnamefont {I.~A.}\ \bibnamefont {Finneran}}, \bibinfo {author}
  {\bibfnamefont {L.}~\bibnamefont {Anderegg}}, \bibinfo {author}
  {\bibfnamefont {B.~L.}\ \bibnamefont {Augenbraun}}, \bibinfo {author}
  {\bibfnamefont {J.~M.}\ \bibnamefont {Doyle}},\ and\ \bibinfo {author}
  {\bibfnamefont {J.}~\bibnamefont {Ye}},\ }\bibfield  {title} {\bibinfo
  {title} {{3D Magneto-Optical Trap of Yttrium Monoxide}},\ }\href
  {https://doi.org/10.1103/PhysRevLett.121.213201} {\bibfield  {journal}
  {\bibinfo  {journal} {Phys. Rev. Lett.}\ }\textbf {\bibinfo {volume} {121}},\
  \bibinfo {pages} {213201} (\bibinfo {year} {2018})}\BibitemShut {NoStop}%
\bibitem [{\citenamefont {Auzinsh}\ \emph {et~al.}(2010)\citenamefont
  {Auzinsh}, \citenamefont {Budker},\ and\ \citenamefont
  {Rochester}}]{Auzinsh2010}%
  \BibitemOpen
  \bibfield  {author} {\bibinfo {author} {\bibfnamefont {M.}~\bibnamefont
  {Auzinsh}}, \bibinfo {author} {\bibfnamefont {D.}~\bibnamefont {Budker}},\
  and\ \bibinfo {author} {\bibfnamefont {S.}~\bibnamefont {Rochester}},\
  }\href@noop {} {\emph {\bibinfo {title} {Optically Polarized Atoms}}}\
  (\bibinfo  {publisher} {Oxford University Press},\ \bibinfo {year}
  {2010})\BibitemShut {NoStop}%
\bibitem [{\citenamefont {Devlin}\ and\ \citenamefont
  {Tarbutt}(2018)}]{Devlin2018}%
  \BibitemOpen
  \bibfield  {author} {\bibinfo {author} {\bibfnamefont {J.~A.}\ \bibnamefont
  {Devlin}}\ and\ \bibinfo {author} {\bibfnamefont {M.~R.}\ \bibnamefont
  {Tarbutt}},\ }\bibfield  {title} {\bibinfo {title} {Laser cooling and
  magneto-optical trapping of molecules analyzed using optical {Bloch}
  equations and the {Fokker-Planck-Kramers} equation},\ }\href
  {https://doi.org/10.1103/PhysRevA.98.063415} {\bibfield  {journal} {\bibinfo
  {journal} {Phys. Rev. A}\ }\textbf {\bibinfo {volume} {98}},\ \bibinfo
  {pages} {063415} (\bibinfo {year} {2018})}\BibitemShut {NoStop}%
\bibitem [{\citenamefont {Aggarwal}\ \emph {et~al.}(2019)\citenamefont
  {Aggarwal}, \citenamefont {Marshall}, \citenamefont {Bethlem}, \citenamefont
  {Boeschoten}, \citenamefont {Borschevsky}, \citenamefont {Denis},
  \citenamefont {Esajas}, \citenamefont {Hao}, \citenamefont {Hoekstra},
  \citenamefont {Jungmann}, \citenamefont {Meijknecht}, \citenamefont {Mooij},
  \citenamefont {Timmermans}, \citenamefont {Touwen}, \citenamefont {Ubachs},
  \citenamefont {Vermeulen}, \citenamefont {Willmann}, \citenamefont {Yin},\
  and\ \citenamefont {Zapara}}]{Aggarwal2019}%
  \BibitemOpen
  \bibfield  {author} {\bibinfo {author} {\bibfnamefont {P.}~\bibnamefont
  {Aggarwal}}, \bibinfo {author} {\bibfnamefont {V.~R.}\ \bibnamefont
  {Marshall}}, \bibinfo {author} {\bibfnamefont {H.~L.}\ \bibnamefont
  {Bethlem}}, \bibinfo {author} {\bibfnamefont {A.}~\bibnamefont {Boeschoten}},
  \bibinfo {author} {\bibfnamefont {A.}~\bibnamefont {Borschevsky}}, \bibinfo
  {author} {\bibfnamefont {M.}~\bibnamefont {Denis}}, \bibinfo {author}
  {\bibfnamefont {K.}~\bibnamefont {Esajas}}, \bibinfo {author} {\bibfnamefont
  {Y.}~\bibnamefont {Hao}}, \bibinfo {author} {\bibfnamefont {S.}~\bibnamefont
  {Hoekstra}}, \bibinfo {author} {\bibfnamefont {K.}~\bibnamefont {Jungmann}},
  \bibinfo {author} {\bibfnamefont {T.~B.}\ \bibnamefont {Meijknecht}},
  \bibinfo {author} {\bibfnamefont {M.~C.}\ \bibnamefont {Mooij}}, \bibinfo
  {author} {\bibfnamefont {R.~G.~E.}\ \bibnamefont {Timmermans}}, \bibinfo
  {author} {\bibfnamefont {A.}~\bibnamefont {Touwen}}, \bibinfo {author}
  {\bibfnamefont {W.}~\bibnamefont {Ubachs}}, \bibinfo {author} {\bibfnamefont
  {S.~M.}\ \bibnamefont {Vermeulen}}, \bibinfo {author} {\bibfnamefont
  {L.}~\bibnamefont {Willmann}}, \bibinfo {author} {\bibfnamefont
  {Y.}~\bibnamefont {Yin}},\ and\ \bibinfo {author} {\bibfnamefont
  {A.}~\bibnamefont {Zapara}} (\bibinfo {collaboration} {NL-eEDM
  Collaboration}),\ }\bibfield  {title} {\bibinfo {title} {{Lifetime
  measurements of the $A$ ${}^{2}{\mathrm{\ensuremath{\Pi}}}_{1/2}$ and $A$
  ${}^{2}{\mathrm{\ensuremath{\Pi}}}_{3/2}$ states in BaF}},\ }\href
  {https://doi.org/10.1103/PhysRevA.100.052503} {\bibfield  {journal} {\bibinfo
   {journal} {Phys. Rev. A}\ }\textbf {\bibinfo {volume} {100}},\ \bibinfo
  {pages} {052503} (\bibinfo {year} {2019})}\BibitemShut {NoStop}%
\end{thebibliography}%

\section*{Appendix}

\subsection{King plot analysis procedure}
We summarize here the procedure used to extract from the King plot analysis~\cite{Athanasakis2023,FelixKogel}.

\subsubsection{King plots for atoms}
In atomic systems, frequency shifts between the same transitions of two isotopes $A$ and $A'$ are well understood to arise from two primary effects: the mass shift, which originates from changes in the center-of-mass motion, and the field shift, which results from variations in the nucleus' electromagnetic field seen by core-penetrating electrons.
While the former is proportional to the relative difference
\begin{align}
	\frac{1}{\tilde{M}^{A',A}} = \frac{M_{A'}-M_A}{M_{A'}M_A}
\end{align}
of the atomic masses $M_{A'}$ and $M_A$,
the latter contribution depends on the nuclear charge radius $\delta\langle r^2\rangle^{A,A'}$.
Taken together, the isotope shift of a certain transition $i$ can be expressed as a sum of both contributions with the $F$ and $K$ being the field-shift and mass-shift factors, respectively:
\begin{align}\label{eq:isotopeshift}
	\delta\nu^{A,A'}_i = F_i\, \delta\langle r^2\rangle^{A,A'} + K_i\, \frac{1}{\tilde{M}^{A',A}} .
\end{align}
By comparing the shifts of two different transitions $i$ and $j$, a King plot analysis
allows for calculating the nuclear charge radius~\cite{King2013} when combining and rearranging the equation to
\begin{align}\label{eq:kingplotlinear}
	\tilde{M}^{A',A}\, \delta\nu^{A,A'}_i  =
	\frac{F_i}{F_j}\, \tilde{M}^{A',A}\, \delta\nu^{A,A'}_j
	+ (K_i - \frac{F_i}{F_j} K_j) .
\end{align}
To first order, the resulting plot is linear, with the slope and intercept directly linked to the mass and field shift parameters.

\subsubsection{Nuclear charge radius from molecular transitions}
Recent work has extended this concept to molecular systems, demonstrating that isotope shifts in molecules can also be described as the sum of mass and field shifts in Eq.~\ref{eq:isotopeshift}~\cite{Athanasakis2023}. This insight enables a comprehensive analysis of the isotope shifts observed in BaF, based on our experimental results.

By plotting the isotope shifts of the isotopologues \BaFfour, \BaFfive, \BaFsix, and \BaFseven\ against the transition $T_{00}$ of \BaFeight\ as a reference for the molecular transitions $T_{01}$, $T_{11}$, and $T_{10}$, reveals the expected linear behavior in Eq.~\ref{eq:kingplotlinear} for purely molecular king plots in Fig.~\ref{fig:kingplot}(a). Next, all aforementioned transitions $T_{v,v'}$ are compared with atomic data for the \SI{553.6}{\nano\meter} transition known from barium spectroscopy~\cite{FrickeDatabase}, as described in~\cite{Athanasakis2023}. The data points in Fig.~\ref{fig:kingplot}(b) also showing a perfectly linear behavior confirm our extracted molecular constants for all isotopologues.

In addition, utilizing established field and mass shift constants, $F_{\SI{553.6}{\nano\meter}} =  \SI{-4.48+-0.26}{\giga\hertz\per\square\femto\meter}$ and $M_{\SI{553.6}{\nano\meter}} = \SI{1207+-186}{\giga\hertz}$, along with the slope and $y$-axis offsets according to Eq.~\ref{eq:kingplotlinear}, the relative change in the nuclear charge radius of the barium nuclei with respect to \BaFeight\ can be determined~\cite{FrickeDatabase}. Here, the error bars of the atomic data (shaded area) are determined from respective uncertainties in the atomic isotope, mass and field shifts. This analysis allows for the exploration of a nuclei containing $56$ protons, and a varying number of neutrons from $78$ in \BaFfour, to $82$ in \BaFeight.

\subsection{Description of the Molecular Hyperfine Interaction}

In the following, we present the effective Hamiltonian employed to describe the rotational and hyperfine structure of the odd isotopologues of BaF. We start by introducing the appropriate basis states for the individual electronic levels to be studied. 

\subsubsection{Basis States}
\paragraph{$X ^2\Sigma^+$ state}

The $X ^2\Sigma^{+}$ states of $^{137}$BaF and $^{135}$BaF are best described in the Hund's case (b) basis. Here, the effective rotational angular momentum $\mathbf{N}$ is given by $\mathbf{N} = \mathbf{R} + \mathbf{L}$, where $\mathbf{R}$ is the pure rotational angular momentum and $\mathbf{L}$ is the electron orbital angular momentum. 
In sub-case ($b_{\beta S}$) the magnetic dipole hyperfine interaction with the Ba nucleus couples the electron spin $\mathbf{S}$ strongly with the nuclear spin $\mathbf{I}^{\rm (Ba)}$ to form the intermediate angular momentum $\mathbf{G} = \mathbf{S} + \mathbf{I}^{\rm (Ba)}$ \cite{Steimle2011}. Then, $\mathbf{G}$ couples with  $\mathbf{N}$ to form $\mathbf{F_1}=\mathbf{G}+\mathbf{N}$. This quantity finally couples with the $^{19}$F nuclear spin, $\mathbf{I}^{\rm (F)}$, to produce the total angular momentum $\mathbf{F}=\mathbf{F_1}+\mathbf{I}^{\rm (F)}$. 
We write the basis states for $X ^2\Sigma^{+}$ as
\begin{equation*}
    \ket{\eta\Lambda;S, I^{\rm (Ba)}, G, N, F_1, I^{\rm (F)},F,m_F;P},
\end{equation*}
where 
$\eta$ represents all non-angular momentum quantum numbers; $\Lambda = 0$ is the projection of $\mathbf{L}$ on the internuclear axis; $S=1/2$, $G,N,F_1,F$ are the quantum numbers associated with $\mathbf{G},\mathbf{N},\mathbf{F_1}$, and $\mathbf{F}$ respectively, and $m_F$ is the projection of $\mathbf{F}$ on the laboratory $Z$-axis. These states have parity quantum number $P = (-1)^{N}$.

\paragraph{$A ^2\Pi_{1/2}$ state}
The Hund's case (a) basis is a good representation for the sublevels of the $A ^2\Pi_{1/2}$ states \cite{Steimle2011} in BaF. Here, $\mathbf{S}$ and $\mathbf{L}$ couple with $\mathbf{R}$ to produce the total angular momentum excluding nuclear spins, $\mathbf{J}=\mathbf{L} + \mathbf{S} + \mathbf{R}$. 
Next, $\mathbf{I}^{\rm (Ba)}$ couples with $\mathbf{J}$ to produce an intermediate angular momentum, $\mathbf{F_1}=\mathbf{J}+\mathbf{I}^{\rm (Ba)}$. The coupling of $\mathbf{F_1}$ with $\mathbf{I}^{\rm (F)}$ generates the total angular momentum, $\mathbf{F}=\mathbf{F_1}+\mathbf{I}^{\rm (F)}$. 
We describe the $A ^2\Pi_{1/2}$ states using a basis with quantum numbers 
$\Lambda=\pm 1$, $\Sigma = \pm 1/2$, (the signed projections of $\mathbf{L}$ and $\mathbf{S}$ on the intermolecular axis, respectively), plus $\Omega \equiv \Lambda + \Sigma = \pm 1/2$, $F_1$, $F$, and $m_F$. Finally, we write the basis states with signed values of $\Omega$ as
\begin{equation}
    \ket{\psi}_{\Omega} = \ket{\eta\Lambda,S,\Sigma,J,\Omega; J, I^{(\rm Ba)},F_1,I^{(\rm F)},F,m_F}.
\end{equation}
The energy and parity eigenstates, $\ket{\psi}_{\rm P}$, can be written as a superposition of the signed-$\Omega$ basis states:
\begin{equation}
    \begin{aligned}
        \ket{\psi}_{\rm P} = \dfrac{1}{\sqrt{2}}& 
        \Bigl( \ket{\Lambda ,S,\Sigma,\Omega; J,I^{(\rm Ba)},F_1,I^{(\rm F)},F,m_F} + \nonumber \\
        (-1)^p P & \ket{-\Lambda,S,-\Sigma,-\Omega; J, I^{(\rm Ba)},F_1,I^{(\rm F)},F,m_F} \Bigr),
%        \nonumber \\
%        \equiv \ket{\psi}_{\rm P}\ket{|\Lambda|,S,|\Sigma|,|\Omega|,\;J,I^{\rm{(Ba)}},F_1,I^{\rm{(F)}},F,m_F;P}.
    \end{aligned}
    \label{eq:parity_rep}
\end{equation}
where $p=J-S$ and $P = \pm 1$ is the parity of the state. For the $\Pi_{1/2}$ states, $\Lambda = \pm 1$, $\Sigma = \mp 1/2$, and $\Omega = \Lambda + \Sigma = \pm 1/2$.

\subsubsection{Effective Hyperfine/Rotation Hamiltonians}

Effective Hamiltonians describing hyperfine and rotational motion in a single vibronic level of $^2\Sigma^+$ and $^2\Pi_{1/2}$ states with two nuclear spins have been summarized in Ref.~\cite{Brown2003}. The effective Hamiltonian for the  $X ^2\Sigma^+$ state is given by

\begin{eqnarray}
    H_{^2\Sigma^+} =& B \mathbf{N}^2 - D \mathbf{N}^4 + \gamma \mathbf{N}\cdot\mathbf{S} + \nonumber \\
    & b^{\rm (Ba)} \mathbf{I}^{\rm (Ba)}\cdot \mathbf{S} + c^{\rm (Ba)} \Bigl(I_z^{\rm (Ba)} S_z - \dfrac{1}{3}\mathbf{I}^{\rm (Ba)} \cdot \mathbf{S}\Bigr) + \nonumber \\
    & e q_0 Q^{\rm (Ba)}\dfrac{3 I^{\rm (Ba)}_{z}-\mathbf{I}^{\rm (Ba)}\cdot \mathbf{I}^{\rm (Ba)}}{4 I^{\rm (Ba)}(2 I^{\rm (Ba)}-1)} +\nonumber \\
     & b^{\rm (F)} \mathbf{I}^{\rm (F)}\cdot \mathbf{S} + c^{\rm (F)} \Bigl(I_z^{\rm (F)} S_z - \dfrac{1}{3}\mathbf{I}^{\rm (F)} \cdot \mathbf{S}\Bigr). \nonumber \\
\end{eqnarray}

Here, $B$ and $D$ parameterize the rotational and the centrifugal distortion corrections energies, and $\gamma$ describes the strength of the electron spin coupling with molecular rotation. The terms with coefficients $b=b_F-c/3$ and $c$, indexed by the nucleus in the superscript, represent the Fermi contact interaction and the electron spin-nuclear spin dipolar interaction, respectively. The subscript $z$ refers to the direction along the internuclear axis in the molecule-fixed frame. The nuclear quadrupole moment of the Ba nucleus, $Q^{\rm (Ba)}$, couples to the electric field gradient caused by the valence electron, $eq_0$, which contributes an electric quadrupole term parametrized by $eq_0Q^{\rm (Ba)}$. The values of these parameters were already established with high accuracy via microwave spectroscopy \cite{Ryzlewicz1982,Preston2025}, and we treat them as fixed in our analysis.

We write the $^2\Pi_{1/2}$ state Hamiltonian as 
\begin{eqnarray}
    H_{^2\Pi_{1/2}} = &\dfrac{1}{2}A_D \{L_zS_z,\mathbf{R}\} + B \mathbf{R}^2 - D \mathbf{R}^4 + \nonumber \\
    &\dfrac{1}{2}(p+2q) (e^{-2 i \phi} J_{+}S_{+} + h.c.) +  H_{\Pi_{1/2}}^{\rm HF}.
\end{eqnarray}
Here, $A_D$ characterizes the centrifugal distortion of the spin-orbit interaction; $\{\}$ denotes the anti-commutator; $p+2q$ characterizes the $\Lambda$-doubling; the subscripts $\pm$ correspond to spin raising/lowering operators respectively, in the molecule's body fixed frame; $h.c.$ denotes the Hermitian conjugate of the preceding expression; and $H_{\Pi_{1/2}}^{\rm HF}$ describes the hyperfine interaction. Values of the parameters $A_D, B, D,$ and $p+2q$ have been measured by Steimle \textit{et al.} \cite{Steimle2011} and also by absorption spectroscopy in the current work. We fix these parameters to the values determined in the current work for our analysis of the hyperfine structure. 
Following the form of the Hamiltonian used in Ref.~\cite{Denis2022} for the magnetic hyperfine terms and Ref.~\cite{Brown2003} for the quadrupole term, we write the hyperfine Hamiltonian as:
\begin{eqnarray}
H_{\Pi_{1/2}}^{\rm HF} =  H_{h_{1/2}}^{\rm (F)} +H_d^{\rm (F)}+H_{h_{1/2}}^{\rm (Ba)} + H_d^{\rm (Ba)}+H^{\rm (Ba)}_{\mathrm{Q}} \nonumber \\
= h_{1 / 2}^{\rm (F)} I^{\rm (F)}_{z}+\frac{1}{2} d^{\rm (F)}\left[e^{2 i \phi} I^{\rm (F)}_{-} S_{-}+e^{-2 i \phi} I^{\rm (F)}_{+} S_{+}\right] \nonumber \\
+  h_{1 / 2}^{\rm (Ba)} I^{\rm (Ba)}_{z}+\frac{1}{2} d^{\rm (Ba)}\left[e^{2 i \phi} I^{\rm (Ba)}_{-} S_{-}+e^{-2 i \phi} I^{\rm (Ba)}_{+} S_{+}\right] \nonumber \\ + 
e q_0 Q^{\rm (Ba)}\dfrac{3 I^{\rm (Ba)}_{z}-\mathbf{I}^{\rm (Ba)}\cdot \mathbf{I}^{\rm (Ba)}}{4 I^{\rm (Ba)}(2 I^{\rm (Ba)}-1)}.~~~~
\end{eqnarray}
\label{eq:Htotal}
The magnetic hyperfine coefficient $h_{1/2}$ can be expressed in terms of the Frosch and Foley $a$, $b$, and $c$ parameters~\cite{Frosch1952} as 
\begin{equation}
    h_{1/2}= a - \dfrac{1}{2} (b + c) = a - \dfrac{b_F}{2} - \dfrac{c}{3}.
\end{equation}
Note that both parameters $h_{1/2}^{\rm(Ba)}$ and $d^{\rm(Ba)}$ are proportional to the nuclear magnetic moment of the relevant isotope, $\mu^{\rm(Ba)}$, while the quadrupole term $e q_0 Q^{\rm (Ba)}$ is proportional to their electric quadrupole moments $Q^{\rm (Ba)}$.\\ 

\subsection{Matrix elements of the hyperfine Hamiltonian for the \exs\, state} \label{appendix:mat_elts}

We provide here expressions for the matrix elements of the magnetic hyperfine and nuclear electric quadrupole interaction terms in the Hund's case (a) basis. 

\begin{widetext}

\paragraph{Matrix elements of $H_{h_{1/2}}^{\rm (Ba)}$}
\begin{equation}
    \begin{aligned}
        &\bra{\eta\Lambda,S,\Sigma,J,\Omega; J, I_1,F_1,I_2,F,m_F} H_{h_{1/2}}^{\rm (Ba)}\ket{\eta\Lambda',S,\Sigma',J',\Omega'; J', I_1,F_1',I_2,F',m_F'} \\
        = &h_{1/2}^{\rm (Ba)} \Lambda \delta_{\Lambda,\Lambda'} (-1)^{J - \Omega}  \threejsymbol{J}{1}{J'}{-\Omega}{0}{\Omega'} [JJ']^{1/2} \\
        & \delta_{m_F,m_F'} \delta_{F,F'} \delta_{F_1,F_1'} (-1)^{J'+F_1+I_1} \{I_1\}^{1/2}
        \sixjsymbol{I_1}{J'}{F_1}{J}{I_1}{1}
    \end{aligned}
\end{equation}

\paragraph{Matrix elements of $H_{h_{1/2}}^{\rm (F)}$}
\begin{equation}
    \begin{aligned}
        &\bra{\eta\Lambda,S,\Sigma,J,\Omega; J, I_1,F_1,I_2,F,m_F} H_{h_{1/2}}^{\rm (F)}\ket{\eta\Lambda',S,\Sigma',J',\Omega'; J', I_1,F_1',I_2,F',m_F'} \\
        = &h_{1/2}^{\rm (F)} \Lambda \delta_{\Lambda,\Lambda'} (-1)^{J - \Omega}  \threejsymbol{J}{1}{J'}{-\Omega}{0}{\Omega'} [JJ']^{1/2} \\
         & \delta_{m_F,m_F'} \delta_{F,F'} \delta_{F_1,F_1'} (-1)^{2F_1'+F+I_2+J+I_1+1}
        [F_1F_1']^{1/2} 
        \sixjsymbol{I_2}{F_1'}{F}{F_1}{I_2}{1}
        \sixjsymbol{J'}{F_1'}{I_1}{F_1}{J}{1}
        \{I_2\}^{1/2}
    \end{aligned}
\end{equation}

\paragraph{Matrix elements of $H_d^{\rm (Ba)}$}
\begin{equation}
    \begin{aligned}
        &\bra{\eta\Lambda,S,\Sigma,J,\Omega; J, I_1,F_1,I_2,F,m_F} H_d^{\rm (Ba)}\ket{\eta\Lambda',S,\Sigma',J',\Omega'; J', I_1,F_1',I_2,F',m_F'} \\
        = &\left[\delta_{\Lambda,\Lambda'+2} 
        \threejsymbol{S}{1}{S}{-\Sigma}{-1}{\Sigma'} 
        \threejsymbol{J}{1}{J'}{-\Omega}{1}{\Omega'} +
        \delta_{\Lambda,\Lambda'-2} 
        \threejsymbol{S}{1}{S}{-\Sigma}{1}{\Sigma'} 
        \threejsymbol{J}{1}{J'}{-\Omega}{-1}{\Omega'}
        \right] 
        d^{\rm (Ba)} (-1)^{S - \Sigma + J -\Omega} [JJ']^{1/2} \{S\}^{1/2}\\
        &\delta_{m_F,m_F'} \delta_{F,F'} \delta_{F_1,F_1'} (-1)^{J'+F_1+I_1}
        \sixjsymbol{I_1}{J'}{F_1}{J}{I_1}{1}
        \{I_1\}^{1/2}
    \end{aligned}
\end{equation}

\paragraph{Matrix elements of $H_d^{\rm (F)}$}
\begin{equation}
    \begin{aligned}
        &\bra{\eta\Lambda,S,\Sigma,J,\Omega; J, I_1,F_1,I_2,F,m_F} H_d^{\rm (F)}\ket{\eta\Lambda',S,\Sigma',J',\Omega'; J', I_1,F_1',I_2,F',m_F'} \\
        = &\left[\delta_{\Lambda,\Lambda'+2} 
        \threejsymbol{S}{1}{S}{-\Sigma}{-1}{\Sigma'} 
        \threejsymbol{J}{1}{J'}{-\Omega}{1}{\Omega'} +
        \delta_{\Lambda,\Lambda'-2} 
        \threejsymbol{S}{1}{S}{-\Sigma}{1}{\Sigma'} 
        \threejsymbol{J}{1}{J'}{-\Omega}{-1}{\Omega'}
        \right] 
        d^{\rm (F)} (-1)^{S - \Sigma + J -\Omega} [JJ']^{1/2} \{S\}^{1/2}\\
        & \delta_{m_F,m_F'} \delta_{F,F'} (-1)^{2F_1'+F+I_2+J+I_1+1}
        [F_1F_1']^{1/2} 
        \sixjsymbol{I_2}{F_1'}{F}{F_1}{I_2}{1}
        \sixjsymbol{J'}{F_1'}{I_1}{F_1}{J}{1}
        \{I_2\}^{1/2}
    \end{aligned}
\end{equation}

\paragraph{Matrix element of $H_Q$}
\begin{equation}
    \begin{aligned}
        &\bra{\eta\Lambda,S,\Sigma,J,\Omega; J, I_1,F_1,I_2,F,m_F} H_Q\ket{\eta\Lambda',S,\Sigma',J',\Omega'; J', I_1,F_1',I_2,F',m_F'} \\
        = &\dfrac{eq_0Q^{\rm (Ba)}}{4} \delta_{\Sigma,\Sigma'} 
        (-1)^{J - \Omega}\threejsymbol{J}{2}{J'}{-\Omega}{0}{\Omega'} [JJ']^{1/2} \\
        & \delta_{m_F,m_F'} \delta_{F,F'}\delta_{F_1,F_1'} (-1)^{J'+F_1+I_1}
        \sixjsymbol{I_1}{J}{F_1}{J'}{I_1}{2}
        \threejsymbol{I_1}{2}{I_1}{-I_1}{0}{I_1}^{-1}
    \end{aligned}
\end{equation}

\end{widetext}

Here we have used the following short hand notations as in Ref \cite{Chen2016}:
\begin{equation*}
    \begin{aligned}
    \{A \} &= A(A+1)(2A+1) \\
    [A B] &= (2A+1)(2B+1)
    \end{aligned}
\end{equation*}

\subsection{Diagonalization and state mixing}
From these expressions, it can be seen that the magnetic hyperfine interaction mixes different J states with the selection rule $\Delta J = \pm 1$. Since we find mixing matrix elements as large as $\sim 100$ MHz (based on the parameters extracted in the current work), and with the separation between consecutive $J$ levels of $\sim 10$ GHz, we expect a mixing of order $1\%$ between nearby J states. The quadrupole Hamiltonian, on the other hand, mixes J states with the selection rule $\Delta J = \pm 2$.  For our final analysis, we diagonalize the Hamiltonian for the $A ^2\Pi_{1/2}$ state by including all basis states with values of $J$ from 1/2 to 7/2, and both parities.

This hyperfine mixing between the states $J'=1/2^+$ and $J'=3/2^+$ of \exs\ also has important consequences for laser cooling, which relies on selection rules to limit rotational branching. Whereas the pure state $J'=1/2^+$ only branches into levels of the rotational level $N=1$ of the ground state \gs, the state $J'=3/2^+$ decays with similar probabilities into $N=1$ and $N=3$. The small mixing of these two $J'$ states therefore creates a leakage of approximately \SI{0.05}{\percent} into $N=3$ that allows the scattering of approximately $2000$ photons. We note that this value is of the same order of magnitude as the branching ratio into the intermediate $A'\Delta$ state previously studied in the bosonic isotopologues~\cite{Albrecht2020, Hao2019}.

To recover these molecules, neighboring rotational states are typically mixed by applying microwave radiation~\cite{Yeo2015}. As the number of ground states involved in the optical cycle increases significantly, the maximum scattering rate is substantially reduced. To address this, an alternative approach is to drive optical transitions, specifically, the \gs$(N=3)$ to \exs$(J'=3/2^+)$ transition in this case~\cite{Collopy2018, Zeng2024}.

\subsection{Numerical modeling of the spectra}
\begin{figure}[tb]
\centering
\includegraphics[width=0.48\textwidth]{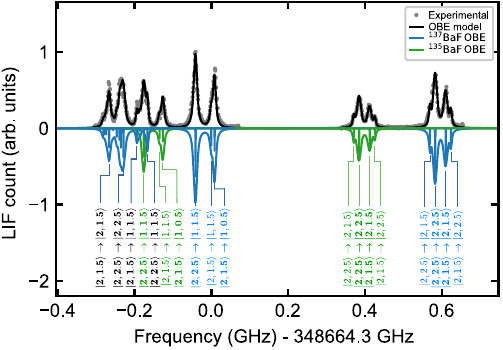}
\caption{Experimental (gray dots) and modeled (black curve) spectra for the $X\ket{N=0, G=2, F_1,F} \to A\ket{J'^{ P} = 1/2^-,F_1',F'}$ transitions in $^{135}$BaF and $^{137}$BaF (with individual modeled spectra and quantum numbers $\ket{F_1,F} \to \ket{F_1',F'}$ labeled in green and blue, respectively). Also shown is the $X\ket{N=1, G=1,F_1,F} \to A\ket{J'^{ P} = 1/2^+,F_1',F'}$ transition in $^{137}$BaF (with quantum numbers in blue), which overlaps partially with this frequency range.  Among the transitions labeled in the diagram, only the ones in bold were used in the quantitative analysis.  A stick plot depicting all transitions, with strength proportional to the square of the transition dipole matrix elements, is overlaid on the full modeled spectrum generated using simulations based
on the optical Bloch equations. Note that for the simulated spectra shown here (and in all figures), the center of mass of each isotopologue and rotation state was adjusted to maximize overlap with the data.  The resulting adjustments were well within the range of uncertainties for the rotational and isotope splittings determined via absorption spectroscopy.} 
\label{fig:N0_upper_new}
\end{figure}

Transition dipole matrix elements were computed by diagonalizing the hyperfine and rotational Hamiltonians, transforming $X$ state kets into the Hund's case (a) basis, and applying standard formulae for line strengths between case (a) states. This information was sufficient to produce the ``stick plots'' in Figs.~\ref{fig:N0_upper_new}--\ref{fig:R0_G1_137_new}. 

To reproduce the measured spectra and to make estimates of systematics, we model the light-atom interaction with the optical Bloch equations (OBEs) \cite{Auzinsh2010,Devlin2018}. This approach is required to account for saturation and (partial) optical cycling effects, which affect the relative line strengths and widths. For this, absolute line strengths were computed using the known spontaneous decay rate $\Gamma  = 2\pi\, \times\, 2.7 $ MHz of the A$^2\Pi_{1/2}$ state \cite{Aggarwal2019}. We numerically integrate the OBEs taking into account the Gaussian shape of the laser beam and associated spatial and temporal distribution of intensity experienced by the ensemble of molecules in the beam. The effect of the lab magnetic field was included to account for remixing of dark states on transitions that could accommodate some optical cycling.  We also account for Doppler broadening from the transverse velocity distribution of the molecular ensemble, computed from the known collimation geometry and mean forward velocity of the beam.

Note that similar modeling for the absorption spectroscopy data taken inside a buffer gas cell is challenging. The high rate of collisions inside the buffer gas cell means that the coherences revealed by the OBEs (and associated optical pumping, optical cycling, saturation of signals with laser power, etc.) will be strongly altered. The corresponding complex interplay of collisions and laser light will be investigated in future work.

\subsection{Line assignments}
In the following, we outline how we use the knowledge of the spectra to reliably assign individual lines in the recorded fluorescence spectroscopy data.

\subsubsection{\label{sec:Q0}$\mathbf{X(N=0,G=2) \to A(J'^{P}=1/2^-)}$ spectrum}
The spectrum for the transition $X(N=0,  G=2) \to A(J'^{P} = 1/2^-)$ is shown in Fig.~\ref{fig:N0_upper_new}.
This spectrum is relatively simple, with a small number of sublevels. Moreover, the $J'=1/2$ excited states experience no first-order shift from the quadrupole hyperfine interaction, $H_Q^{\rm(Ba)}$. Since the magnetic hyperfine interaction is small compared to the rotational energy splitting, the spectrum can be approximately described using an effective Hamiltonian that operates only on levels with the same value of $J'^{P}$.  
This effective Hamiltonian can be written as
\begin{equation}
\begin{aligned}
    &~~H^{\rm{eff}}_{J'=1/2}\\
    =&\left[h_{1 / 2}^{\rm(Ba)}-d^{\rm(Ba)} \cdot P(-1)^p \cdot \frac{1}{2}(2 J+1)\right] \frac{\mathbf{I^{\rm (Ba)}}\cdot \mathbf{J}}{2 J(J+1)}\\
    +&\left[h_{1 / 2}^{\rm (F)}-d^{\rm (F)} \cdot P(-1)^p \cdot \frac{1}{2}(2 J+1)\right] \frac{\mathbf{I^{\rm (F)}}\cdot \mathbf{J}}{2 J(J+1)}.
\end{aligned}
\label{dh_expansion}
\end{equation}
Hence, to first approximation the hyperfine splittings in the $J'^{P}=1/2^-$ state depend only on the parameters $h_{1 / 2}^{\rm (Ba)}+d^{\rm (Ba)}$ and $h_{1 / 2}^{\rm (F)}+d^{\rm (F)}$.  Moreover, $h_{1/2}^{\rm (F)}$ and $d^{\rm (F)}$ are known from prior data on $^{138}$BaF \cite{Denis2022} and give only small ($\lesssim 10$ MHz) splittings.

Lines in the high-frequency region ($\approx\! 350$-$650$ MHz in Fig.~\ref{fig:N0_upper_new}) clearly originated from the $X(N=0,G=2)$ sublevels. From their relative position and amplitude, the higher- (lower-)frequency cluster of four lines here were assigned to $^{137}$BaF ($^{135}$BaF). The smaller (nearly overlapped) doublet splittings corresponded to expected splittings from the known values of $h^{\rm (F)}_{1/2}$ and $d^{\rm (F)}$ \cite{Denis2022}, and the average of the larger doublet splittings corresponded to the known $^{19}$F hyperfine splittings in the $X ^2\Sigma^+$ state. All of these high-frequency lines were assigned to transitions to the maximum value of $F_1'=2$ in the excited state with $J'=1/2$. Next, the largest-amplitude lines in the low-frequency region of this spectrum (the two unresolved doublets near 0 frequency in Fig.~\ref{fig:N0_upper_new}) were assigned to $^{137}$BaF, and to transitions to excited states with $F'=1$. Then, peaks were fit to each of the lines assigned to $^{137}$BaF to find a preliminary location of the line centers.

Next, pairs of large and relatively isolated lines were identified that originated from a common X state sublevel, but terminated in different excited state sublevels.  In Fig.~\ref{fig:N0_upper_new}, these lines, labeled in bold, corresponded to the transitions $X\ket{F_1=2,F=2.5}\to A\ket{F_1=2,F=2.5}$ and $\to A\ket{F_1=1,F=1.5}$; and $X\ket{F_1=2,F=1.5}\to A\ket{F_1=2,F=1.5}$ and $\to A\ket{F_1=1,F=0.5}$. 
By fitting to $H^{\rm{eff}}_{J'=1/2}$, these splittings yielded a preliminary value of $h^{\rm{(Ba)}}_{1/2}+d^{\rm{(Ba)}}$ for $^{137}$BaF, accurate to a few MHz.

From the expected scaling of $h^{\rm(Ba)}_{1/2} + d^{\rm(Ba)}$ with the nuclear magnetic moment $\mu^{\rm (Ba)}$, we could then estimate the splittings of the $^{135}$BaF lines for transitions to excited states with $F'=1$. In the spectrum of Fig.~\ref{fig:N0_upper_new} these clearly overlapped with another set of lines, which were later assigned to transitions $X(N=1,G=1) \to A(J'^{P}=1/2^+)$. However, two pairs of sufficiently large and isolated lines were identified to enable preliminary estimates of $h^{\rm(Ba)}_{1/2} + d^{\rm(Ba)}$ for $^{135}$BaF (independent of the values from $^{137}$BaF) from this spectrum as well.

\begin{figure}[tb]
\centering
\includegraphics[width=0.48\textwidth]{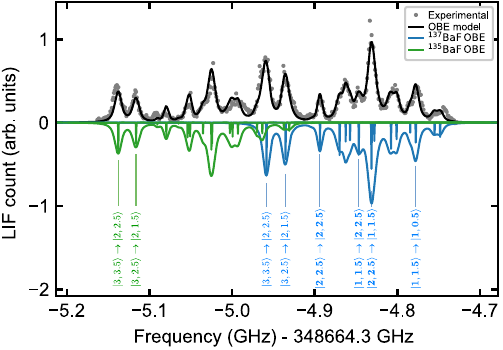}

\caption{Spectra for the $X\ket{N=1, G=2, F_1,F} \to A\ket{J'^{ P} = 1/2^+,F_1',F'}$ transitions in $^{135/137}$BaF. All notations are as in Fig.~\ref{fig:N0_upper_new}.}
\label{fig:N1_G2_new}
\end{figure}

\subsubsection{$\mathbf{X(N=1, G=2) \to A(J'^{ P}=1/2^+)}$ spectrum}
%The spectrum of $X(N = 1, G =2) \to A(J'^{ P}=1/2^+)$ transitions is shown in Fig.~\ref{fig:N1_G2_new}.  
The $X(N = 1, G =2) \to A(J'^{ P}=1/2^+)$ transitions with lines identified by their quantum numbers are shown in Fig.~\ref{fig:N1_G2_new}. The largest lines on the high-frequency side of the spectrum were tentatively assigned to transitions in $^{137}$BaF with the strongest transition dipole matrix elements. Line positions and further line assignments for both isotopologues then proceeded as described above.  This analysis was used to determine a preliminary value of $h^{\rm (Ba)}_{1/2} - d^{\rm (Ba)}$ for $^{137}$BaF.  
(Lines in this spectrum were only used for estimation of hyperfine parameters in $^{137}$BaF, not in $^{135}$BaF.) With the preliminary value of $h^{\rm (Ba)}_{1/2}-d^{\rm (Ba)}$, it was possible to estimate the positions of the $X(N=1,G=1) \to A(J'^{ P}=1/2^+)$ transitions for $^{137}$BaF and assign the ``extra'' lines in Fig.~\ref{fig:N0_upper_new} to them.

\begin{figure}[tb]
\centering
\includegraphics[width=0.48\textwidth]{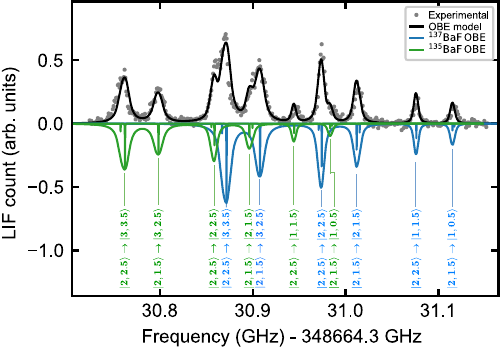}
\caption{Spectra for the $X(N=0,G=2)\to A(J'^{P}=3/2^-)$ transitions in $^{135/137}$BaF. All notations are as in Fig.~\ref{fig:N0_upper_new}.}
\label{fig:R0_G2_new}
\end{figure}

\subsubsection{$\mathbf{X(N=0,G=2,1) \to A(J'^{ P}=3/2^-)}$ spectra}
\begin{figure}[tb]
\centering
\includegraphics[width=0.48\textwidth]{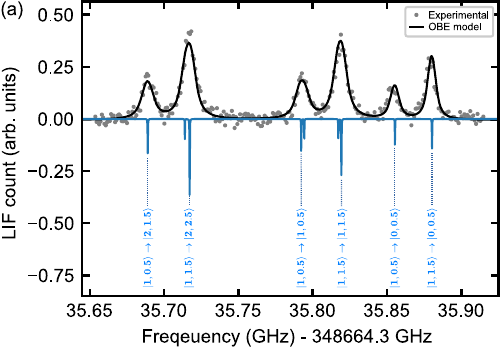}
~\\
\includegraphics[width=0.48\textwidth]{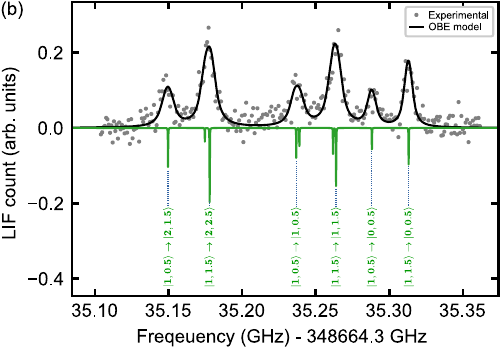}

\caption{Spectra for the $X(N=0,G=1)\to A(J'^{P}=3/2^-)$ transition. All notations are as in Fig.~\ref{fig:N0_upper_new}. For this transition, the spectral lines of \(^{137}\mathrm{BaF}\) presented in~(a) are well separated from the corresponding \(^{135}\mathrm{BaF}\) features shown in~(b).
} 
\label{fig:R0_G1_137_new}
\end{figure}

The spectrum of $X(N=0,G=2)\to A(J'^{ P}=3/2^-)$ transitions is shown in Fig.~\ref{fig:R0_G2_new} for both isotopologues. The $X(N=0,G=1)\to A(J'^{ P}=3/2^-)$ spectra for $^{137}$BaF and $^{135}$BaF are shown in Figs.~\ref{fig:R0_G1_137_new}a,b respectively.  
These measurements, together with the preliminary evaluations of $h_{1/2}^{\rm{(Ba)}}$ and $d^{\rm{(Ba)}}$, made it possible to determine the values of $eq_0Q^{\rm (Ba)}$ for the two isotopologues. 

\subsubsection{Global analysis of the hyperfine structure and uncertainty budget}

After all lines were assigned and preliminary fits performed to find values of hyperfine Hamiltonian parameters, a final global fit to all line pairs used in the preliminary analysis of each spectrum scan was performed, including the effect of Hamiltonian matrix elements off-diagonal in $J'$.  

For each fitted level splitting, a statistical uncertainty (typically $1$-$4$ MHz) was assigned from the standard error in the mean between nominally identical spectrum scans. The reduced chi-squared value of this global fit was $\chi^2_\nu = 1.7~(2.1)$ for $^{137}$BaF ($^{135}$BaF). The final statistical uncertainties for each splitting were assigned by multiplying the raw uncertainty by the corresponding value of $\sqrt{\chi^2_\nu}$ for each isotopologue.  An systematic uncertainty of up to $0.7$ MHz was added to some lines to account for the effect of unresolved overlaps with smaller nearby lines.  An additional systematic uncertainty, in the range $0.4-2.8$ MHz, was assigned to each line due to possible Zeeman shifts from the laboratory magnetic field (roughly earth's field).  

\begin{figure}[tb]
    \centering
    \includegraphics[width=0.45\textwidth]{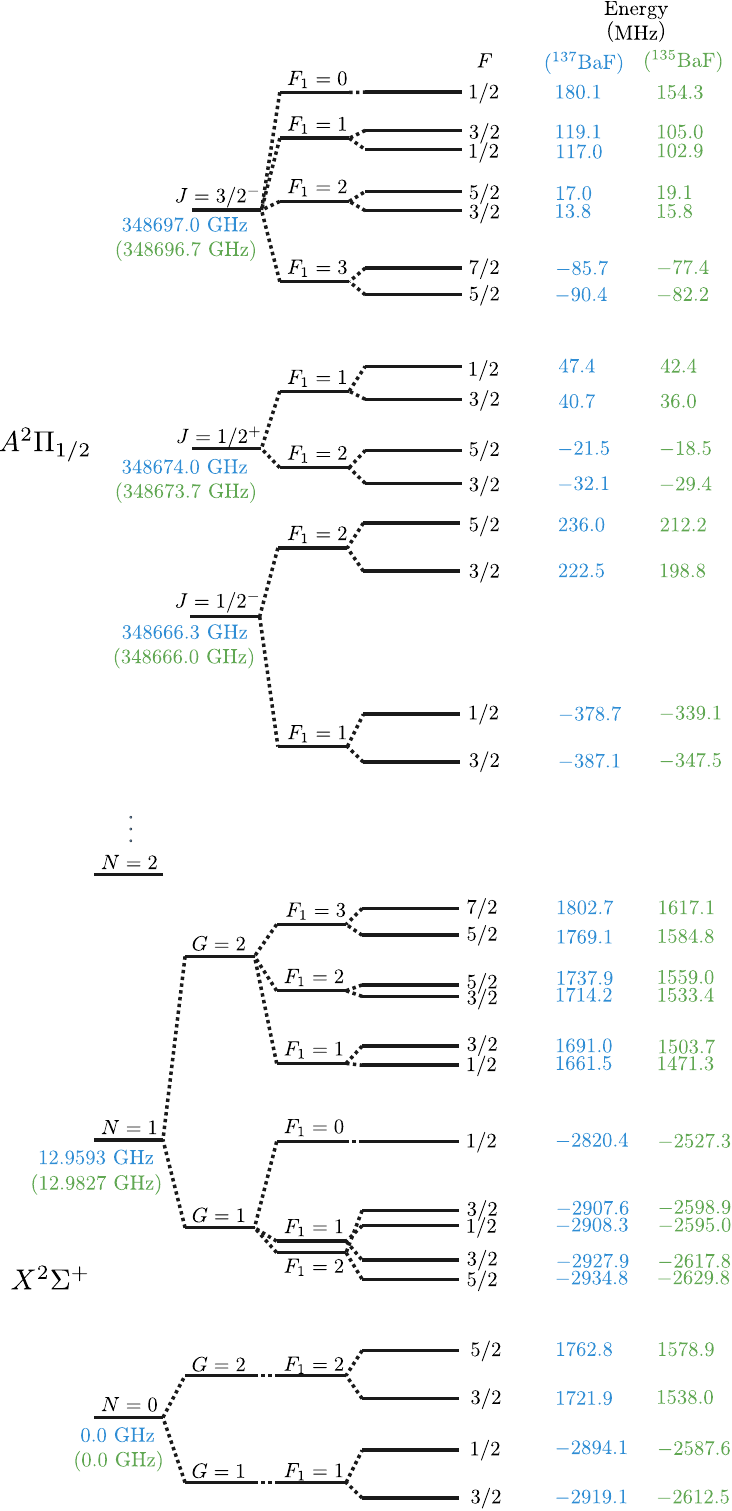}
    \caption{Energy levels in the lower hyperfine-rotational levels of the $X ^2\Sigma^{+}$ and $A ^2\Pi_{1/2}$ states, for the isotopologues $^{137}$BaF (blue) $^{135}$BaF (green). The hyperfine splittings within the $A ^2\Pi_{1/2}$ states are based on the parameters estimated in the current work. The two isotopologues have similar structure of the energy splittings, except for the $\ket{X, N\!=\!1, G\!=\!1, F_1\!=\!1, F\!=\!1/2}$ and $\ket{X, N\!=\!1, G\!=\!1, F_1\!=\!2, F\!= 3/2}$ states, where the order of these levels is reversed in the case of $^{135}$BaF.}
    \label{fig:energy_levels}
\end{figure}

Fig.~\ref{fig:energy_levels} shows the final assigned energies of all states used in the analysis of hyperfine splittings, complementing Fig.~\ref{fig:levelstructure} in the main text.

All transition pairs used to determine the parameters of the hyperfine Hamiltonian are shown in Tabs. \ref{tab:Overall_135} and \ref{tab:Overall_137}, for $^{135}$BaF and $^{137}$BaF, respectively. These tables show the quantum numbers for each line, the value of the splitting between excited states in the transition pair, and the statistical uncertainty (after correction for $\chi^2_\nu >1$) of each splitting.

We also account for two sources of systematic errors. The first arises from apparent line shifts due to unresolved overlapping spectral lines.  To account for this, we fit the OBE-modeled spectra using the same procedure as for the real data, then calculate the difference between the fitted level splitting and the splitting known directly from the Hamiltonian eigenvalues. Note that some lines are sufficiently isolated so that this results in no error.  For lines with substantial overlap, the resulting errors are in the range $0.2-0.7$ MHz.

The second source of possible systematic error arises from Zeeman shifts due to the nonzero magnetic field in the lab used for the fluorescence measurements, which was measured to be about $0.45\,$G ($0.17\,$G) perpendicular (parallel) to the linear polarization of the laser beam. Using known magnetic $g$-factors of the $X ^2\Sigma$ \cite{Cahn2014} and $A ^2\Pi_{1/2}$ \cite{Steimle2011} states and the diagonalized hyperfine/rotation Hamiltonian, we calculated Zeeman shifts in the energy of each state used in the analysis. We conservatively assume that the error in the observed state energy could be as large as half the maximum splitting between $m_F$ sublevels in the state. The total systematic error in the measurement of the position of each of the transitions is assigned by linearly adding errors from Zeeman shifts in the ground and excited states. 

%\begin{widetext}
\begin{flushleft}
\begin{table*}[tb]
\centering
%\begin{NiceTabular}{|c|c|c|c|c|c|c|c|c|c|}
\begin{NiceTabular}{cccccccccc}
\noalign{\hrule height 0.4mm \vspace{-7pt}}
~\\
Transitions in $^{135}$BaF & $\Delta$  & $\sigma_{\rm st}$ & $\sigma_{\rm f}$ &$\sigma_{\rm f}^{\Delta}$ & $\sigma_{\rm Z}$ & $\sigma_{\rm Z}^{\Delta}$ & $\sigma_{\rm tot}$ & $\sigma_{\rm tot}^{\rm wt}$ & $\Delta^{\rm wt}$ \\ 
$X\ket{N,G,F_1,F}\to A\ket{J'^P,F'_1,F'}$ & (MHz) & (MHz)& (MHz) &(MHz) & (MHz) & (MHz) & (MHz) & (MHz) & (MHz)\\
\hline\hline
$\ket{ 0,    2,  2,   2.5} \to \ket{3/2^-, 3,  3.5}$ & \multirow[c]{2}{*}{98.62}  & \multirow{2}{*}{1.55}  & 
0.42 & \multirow{2}{*}{0.47} &  0.76 & \multirow{2}{*}{1.48} & \multirow{2}{*}{2.19} &\multirow{2}{*}{2.19} & \multirow{2}{*}{98.62} \\
$\ket{ 0,    2,  2,   2.5} \to \ket{  3/2^-,  2,  2.5}$ &  &  & 0.22 & & 0.72 &  &   &  &  \\

\hline\hline

$\ket{ 0,    2,  2,   1.5} \to \ket{  3/2^-,  3,  2.5}$ & \multirow{2}{*}{97.13}  & \multirow{2}{*}{1.42} & --  & \multirow{2}{*}{0.74} & 0.70  & \multirow{2}{*}{1.34} & \multirow{2}{*}{2.09} & \multirow{2}{*}{2.09} & \multirow{2}{*}{97.13} \\ 
$\ket{ 0,    2,  2,   1.5} \to \ket{  3/2^-,  2,  1.5}$ &   &  & 0.74 & & 0.65 &  &   &  & \\ 

\hline\hline

$\ket{ 0,    2,  2,   2.5} \to \ket{  3/2^-,  2,  2.5}$ & \multirow{2}{*}{83.61}  & \multirow{2}{*}{1.54} & 0.22 & \multirow{2}{*}{0.22} & 0.72 & \multirow{2}{*}{1.39}  & \multirow{2}{*}{2.08}  & \multirow{4}{*}{1.50} & \multirow{4}{*}{84.89} \\

$\ket{ 0,    2,  2,   2.5} \to \ket{  3/2^-,  1,  1.5}$ &    &   & --  &  & 0.67 &   &   &   &   \\

\cline{1-8}
$\ket{ 0,    1,  1,   1.5} \to \ket{  3/2^-,  2,  2.5}$ & \multirow{2}{*}{86.30} & \multirow{2}{*}{1.90} & 0.52 &\multirow{2}{*}{0.71} & 0.41 & \multirow{2}{*}{0.78} & \multirow{2}{*}{2.17} &   &   \\

$\ket{ 0,    1,  1,   1.5} \to \ket{  3/2^-,  1,  1.5}$ &    &   & 0.48 &  & 0.37 &    &   &   &   \\

\hline\hline
$\ket{ 0,    2,  2,   1.5} \to \ket{  3/2^-,  2,  1.5}$ & \multirow{2}{*}{88.12}  & \multirow{2}{*}{1.51} & 0.74 &\multirow{2}{*}{0.96} & 0.65 & \multirow{2}{*}{1.24} & \multirow{2}{*}{2.17}  & \multirow{4}{*}{1.57} & \multirow{4}{*}{88.29} \\

$\ket{ 0,    2,  2,   1.5} \to \ket{  3/2^-,  1,  0.5}$ &  &   & 0.61 & & 0.59 &   &   &  &   \\

\cline{1-8}
$\ket{ 0,    1,  1,   0.5} \to \ket{  3/2^-,  2,  1.5}$ & \multirow{2}{*}{88.49}  & \multirow{2}{*}{2.11} & --  & \multirow{2}{*}{0.7} & 0.29 & \multirow{2}{*}{0.53} & \multirow{2}{*}{2.29} &  &   \\

$\ket{ 0,    1,  1,   0.5} \to \ket{  3/2^-,  1,  0.5}$ &   &  & 0.70 &   & 0.24 &  &   &  &   \\
\hline\hline

$\ket{ 0,    1,  1,   0.5} \to \ket{  3/2^-,  1,  0.5}$ & \multirow{2}{*}{50.25}  & \multirow{2}{*}{0.84} & 0.7 & \multirow{2}{*}{0.7} & 0.24 & \multirow{2}{*}{0.46} & \multirow{2}{*}{1.18} & \multirow{2}{*}{1.18} & \multirow{2}{*}{50.25} \\

$\ket{ 0,    1,  1,   0.5} \to \ket{  3/2^-,  0,  0.5}$ &   &  & -- & & 0.22 &  &  &  &  \\
\hline\hline

$\ket{ 0,    1,  1,   1.5} \to \ket{  3/2^-,  1,  1.5}$ & \multirow{2}{*}{49.48}  & \multirow{2}{*}{1.41} & 0.48 &\multirow{2}{*}{0.48} & 0.37 & \multirow{2}{*}{0.69} & \multirow{2}{*}{1.64} & \multirow{2}{*}{1.64} & \multirow{2}{*}{49.48} \\ 

$\ket{ 0,    1,  1,   1.5} \to \ket{  3/2^-,  0,  0.5}$ &   &   & --  &  & 0.33 &   &   &   &   \\ 
\hline\hline

$\ket{ 0,    2,  2,   2.5} \to \ket{  1/2^-,  2,  2.5}$ & \multirow{2}{*}{560.22}  & \multirow{2}{*}{5.01} & -- & \multirow{2}{*}{--} & 1.33 & \multirow{2}{*}{2.6} & \multirow{2}{*}{5.64} & \multirow{2}{*}{5.64} & \multirow{2}{*}{560.22} \\

$\ket{ 0,    2,  2,   2.5} \to \ket{  1/2^-,  1,  1.5}$ &   &   & --  &  & 1.28 &  &   &   &   \\
\hline\hline

$\ket{ 0,    2,  2,   1.5} \to \ket{  1/2^-,  2,  1.5}$ & \multirow{2}{*}{537.94}  & \multirow{2}{*}{3.82} & -- & \multirow{2}{*}{--} & 1.27  & \multirow{2}{*}{2.47} & \multirow{2}{*}{4.55} & \multirow{2}{*}{4.55} & \multirow{2}{*}{537.94} \\ 

$\ket{ 0,    2,  2,   1.5} \to \ket{  1/2^-,  1,  0.5}$ &  &   &  -- &  & 1.20 &   &   &   &   \\ 
%\hline\hline
\noalign{\hrule height 0.4mm}

\end{NiceTabular}
\caption{List of all transition pairs contributing to extraction of hyperfine Hamiltonian parameters in $^{135}$BaF. In all cases, a common initial state connects to two different excited states. Here, $\Delta$ is splitting between the two excited states; $\sigma_{\rm st}$ is the purely statistical uncertainty in $\Delta$, after rescaling to account for the global value of $\chi^2_\nu$; $\sigma_{\rm f}$ is the uncertainty in each line position associated with unresolved overlapping lines and $\sigma_{\rm f}^\Delta$ is the associated uncertainty in $\Delta$, obtained by adding contributions for each line in quadrature; $\sigma_{\rm Z}$ is the uncertainty in each line position associated with Zeeman shifts in the uncalibrated laboratory magnetic field and $\sigma_{\rm Z}^\Delta$ is the associated uncertainty in $\Delta$, obtained by adding contributions for each line linearly (since they are correlated); $\sigma_{\rm tot}$ is the total uncertainty in $\Delta$ for each transition pair, obtained by adding $\sigma_{\rm st}$, $\sigma_{\rm Z}^\Delta$, and $\sigma_{\rm Z}^\Delta$ in quadrature; $\sigma_{\rm tot}^{\rm wt}$ is the total uncertainty in $\Delta$ from a weighted average over all transition pairs that probed the same pair of excited states; and $\Delta^{\rm wt}$ is the weighted average value of $\Delta$, used in the final global fit for hyperfine parameters.}
\label{tab:Overall_135}
\end{table*}
\end{flushleft}

%For 137 %%%%%%%%%%%%%%%%%%%%%%%%%%%
%%%%%%%%%%%%%%%%%%%%%%%%%%%%%%%%%%%%

\begin{flushleft}
\begin{table*}[tb]
\centering
%\begin{NiceTabular}{|c|c|c|c|c|c|c|c|c|c|}
\begin{NiceTabular}{cccccccccc}
\noalign{\hrule height 0.4mm \vspace{-7pt}}
~\\
Transition in $^{137}$BaF & $\Delta$ & $\sigma_{\rm st}$ & $\sigma_{\rm f}$ &$\sigma_{\rm f}^{\Delta}$ & $\sigma_{\rm Z}$ & $\sigma_{\rm Z}^{\Delta}$ & $\sigma_{\rm tot}$ & $\sigma_{\rm tot}^{\rm wt}$ & $\Delta^{\rm wt}$ \\ 
$X\ket{N,G,F_1,F}\to A\ket{J'^P,F'_1,F'}$ & (MHz) & (MHz)& (MHz) &(MHz) & (MHz) & (MHz) & (MHz) & (MHz) & (MHz) \\ 

\hline\hline
$\ket{ 0,    2,  2,   2.5} \to \ket{3/2^-, 3,  3.5}$ & \multirow[c]{2}{*}{101.59}  & \multirow{2}{*}{1.11}  & 
0.44 & \multirow{2}{*}{0.51} &  0.76 & \multirow{2}{*}{1.48} & \multirow{2}{*}{1.92} &\multirow{2}{*}{1.92} & \multirow{2}{*}{101.59} \\
$\ket{ 0,    2,  2,   2.5} \to \ket{  3/2^-,  2,  2.5}$ &  &  & 0.25 & & 0.72 &  &   &  &  \\

\hline\hline

$\ket{ 0,    2,  2,   1.5} \to \ket{  3/2^-,  3,  2.5}$ & \multirow{2}{*}{104.67} & \multirow{2}{*}{1.07} & --  & \multirow{2}{*}{0.5} & 0.70  & \multirow{2}{*}{1.34} & \multirow{2}{*}{1.78} & \multirow{2}{*}{1.78} & \multirow{2}{*}{104.67} \\ 
$\ket{ 0,    2,  2,   1.5} \to \ket{  3/2^-,  2,  1.5}$ &   &  & 0.5 & & 0.65 &  &   &  & \\ 

\hline\hline

$\ket{ 0,    2,  2,   2.5} \to \ket{  3/2^-,  2,  2.5}$ & \multirow{2}{*}{102.17}  & \multirow{2}{*}{0.93} & 0.25 & \multirow{2}{*}{0.25} & 0.72 & \multirow{2}{*}{1.39}  & \multirow{2}{*}{1.69}  & \multirow{4}{*}{1.14} & \multirow{4}{*}{101.60} \\

$\ket{ 0,    2,  2,   2.5} \to \ket{  3/2^-,  1,  1.5}$ &     &   & --  &  & 0.67 &   &   &   &   \\

\cline{1-8}
$\ket{ 0,    1,  1,   1.5} \to \ket{  3/2^-,  2,  2.5}$ & \multirow{2}{*}{101.13} & \multirow{2}{*}{1.15} & 0.48 &\multirow{2}{*}{0.66} & 0.41 & \multirow{2}{*}{0.78} & \multirow{2}{*}{1.54} &   &   \\

$\ket{ 0,    1,  1,   1.5} \to \ket{  3/2^-,  1,  1.5}$ &     &   & 0.45 &  & 0.37 &    &   &   &   \\

\hline\hline
$\ket{ 0,    2,  2,   1.5} \to \ket{  3/2^-,  2,  1.5}$ & \multirow{2}{*}{103.34}  & \multirow{2}{*}{0.88} & 0.5 &\multirow{2}{*}{0.66} & 0.65 & \multirow{2}{*}{1.24} & \multirow{2}{*}{1.66}  & \multirow{4}{*}{1.02} & \multirow{4}{*}{103.68} \\

$\ket{ 0,    2,  2,   1.5} \to \ket{  3/2^-,  1,  0.5}$ &    &   & 0.43 & & 0.59 &   &   &  &   \\

\cline{1-8}
$\ket{ 0,    1,  1,   0.5} \to \ket{  3/2^-,  2,  1.5}$ & \multirow{2}{*}{103.89}  & \multirow{2}{*}{0.97} & --  & \multirow{2}{*}{0.67} & 0.29 & \multirow{2}{*}{0.52} & \multirow{2}{*}{1.29} &  &   \\

$\ket{ 0,    1,  1,   0.5} \to \ket{  3/2^-,  1,  0.5}$ &    &  & 0.67 &   & 0.23 &  &   &  &   \\
\hline\hline

$\ket{ 0,    1,  1,   0.5} \to \ket{  3/2^-,  1,  0.5}$ & \multirow{2}{*}{61.57}  & \multirow{2}{*}{0.66} & 0.67 & \multirow{2}{*}{0.67} & 0.23 & \multirow{2}{*}{0.44} & \multirow{2}{*}{1.04} & \multirow{2}{*}{1.04} & \multirow{2}{*}{61.57} \\

$\ket{ 0,    1,  1,   0.5} \to \ket{  3/2^-,  0,  0.5}$ &   &  & -- & & 0.21 &  &  &  &  \\
\hline\hline

$\ket{ 0,    1,  1,   1.5} \to \ket{  3/2^-,  1,  1.5}$ & \multirow{2}{*}{60.97} & \multirow{2}{*}{0.41} & 0.45 &\multirow{2}{*}{0.45} & 0.37 & \multirow{2}{*}{0.69} & \multirow{2}{*}{0.91} & \multirow{2}{*}{0.91} & \multirow{2}{*}{60.97} \\ 

$\ket{ 0,    1,  1,   1.5} \to \ket{  3/2^-,  0,  0.5}$ &    &   & --  &  & 0.32 &   &   &   &   \\ 
\hline\hline

$\ket{ 0,    2,  2,   2.5} \to \ket{  1/2^-,  2,  2.5}$ & \multirow{2}{*}{625.38} & \multirow{2}{*}{4.76} & -- & \multirow{2}{*}{--} & 1.46 & \multirow{2}{*}{2.86} & \multirow{2}{*}{5.55} & \multirow{2}{*}{5.55} & \multirow{2}{*}{625.38} \\

$\ket{ 0,    2,  2,   2.5} \to \ket{  1/2^-,  1,  1.5}$ &    &   & --  &  & 1.41 &  &   &   &   \\
\hline\hline

$\ket{ 0,    2,  2,   1.5} \to \ket{  1/2^-,  2,  1.5}$ & \multirow{2}{*}{602.78} & \multirow{2}{*}{4.19} & -- & \multirow{2}{*}{--} & 1.39  & \multirow{2}{*}{2.72} & \multirow{2}{*}{4.99} & \multirow{2}{*}{4.99} & \multirow{2}{*}{602.78} \\ 

$\ket{ 0,    2,  2,   1.5} \to \ket{  1/2^-,  1,  0.5}$ &    &   &  -- &  & 1.33 &   &   &   &   \\ 
\hline\hline

$\ket{ 1,    2,  3,   2.5} \to \ket{  1/2^+,  2,  1.5}$ & \multirow{2}{*}{74.41}  & \multirow{2}{*}{2.24} & --  & \multirow{2}{*}{--} & 0.72 & \multirow{2}{*}{1.38} & \multirow{2}{*}{2.63} & \multirow{4}{*}{1.50} & \multirow{4}{*}{74.25} \\

$\ket{ 1,    2,  3,   2.5} \to \ket{  1/2^+,  1,  1.5}$ &    &   & --  &  &  0.66 &   &   &   \\

\cline{1-8}
$\ket{ 1,    1,  2,   1.5} \to \ket{  1/2^+,  2,  1.5}$ & \multirow{2}{*}{74.18}  & \multirow{2}{*}{1.16} & 0.45 & \multirow{2}{*}{1.32} & 0.28 & \multirow{2}{*}{0.51} & \multirow{2}{*}{1.83} &   &   \\

$\ket{ 1,    1,  2,   1.5} \to \ket{  1/2^+,  1,  1.5}$ &  &   & 1.24 &  & 0.23 &   &   &   &   \\
\hline\hline

$\ket{ 1,    2,  2,   2.5} \to \ket{  1/2^+,  2,  2.5}$ & \multirow{2}{*}{61.8} & \multirow{2}{*}{1.40} & --  & \multirow{2}{*}{--} & 0.62 & \multirow{2}{*}{1.16}  & \multirow{2}{*}{1.82}  & \multirow{4}{*}{1.24} & \multirow{4}{*}{62.35} \\ 

$\ket{ 1,    2,  2,   2.5} \to \ket{  1/2^+,  1,  1.5}$ &    &   & --  &  & 0.54 &   &   &   &   \\ 

\cline{1-8}
$\ket{ 1,    1,  2,   2.5} \to \ket{  1/2^+,  2,  2.5}$ & \multirow{2}{*}{62.83} & \multirow{2}{*}{1.47} & --  & \multirow{2}{*}{0.25} & 0.44  & \multirow{2}{*}{0.80} & \multirow{2}{*}{1.69} &  & \\ 

$\ket{ 1,    1,  2,   2.5} \to \ket{  1/2^+,  1,  1.5}$ &   &   & 0.25 &  & 0.36 &  &   &   &   \\

\hline\hline
$\ket{ 1,    2,  1,   1.5} \to \ket{1/2^+, 2,  2.5}$ & \multirow[c]{2}{*}{68.48} & \multirow{2}{*}{1.32}  & 
-- & \multirow{2}{*}{--} &  0.52 & \multirow{2}{*}{0.95} & \multirow{2}{*}{1.63} &\multirow{2}{*}{1.63} & \multirow{2}{*}{68.48} \\
$\ket{ 1,    2,  1,   1.5} \to \ket{  1/2^+,  1,  0.5}$ & &  & -- & & 0.43 &  &   &  &  \\

\noalign{\hrule height 0.4mm}

\end{NiceTabular}

\caption{List of all transition pairs contributing to extraction of hyperfine Hamiltonian parameters in $^{137}$BaF. All notations are the same as in Tab.~\ref{tab:Overall_135}.}
\label{tab:Overall_137}
\end{table*}
\end{flushleft}

%\end{widetext}

\subsection{Analysis of the absorption spectroscopy data}
The analysis of absorption spectroscopy data has been discussed in detail in previous work~\cite{Albrecht2020,Rockenhaeuser2023}. Here, we follow the same procedures, with adaptations as needed for the detection of low-abundance isotopologues.

\subsubsection{Data acquisition}
As molecular formation is based on a pulsed ablation process, a continuous scan of the probe laser --- as common in room-temperature vapor cell experiments --- is not possible, and thus an incremental frequency scan is used.
For each frequency step, a time-resolved absorption trace is recorded, the signal is integrated over specific time intervals, and averaged over typically four repetitions to reduce signal fluctuations~\cite{Albrecht2020}. By combining these data and applying a moving average filter for smoothing, we obtain distinct spectra across multiple time intervals, with later times corresponding to well-thermalized molecules. We then choose the latest time interval for which the signal-to-noise ratio of the decaying absorption signal still allows for detecting all smaller pronounced peaks, which are later associated with lower-abundance isotopologues. 

For frequency ranges that span more than a few \si{\giga\hertz}, a single scan is not feasible due to the limited time window during which the dewar-style cryostat employed for the absorption spectroscopy measurements can maintain cold temperatures, the restricted mode-hop-free tuning range of the laser, and signal degradation caused by prolonged ablation at a fixed target position. To overcome these limitations, successive scans were performed over smaller, slightly overlapping frequency ranges. These segments are later stitched together, accounting for variations in the molecular signal arising from target degradation.

The laser frequency is recorded using a wavemeter whose frequency is regularly validated against cesium and rubidium vapor lines, the BaF cooling transition~\cite{Rockenhaeuser2024}, and a frequency-stabilized HeNe laser \cite{Rockenhaeuser2023}, achieving an absolute accuracy of better than \SI{50}{\mega\hertz}, smaller than the uncertainty arising from the inherent molecular dynamics in the buffer gas cell.

\subsubsection{Fitting molecular constants}
The determination of the rovibrational constants begins with fixing the well-established constants of the bosonic isotopologues \BaFeight\ and \BaFsix\ with a simple structure and clearly pronounced transition lines~\cite{Rockenhaeuser2023}. In the next step, these constants are scaled for the less abundant isotopologues (\BaFseven, \BaFfive\ and \BaFfour) and combined with the hyperfine constants of the excited state \exs\ extracted from fluorescence spectroscopy. 

Typically, ten to twenty prominent peaks in the measured spectra can be unambiguously matched to those from the theoretical model. For each isotopologue and each vibrational transition \gs$(\nu) \rightarrow$ \exs$(\nu')$, we then determine an individual energy offset $T_{\nu',\nu}=T'_\nu - T_\nu$ by minimizing the least-squared error between the peak positions between these experimental and theoretical transitions.
Finally, applying these offsets in the relation $T_\nu=T_e +\omega_e(\nu+1/2) -\omega_e\chi_e(\nu+1/2)^2+\omega_e y_e(\nu+1/2)^3$, the vibrational constants $\omega_e,\omega'_e$ and $T'_e$ are extracted using least-square fits, while the remaining constants are fixed from isotope scalings, as summarized in Tab.~\ref{table:constants}.

\end{document}